\RequirePackage{fix-cm}
\documentclass[twocolumn,epjc3]{svjour3}  
\smartqed  
\RequirePackage{graphicx}
\journalname{Eur. Phys. J. C}

\pdfoutput=1

\RequirePackage{xspace}
\RequirePackage{amssymb}
\RequirePackage{url}
\RequirePackage{siunitx}
\RequirePackage{physics} 
\RequirePackage{lscape} 
\RequirePackage{mhchem} 
\RequirePackage[inline]{enumitem} 
\RequirePackage{amssymb}
\RequirePackage{color}
\RequirePackage{lineno}
\usepackage{gensymb}

\newcommand{\cygno}{{\textsc{Cygno}}\xspace}
\newcommand{\idbscan}{{\textsc{Idbscan}}\xspace}

\newcommand{\Et}  {E$_{\mathrm{Transf}}$}
\newcommand{\Ed}  {E$_{\mathrm{Drift}}$}

\newcommand{\Vg}  {V$_{\mathrm{GEM}}$}

\newcommand{\fe} {\ce{^{55}Fe}\xspace}
\newcommand{\Isc} {$\SI{}{I_{SC}}$\xspace}
\newcommand{\St} {$\sigma_\mathrm{T}$\xspace}
\newcommand{\Sl} {$\sigma_\mathrm{L}$\xspace}
\newcommand{\Erec} {\SI{}{E_{rec}}\xspace}

\begin{document}

\title{A 50 liter \cygno prototype overground characterization 
}


\author{
\mbox{Fernando Domingues Amaro\thanksref{1}}
           \and
\mbox{Rita Antonietti\thanksref{12,13}}
           \and
\mbox{Elisabetta Baracchini \thanksref{2,3}}
           \and
\mbox{Luigi  Benussi \thanksref{4}}
           \and
\mbox{Stefano Bianco \thanksref{4}}
           \and
\mbox{Francesco Borra \thanksref{7,8}}
           \and
\mbox{Cesidio Capoccia \thanksref{4}}
           \and
\mbox{Michele Caponero \thanksref{4,5}}
           \and
\mbox{Danilo Santos Cardoso \thanksref{6}}
           \and
\mbox{Gianluca Cavoto \thanksref{7,8,e}}
           \and
\mbox{Igor Abritta Costa \thanksref{12,13}}
           \and
\mbox{Emiliano Dan\'e \thanksref{4}}
           \and
\mbox{Giorgio Dho \thanksref{2,3}}
           \and
\mbox{Flaminia Di Giambattista \thanksref{2,3}}
           \and
\mbox{Emanuele Di Marco \thanksref{7}}
           \and
\mbox{Giulia D'Imperio \thanksref{7}}
           \and
\mbox{Joaquim Marques Ferreira dos Santos \thanksref{1}}
           \and
\mbox{Giovanni Grilli di Cortona \thanksref{4}}
           \and
\mbox{Francesco Iacoangeli \thanksref{7}}
           \and
\mbox{Herman Pessoa Lima J\'unior \thanksref{6}}
           \and
\mbox{Guilherme Sebastiao Pinheiro Lopes \thanksref{9}}
           \and
\mbox{Amaro da Silva Lopes Júnior \thanksref{9}}
           \and
\mbox{Giovanni Maccarrone \thanksref{4}}
           \and
\mbox{Rui Daniel Passos Mano \thanksref{1}}
           \and
\mbox{Robert Renz Marcelo Gregorio \thanksref{11}}
           \and
\mbox{David Jos\'e Gaspar  Marques \thanksref{2,3}}
           \and
\mbox{Giovanni Mazzitelli \thanksref{4}}
           \and
\mbox{Alasdair Gregor McLean \thanksref{11}}
           \and
\mbox{Pietro Meloni\thanksref{12,13}}
           \and
\mbox{Andrea Messina \thanksref{7,8}}
           \and
\mbox{Cristina Maria Bernardes Monteiro \thanksref{1}}
           \and
\mbox{Rafael Antunes Nobrega \thanksref{9}}
           \and
\mbox{Igor Fonseca Pains \thanksref{9}}
           \and
\mbox{Emiliano Paoletti \thanksref{4}}
           \and
\mbox{Luciano Passamonti \thanksref{4}}
           \and
\mbox{Sandro Pelosi \thanksref{7}}
           \and
\mbox{Fabrizio Petrucci \thanksref{12,13}}
           \and
\mbox{Stefano Piacentini \thanksref{7,8}}
           \and
\mbox{Davide Piccolo \thanksref{4}}
           \and
\mbox{Daniele Pierluigi \thanksref{4}}
           \and
\mbox{Davide Pinci \thanksref{7}}
           \and
\mbox{Atul Prajapati \thanksref{2,3}}
           \and
\mbox{Francesco Renga \thanksref{7}}
           \and
\mbox{Rita Cruz Roque \thanksref{1}}
           \and
\mbox{Filippo Rosatelli \thanksref{4}}
           \and
\mbox{Alessandro Russo \thanksref{4}}
           \and
\mbox{Giovanna Saviano \thanksref{4,14}}
           \and
\mbox{Neil John Curwen Spooner \thanksref{11}}
           \and
\mbox{Roberto Tesauro \thanksref{4}}
           \and
\mbox{Sandro Tomassini \thanksref{4}}
           \and
\mbox{Samuele Torelli \thanksref{2,3} 
           \and
\mbox{Donatella Tozzi \thanksref{7,8}}
}
}

\thankstext{e}{e-mail: gianluca.cavoto@roma1.infn.it}


\institute{
LIBPhys, Department of Physics, University of Coimbra, 3004-516 Coimbra, Portugal; \label{1} 
           \and
Istituto Nazionale di Fisica Nucleare, Sezione di Roma TRE, 00146, Roma, Italy; \label{12}
            \and
 Dipartimento di Matematica e Fisica, Universit\`a Roma TRE, 00146, Roma, Italy; \label{13}
            \and
Gran Sasso Science Institute, 67100, L'Aquila, Italy; \label{2}
           \and
 Istituto Nazionale di Fisica Nucleare, Laboratori Nazionali del Gran Sasso, 67100, Assergi, Italy; \label{3}
            \and
 Istituto Nazionale di Fisica Nucleare, Laboratori Nazionali  di Frascati,  00044, Frascati, Italy; \label{4}
           \and
  Istituto Nazionale di Fisica Nucleare, Sezione di Roma, 00185, Rome, Italy; \label{7}
           \and
  Dipartimento di Fisica, Sapienza Universit\`a di Roma, 00185, Roma, Italy; \label{8}
           \and
 ENEA Centro Ricerche Frascati, 00044, Frascati, Italy; \label{5}
           \and
Centro Brasileiro de Pesquisas Físicas, Rio de Janeiro 22290-180, RJ, Brazil; \label{6}
           \and
  Universidade Federal de Juiz de Fora, Faculdade de Engenharia, 36036-900, Juiz de Fora, MG, Brasil; \label{9}
           \and
 Department of Physics and Astronomy, University of Sheffield, Sheffield, S3 7RH, UK; \label{11}
 \and
 Dipartimento di Ingegneria Chimica, Materiali e Ambiente, Sapienza Universit\`a di Roma, 00185, Roma, Italy; \label{14}
}

\date{Received: date / Accepted: date}

\date{Internal v1: 2023-03-02}

\maketitle

\begin{abstract}
The nature of dark matter is still unknown and an experimental program
to look for dark matter particles in our Galaxy should extend its
sensitivity to light particles in the GeV mass range and exploit the
directional information of the DM particle motion
\cite{Vahsen:2020pzb}.  The \cygno project is studying a gaseous time
projection chamber operated at atmospheric pressure with a Gas
Electron Multiplier \cite{Sauli:1997qp} amplification and with an
optical readout as a promising technology for light dark matter and
directional searches.

In this paper we describe the operation of a 50 liter prototype named
LIME (Long Imaging ModulE) in an overground location at Laboratori
Nazionali di Frascati of INFN. This prototype employs the technology
under study for the 1 cubic meter \cygno demonstrator to be installed
at the Laboratori Nazionali del Gran Sasso
\cite{instruments6010006}. We report the characterization of LIME with
photon sources in the energy range from few keV to several tens of keV
to understand the performance of the energy reconstruction of the
emitted electron. We achieved a low energy threshold of few keV and an
energy resolution over the whole energy range of 10-20\%, while
operating the detector for several weeks continuously with very high
operational efficiency. The energy spectrum of the reconstructed
electrons is then reported and will be the basis to identify
radio-contaminants of the LIME materials to be removed for future
\cygno detectors.  \keywords{ dark matter  \and time projection chamber \and
  optical readout } \PACS{PACS  95.35.+d  \and PACS 29.40.Cs  \and 29.40.Gx}

\end{abstract}

\section{Introduction }
\label{sect:intro}

A number of astrophysical and cosmological observations are all
consistent with the presence in the Universe of a large amount of
matter with a very weak interaction with ordinary matter besides the
gravitational force, universally known as Dark Matter (DM). The model
of the Weakly Interacting Massive Particle (WIMP)) has been very
popular in the last decades, predicting a possible DM candidate
produced thermally at an early stage of the Universe with a mass in
the range of 10 to 1000 GeV and a cross section of elastic scattering
with standard matter at the level of that of the weak interactions
\cite{bertone2005particle} \cite{RevModPhys.90.045002}.  Hypothetical
particles of DM would also fill our Galaxy forming a halo of particles
whose density profile is derived from the observed velocity
distribution of stars in the Galaxy. This prediction calls for an
experimental program for finding such DM particles with terrestrial
experiments.  These experiments aim at detecting the scattering of the
elusive DM particle on the atoms of the detectors, inducing as
experimental signature a nucleus or an electron to recoil against the
impinging DM particle. Nowadays most of these experimental activities
are based on ton (or multi-ton) mass detectors where scintillation
light, ionization charge, or heat induced by the recoiling particles
are used - sometime in combination - to detect the recoils
\cite{XENON:2018voc,aalbers2022dark,Bernabei:2013xsa,LUX:2016ggv,PandaX-II:2017hlx}.
 
  Most of these experiments however are largely unable to infer the
  direction of motion of the impinging DM particle. While DM particles
  have a random direction in the Galaxy reference system. on the Earth
  a DM particle would be seen as moving along the direction of motion
  of the Earth in the Galaxy. This motion is given by the composition
  of the motion of the Sun toward the Cygnus constellation and the
  revolution and rotation of the Earth. This is then reflected into
  the average direction of motion of the recoiling particles after the
  DM scattering and it can represent an important signature to be
  exploited to discriminate the signal of a DM particle from other
  background sources \cite{MAYET20161}. Therefore this undoubtedly
  calls for a new class of detectors based on the reconstruction of
  the the recoil direction, such as the gaseous time projection
  chamber (TPC)
  \cite{Battat:2014van,Battat:2016pap,Battat:2016xxe,BATTAT20146,Daw:2013waa,Battat:2015rna,bib:loomba55Fe,JINST:nitec,Ikeda:2020mvr,Riffard:2016mgw,Sauzet:2020dut,Hashimoto:2017hlz,BATTAT20151,ALNER2005173,bib:vahsen}.
  Moreover, while the WIMP model for DM candidates has been tested
  thoroughly by the current detectors down to \SI{10}{GeV}, extensions
  of sensitivity of these detectors to lower masses - down to the GeV
  and below - are deemed fundamental to explore new models predicting
  lighter DM particles
  \cite{petraki2013review,Zurek_2014,PhysRevLett.115.021301}. For this
  scope \cygno proposes the use of light atoms as Helium or Hydrogen
  as target for DM. For a DM in the range of 1 to \SI{10}{GeV} mass
  the elastic scattering of DM particle on these nuclei is producing
  nuclear recoils with the most favourable kinetic energy.
  
  In this respect the \cygno project aims to realize an R\&D program
  to demonstrate the feasibility of a DM search based on gaseous TPC
  at atmospheric pressure. The \cygno TPC will use a He/CF$_4$ gas
  mixture featuring a GEM amplification and with an optical readout of
  the light emitted at the GEM amplification stage
  \cite{FRAGA200388,bib:Fraga} as outlined in
  \cite{instruments6010006}.  Gaseous TPC based on optical readout to
  search for DM were proposed and studied before but with the use of a
  gas pressure well below the atmospheric one (DM-TPC,
  \cite{Battat:2014mka,Ahlen:2010ub,DUJMIC2008327,PhysRevD.95.122002}).
  The \cygno project aims to build a 30--100\,\SI{}{m^3} detector that
  would therefore host a larger target mass than a low pressure
  TPC. Given the presence of fluorine nuclei in the gas mixture \cygno
  would be especially sensitive to a scattering of DM that is
  sensitive to the spin of the nucleus. By profiting of the background
  rejection power of the directionality, competitive limits on the
  presence of DM in the Galaxy can be set, under the assumption of a
  spin dependent coupling of DM with matter.
 
 After a series of explorative small size prototypes
 \cite{NIM:Marafinietal,bib:jinst_orange1,bib:ieee_orange,bib:nim_orange2,bib:jinst_orange2,Pinci:2019hhw,Costa:2019tnu,bib:fe55,bib:stab,baracchini2019cygno}
 proving the principle of detecting electron and nuclear recoils down
 to keV kinetic energy, a staged approach is now foreseen to build a
 detector sensitive to DM induced recoils.
 
 A first step requires the demonstration that all the technological
 choices of the detector are viable. Before the construction a
 \SI{1}{m^3} demonstrator of a DM \cygno-type detector, a 50 liter
 prototype - named LIME (Long Imaging ModulE) - has been built and
 operated in an overground laboratory at the Laboratori Nazionali di
 Frascati (LNF) of INFN.  LIME is featuring a \SI{50}{cm} long drift
 volume
 with the amplification realized with a triple GEM system and the
 light produced in the avalanches readout with a scientific CMOS
 camera and four PMT. A \cygno-type detector will be modular with LIME
 being a prototype for one of its modules.  Most of the materials and
 the detection elements used in LIME are not at the radiopurity level
 required for a real DM search. However they can be produced in a
 radiopure version, treated to become radiopure or replaced with
 radiopure materials without affecting the the mechanical feasibility
 and the detector performance of the \SI{1}{m^3} \cygno demonstrator.

 In this paper we summarize our experience with the LIME prototype
 operated during a long campaign of data-taking, conducted to
 primarily understand the long term operation stability, to collect
 data to develop image analysis techniques and to understand the
 particle energy reconstruction performance. These techniques are
 including the reconstruction of clusters of activated pixels due to
 light detection in the images, optical effects characterizations, and
 noise studies. They were mainly oriented to the detection of electron
 originated from the interaction of photons in the gas volume. We
 usually refer to these electrons as electron recoils.  The energy
 response of LIME was fully characterized in a range of few keV to
 tens of keV electron kinetic energy using different photon sources,
 while a $^{55}$Fe X-ray absorption length in the LIME gas mixture was
 also evaluated.

 Finally we report an analysis of the observed background events, induced by sources both internal to the detector and external, in the overground LNF location.




\section{The LIME prototype}
\label{sect:limedet}

\begin{figure*}[ht]
\centering
\vspace{-3cm}
\includegraphics[width=0.8\textwidth, angle=270]{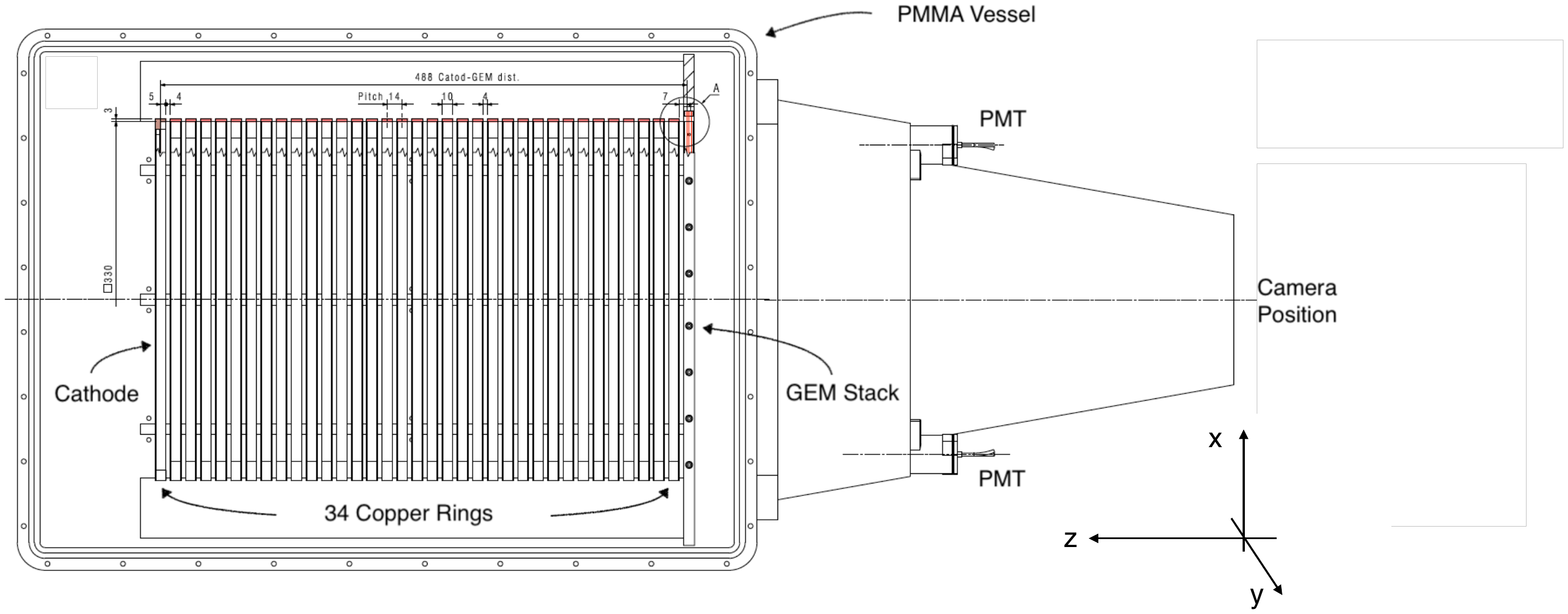}
\vspace{-2.5cm}
\caption{Drawing of  LIME as seen from above.  Square-shaped copper rings are used to create a  field cage closed on one side by the triple-GEM stack. The field cage is closed on the other side with respect to the GEM by a  cathode plane.  The position of the four photomultipliers and  of CMOS optical sensor are indicated. The acrylic gas vessel is enclosing the field cage and the GEM stack.} 
\label{fig:lime}
\end{figure*}

\begin{figure}[ht]
\centering
\includegraphics[width=0.4\textwidth]{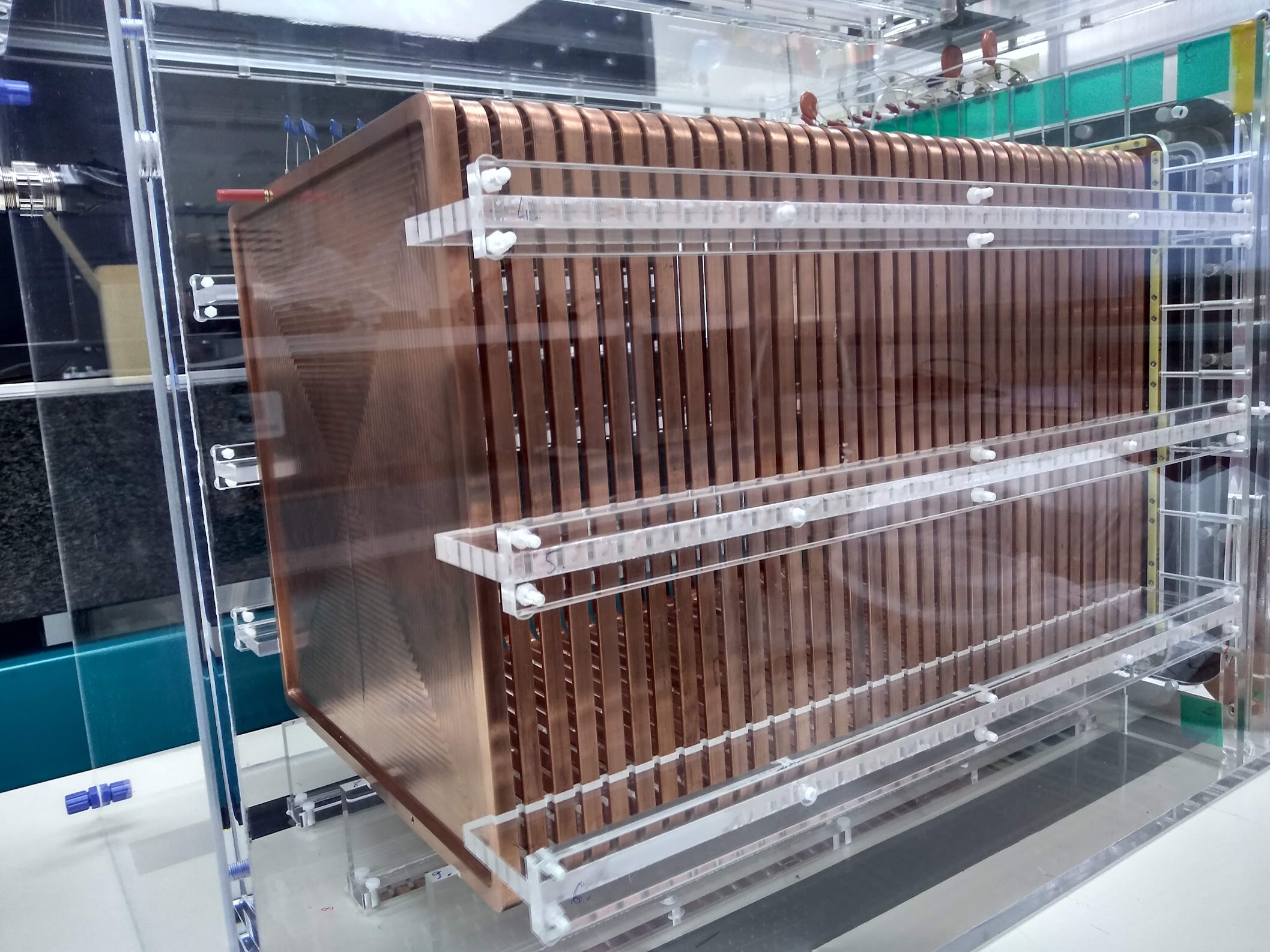}

\caption{LIME vessel:  the field cage  is clearly visible  with all its copper rings mounted on the  PMMA combs to support them and  with the cathode  to close the field region.} 
\label{fig:lime_pict}
\end{figure}

The LIME prototype (as shown in Fig.\ref{fig:lime} and in Fig.\ref{fig:lime_pict}) is composed of a transparent acrylic vessel inside which  the gas mixture is flowed with an over-pressure of about 3 mbar with respect to the external atmospheric pressure. Inside the gas vessel a series of copper rings  are used as electrodes kept at increasing  potential values  from the cathode to define a uniform  electric field directed orthogonal to the cathode plane. This field makes  the ionization electrons (produced by the charged particles in the gas)  to drift towards the anode. A cathode plane is used to define the lower potential of the electric field while on the opposite side a triple GEM stack system  is installed.
When the ionization electrons  reach the GEM, they produce an avalanche of secondary electrons and ions. Interactions of secondary electrons with gas molecules produce also photons whose spectrum and quantity strongly depends on the gas mixture \cite{bib:Fraga}. From the avalanche position  the  light is emitted  towards the exterior of the vessel. A scientific CMOS camera (more details in Sect.~\ref{sec:light_sen}) with a large field-of-view objective is used to collect this light  over a  integration  time that can be set from \SI{30}{ms} to \SI{10}{s} and to yield  an image of the GEM. Four PMT are installed around the camera to detect the same light but with a much faster response time. 
In the following we describe in details the elements of the LIME prototype. The sensitive part of the gas volume of LIME is about 50 liters with a 50 cm long electric field region closed by a 33$\times$33\,cm$^2$ triple-GEM stack. 

\subsection{The gas vessel and the field cage}
The gas vessel is realized with   a  10~mm thick  PMMA box with a total volume of about 100 litres that  is devoted to  contain  the gas mixture used in the operation. Inside the vessel  a field cage  produces a uniform electric field  to drift the primary ionization electrons originated  in  the interaction of charged particles with the gas molecules towards the amplification stage.
The volume is regularly  flushed at a flow rate of \SI{200}{cc/min}.
The field cage has  a square section, with a side of \SI{330}{mm}, a length of \SI{488}{mm}, and consists of:
\begin{itemize}
\item 34 square coils, \SI{10}{mm} wide, placed at a distance of \SI{4}{mm} from each other, with  an effective pitch of \SI{14}{mm} and electrically connected by \SI{100}{M\Omega} resistors;
\item a \SI{0.5}{mm} thin copper cathode with a frame identical in size to the coils described above;
\item a stack of 3 standard GEM (holes with an internal diameter of \SI{50}{\mu m} and pitch of \SI{140}{\mu m}, placed \SI{2}{mm} apart from each other and \SI{7}{mm} from the first coils. 
\end{itemize}

The detector is usually operated with  a He/CF$_4$ gas mixture in proportions of 60/40 kept few millibars above the atmospheric pressure. This is therefore equivalent to a  mass of 87 g in the active volume.

The upper face of the vessel includes   a \SI{5}{cm} wide and \SI{50}{cm} long thin window sealed by a \SI{150}{\mu m} thick ethylene-tetrafluoroethylene (ETFE) layer. This allows low energy photons (down to the keV energy) to enter the gas volume from external artificial   radioactive sources used for calibration purposes.  

 An externally controllable trolley is mounted on the window and  can be moved back and forth along a track. It functions as a source holder and allows to move a radioactive source, kept \SI{18}{cm} above the sensitive volume, along the $z$ axis from \SI{5}{cm} to \SI{45}{cm} far from the GEM.  
 On its base there is a \SI{5}{mm} diameter hole that allows the passage of a beam of photons by collimating it.

The face of the vessel in front of the GEM stack away from the sensitive volume is \SI{1}{mm} thick to allow efficient transmission of light to the outside.

\subsection{The light sensors}
\label{sec:light_sen}
On the same side of the vessel where the GEM stack is installed  a  black PMMA conical structure is fixed to allow the housing of the optical sensors: 
\begin{itemize}
    \item 4 Hamamatsu R7378, 22 mm diameter photo-multipliers;
   \item an Orca Fusion scientific CMOS-based camera   (more dentails on  \cite{hamamatsu}) with 2304~$\times$~2304 pixels with an active area of 6.5~$\times$~6.5~$\mu$m$^2$ each, equipped with a Schneider lens with \SI{25}{mm} focal length and 0.95 aperture at a distance of \SI{623}{mm}. The sCMOS sensor provides a quantum efficiency of about 80\% in the range \SI{450}{nm}-\SI{630}{nm}.
   In this configuration, the sensor faces a surface of $35~\times~35$~cm$^2$ and therefore each pixel at an area of $152 \times 152~\mu$m$^2$. 
    The geometrical acceptance $\epsilon_{\Omega}$ results to be $1.2 \times 10^{-4}$. 
\end{itemize}

According to previous studies \cite{bib:Fraga,bib:Margato1}, 
electro-luminescence spectra of He/CF$_4$ based mixtures show two main maxima: one around  a wavelength of 300~nm and one around  620~nm. This second  wavelength matches the range where the Fusion camera sensor provides thew maximum quantum efficiency. 

\subsection{The Faraday cage}

The entire detector is contained within a 3 mm thick aluminium metal box.  Equipped with feed-through connections for the high voltages required for the GEM, cathode and PMT and for the gas, this box acts as a Faraday cage and guarantees the light tightness of the detector.
A rod is  free to enter through a hole 
 from the rear face to allow  movement of the source holder.
On the front side  a square hole is present on which an optical bellows is mounted, which can then be coupled to the CMOS sensor lens.

\subsection{Data acquisition and trigger systems}
LIME data acquisition is realized  with an integrated system within the Midas framework \cite{midas}.

The PMT signals are sent into a discriminator and a logic module to produce a trigger signal based on a coincidence of the signals of at least two PMT.

A dedicated data acquisition PC is connected via two independent USB 3.0 ports to the camera and to a VME crate that houses I/O register modules for the trigger and controls.


The camera can be  operated with different exposure times. The results presented in this paper are obtained with a 50 ms exposure to minimize the pile-up from natural radioactivity events.




The DAQ system has been designed and built in such a way that it can also integrate digitisers for the acquisition of PMT signal waveforms. In this way, for each interaction in the gas, the light  produced in the GEM stack is simultaneously acquired by the high granularity CMOS sensor and by the four PMT. As it was demonstrated in \cite{bib:jinst_orange2} this will allow a 3D reconstruction of the event in the gas volume within the field cage.

In this paper we report the data analysis of the camera images only.

\subsection{High voltage and gas supply systems}

The gas mixture, obtained from cylinders of  pure gases, is continuously flushed into the detector at a rate of 200 cc/min  and the  output gas  is sent to an exhaust line connected to the external environment via a water filled bubbler ensuring the small (3 mbar) required overpressure. 
Electrical voltages at the various electrodes of the detector are supplied by two generators:
\begin{itemize}
\item an ISEG "HPn 500" provides up to 50 kV and 7 mA with negative polarity and ripple $<0.2\%$ directly to the cathode;
\item CAEN A1515TG board with Individual Floating Channels supplies the voltages (up to 1 kV with 20 mV precision) to the electrodes of the triple GEM stack 
\end{itemize} 

By means of these two suppliers, a constant electric field was generated in the sensitive volume with a standard value of  \Ed~=~0.9~kV/cm  and in the transfer gaps between the GEM (about \Et~= 2.5~kV/cm), while the voltage difference across the two sides of each GEM is  usually set to \Vg~=~440~V for all the three GEM.

\section{Overground run}
\label{sect:overgroundrun}

The measurements reported in this paper were realized at the INFN LNF  during the 2021 summer and autumn. The detector was operated inside an experimental hall where the temperature was varying in a range  between 295 K and 300 K and the atmospheric pressure between 970 and 1000 mbar for the entire duration of the measurements. The typical working conditions of the detector are reported in Table \ref{tab:parameter}.

\begin{table}[]
\begin{center}
\caption{Summary of the typical operating condition of LIME during the data takings.\label{tab:parameter}}
\vspace{1mm}
\begin{tabular}{|l|l|}
\hline
Parameter      & Typical value \\ \hline \hline
Drift Field    & 0.9 kV/cm      \\ \hline
GEM Voltage    & 440 V         \\ \hline
Transfer Field & 2.5 kV/cm     \\ \hline
Gas Flow       & 12 l/h       \\ \hline
PMT Threshold  & 15 mV         \\ \hline
\end{tabular}
\end{center}

\end{table}

\subsection{Instrumental effect studies  }

\label{sect:instreffects}

As a first study, we evaluated  the instrumental non-uniformity  due to the optics system and to the electronic sensor noise.

\subsubsection{Optical vignetting}
\label{sec:vignetting}

\begin{figure}[ht!]
\centering
\includegraphics[width=0.40\textwidth]{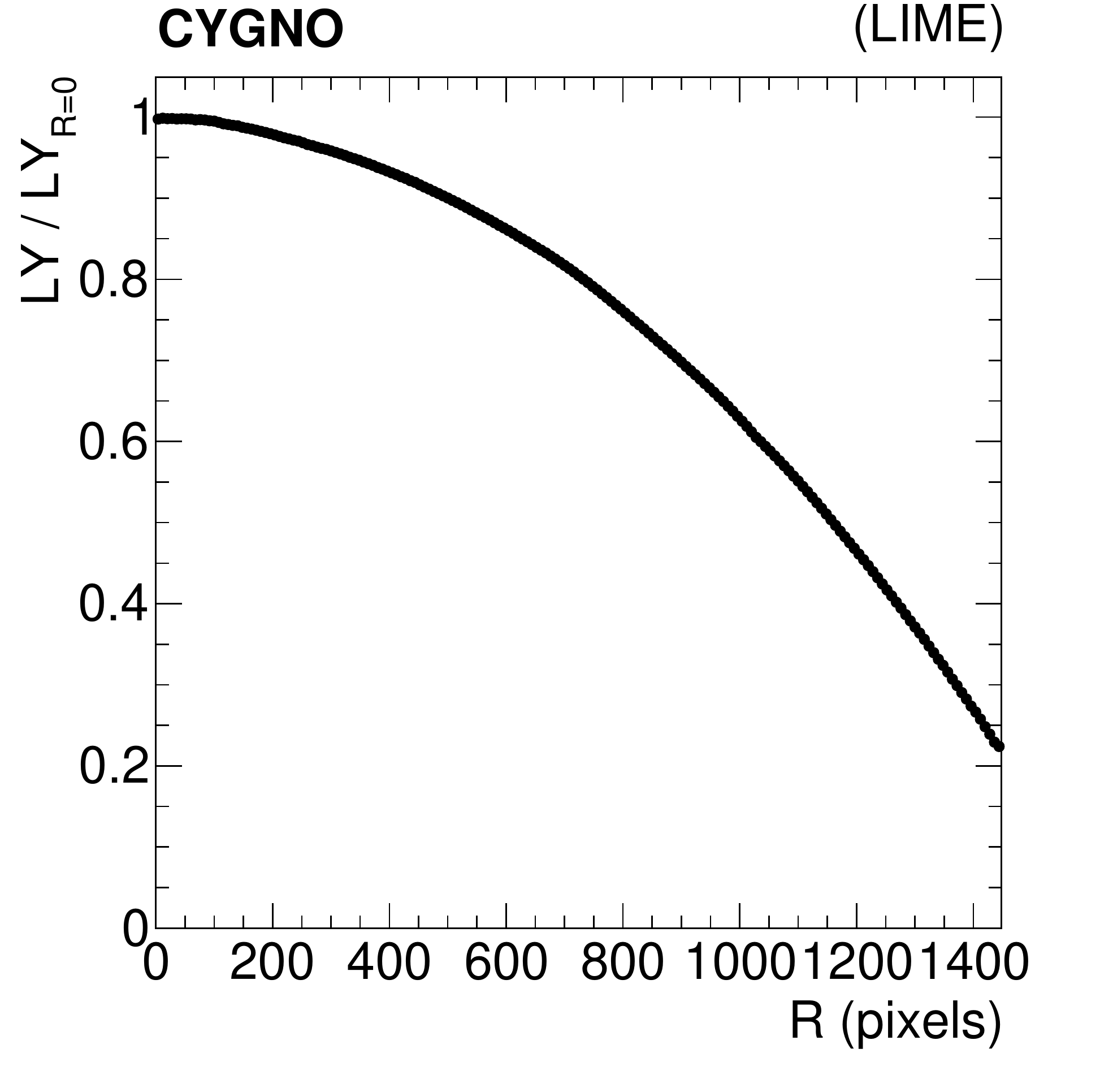}
\caption{Light yield measured as a function of the radial distance
  from the center of the sensor, normalized to the one at the center,
  using pictures of a uniformly illuminated white surface.
\label{fig:vignette1d}}
\end{figure}

With respect to the optics, we evaluated the effects of lens
vignetting, that is the reduction of detected light in the peripheral
region of an image compared to the image center.  For this purpose, we
collected with the same camera images of a uniformly illuminated white
surface. In order to avoid any possible preferiantial direction of the
light impinging the sensor, different images of the same surface are
acquired by rotating the camera around the lens optical axis, and we
obtained a light collection map on the sensor by their average. This
shows a drop of the collected light as a function of the radial
distance from the centre, down to 20\% with respect the center of the
image, as shown in Fig.~\ref{fig:vignette1d}.  The resulting map was
then used to correct all the images collected with the detector.

\subsubsection{Sensor electronic noise}
\label{sec:noise}

A second study consisted in the evaluation of the fluctuations of the
{\it dark offset} of the optical sensor.  These are mainly due to two
different contributions: {\it readout noise} i.e. the electronic noise
of the amplifiers onboard of each pixel (less than 0.7 electrons
r.m.s.) and a {\it dark current} that flows in each camera photo-diode
of about 0.5 electrons/pixel/s \cite{hamamatsu2}.  To obtain this,
dedicated runs were taken throughout the data taking period with the
values of \Vg\ set to \SI{200}{V}. In this way the counts on the
camera pixels were only due to the electronic noise of the sensor
itself and not to any light.  In each of these runs (called pedestal
runs) we collected 100 images and we evaluated, pixel by pixel, the
average value (\SI{}{pix_{ ped}}) and the standard deviation (\SI{}
{pix_{rms}}) of the response.  The light tightness of the detector is
ensured by the Faraday cage. To check its effectiveness, we compared
the values of \SI{}{pix_{ ped}} and \SI{} {pix_{rms}}) in runs
acquired with laboratory lights On and with completely dark laboratory
without finding any significative differences.

In
the reconstruction procedure, described later in
Sec.~\ref{sec:reconstruction}, \SI{}{pix_{ped}} is then subtracted
from the measured value, while \SI{} {pix_{rms}} is used to define
the threshold to retain a pixel, i.e. when it has a number of counts
larger than \SI{1.1}{pix_{rms}} .

\begin{figure}[ht]
\centering
\includegraphics[width=0.30\textwidth]{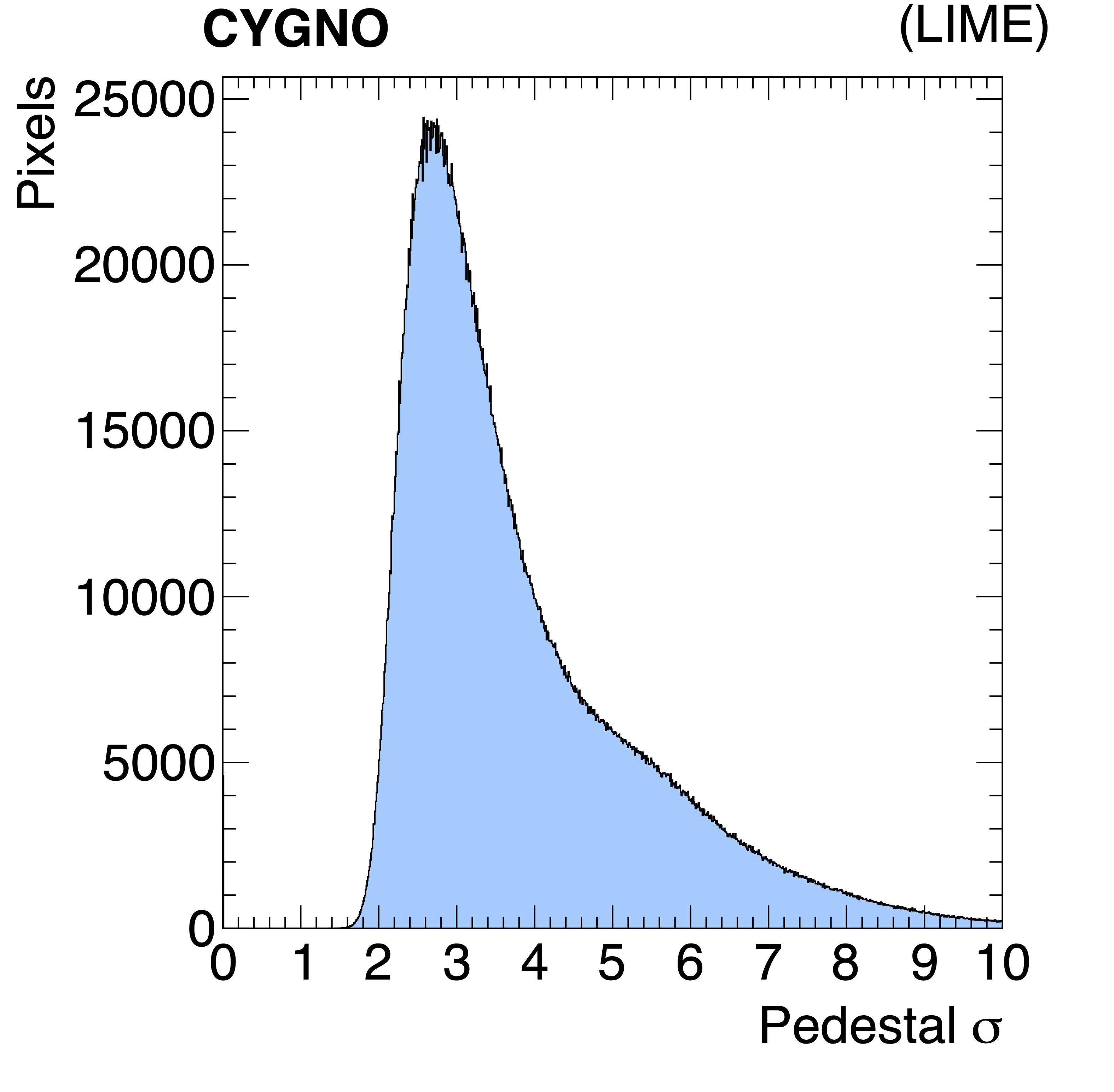}
\includegraphics[width=0.30\textwidth]{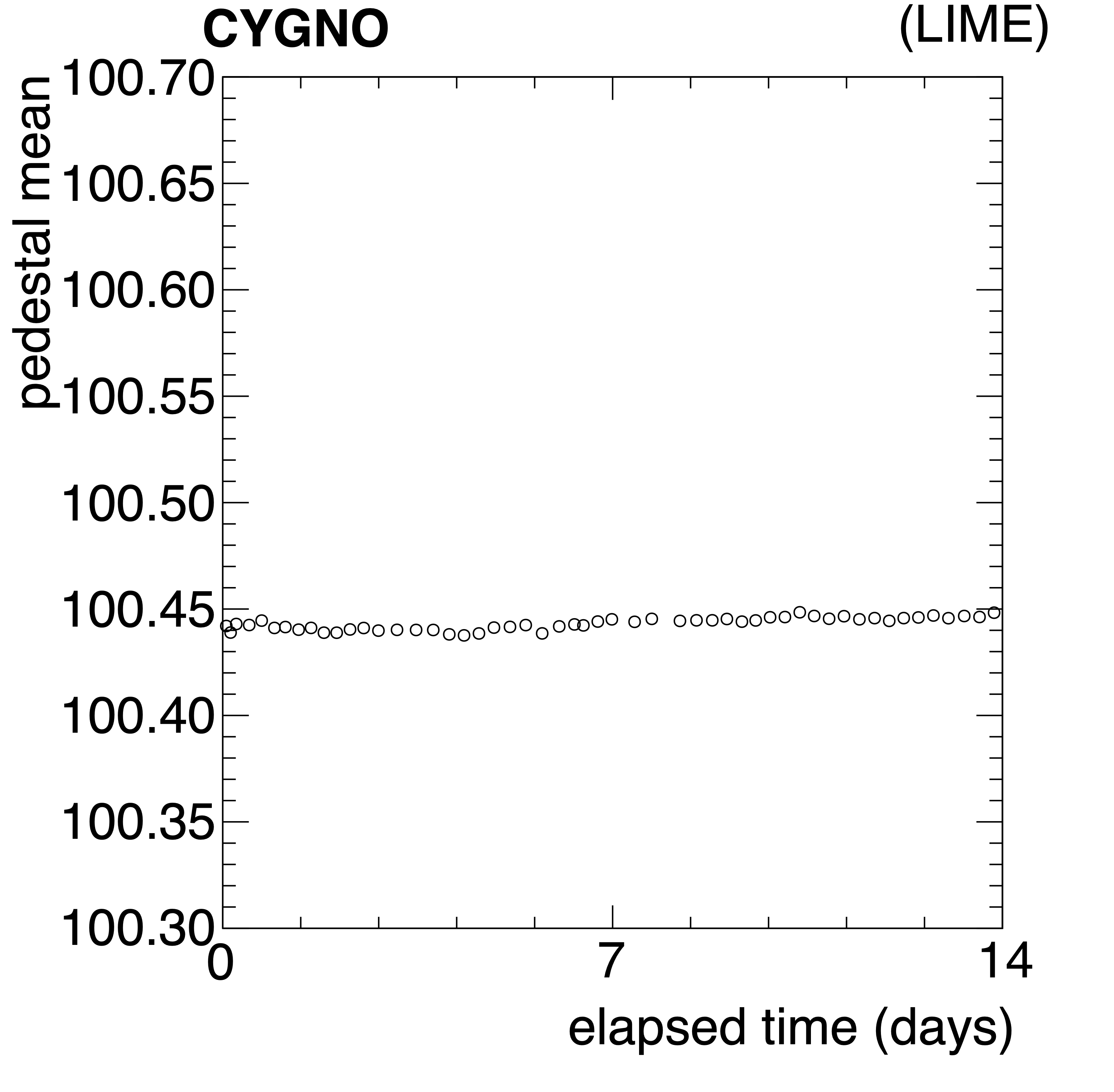}
\includegraphics[width=0.30\textwidth]{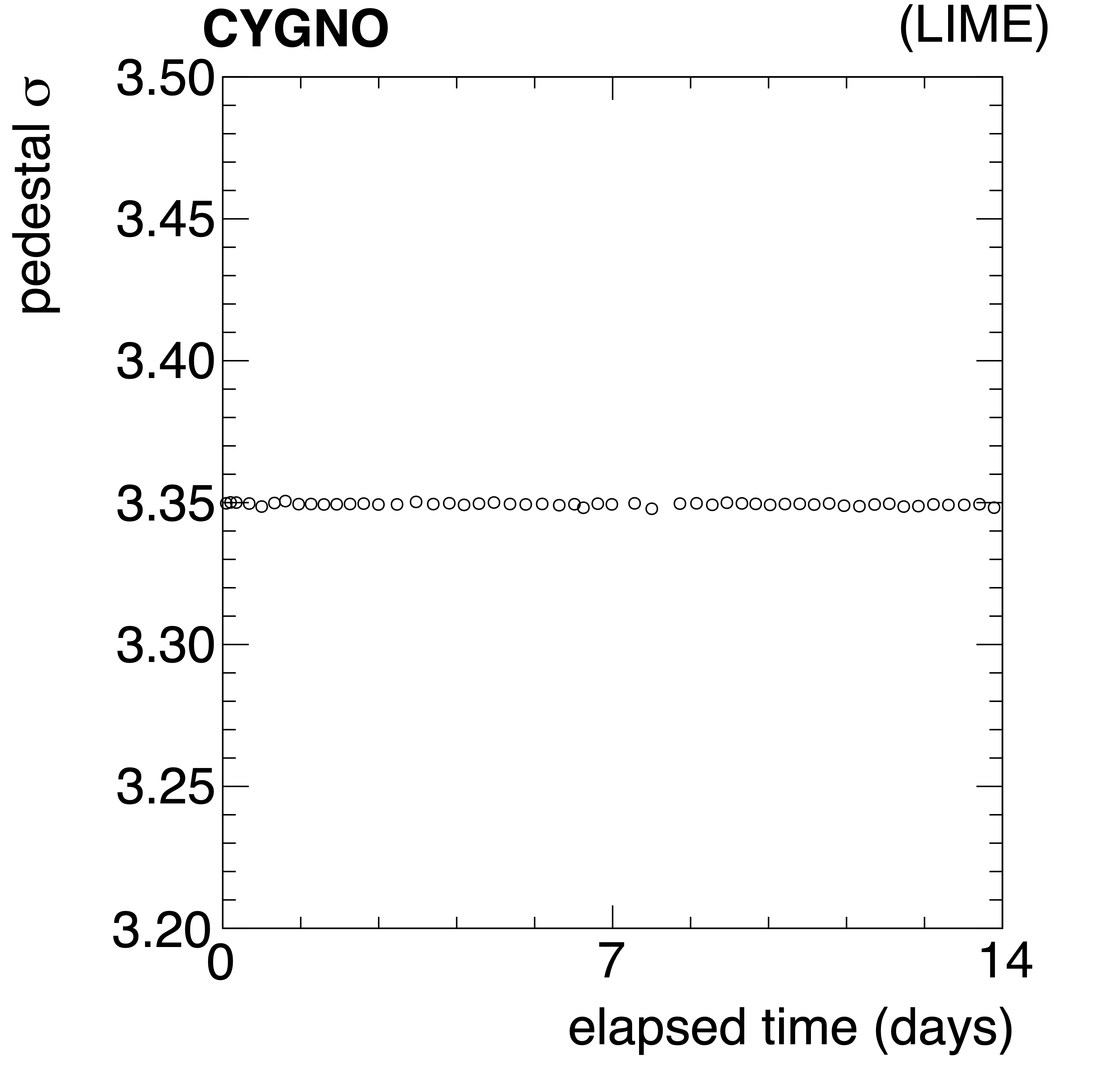}
\caption{Top: distribution of pix$_{\mbox{rms}}$ in one pedestal
  run. Middle and bottom: average of pix$_{\mbox{ped}}$ and
  pix$_{\mbox{rms}}$, respectively, as a function of time, for a
  period of two weeks of data taking, as measured in the regular
  pedestal runs acquired.
\label{fig:noise}}
\end{figure}

The distribution of \SI{} {pix_{rms}} in one pedestal run for all the
pixels of the sensor is shown in Fig.~\ref{fig:noise} (top). The long
tail above the most probable value corresponds to pixels at the top
and bottom boundaries of the sensor, which are slightly noisier than
the wide central part. For this reason 250 pixel rows are excluded
from the reconstruction at the top and 250 pixel rows at the bottom of the
sensor. The stability of the pedestal value and of the electronics
noise has been checked by considering the mean value of the
distribution of \SI{} {pix_{ped}} and of \SI{} {pix_{rms}} as measured
in the regular pedestal runs.  Figure~\ref{fig:noise} middle and bottom
show the distributions of the two quantities in a period of about two
weeks, showing a very good stability of the sensor.

\subsection{Electron recoils  in LIME }

A first standard characterization of the detector response to energy
releases of the order of a few keV utilizes a $^{55}$Fe source with an
activity of \SI{115}{MBq}. \fe decays by electron capture to an
excited \ce{^{55}Mn} nucleus that de-excites by emitting X-rays with an energy
of about \SI{5.9}{keV}, with an additional emission at around
\SI{6.4}{keV}. Given the geometry of the source holder and trolley, the flux of the photons irradiates a cone with an aperture of about 10$^{\circ}$.
This means that in the central region of the detector, the flux is expected to have a gaussian transverse profile with a $\sigma$ of about \SI{1}{cm}.

Moreover, in order to study the energy response for different X-rays
energies, a compact multi-target source was
employed \cite{Amersham}. A sealed $^{241}$Am primary source is
selectively moved in front of different materials. Each material is
presented to the primary source in turn and its characteristic X-ray
is emitted through a 4 mm diameter aperture.  In
Tab. \ref{tab:multitarget} a summary of the materials and energy of
the X-ray lines is reported. The $K_{\beta}$ lines have an intensity
that is about 20\% of corresponding   $K_{\alpha}$ lines.

\begin{table}[h]
\begin{center}
\vspace{1mm}
\caption{X-ray emitted by the multi-target source.\label{tab:multitarget}}
\begin{tabular}{ccc}
\hline
Material       & Energy $K_{\alpha}$ [keV]  & Energy $K_{\beta}$ [keV]\\
\hline
Cu     &  8.04 &  8.91    \\ 
Rb     & 13.37 & 14.97    \\
Mo     & 17.44 & 19.63     \\
Ag     & 22.10 & 24.99      \\
Ba     & 32.06 & 36.55     \\ 
\hline
\end{tabular}
\end{center}
\end{table}

Given the physics interest to the detector response at low energies,
the \fe source X-rays with $E\approx\SI{6}{keV}$ has been used to
induce emissions of lower energy X-rays in two other targets: \ce{Ti}
and \ce{Ca}. The expected $K_{\alpha}$ and $K_{\beta}$ lines are shown
in Table~\ref{tab:multitarget_custom}. 
Given the experimental setup to excite the  \ce{Ti}
and \ce{Ca} lines, also the \SI{6}{keV} X-rays from \fe can reach  the detector active volume,
resulting in the superposition of both contributions.

\begin{table}[h]
\begin{center}
\vspace{1mm}
\caption{X-ray emitted by the additional custom targets excited by the \fe source.\label{tab:multitarget_custom}}
\begin{tabular}{ccc}
\hline
Material       & Energy $K_{\alpha}$ [keV]  & Energy $K_{\beta}$ [keV]\\
\hline
Ti     & 4.51 & 4.93    \\
Ca     & 3.69 & 4.01    \\
\hline
\end{tabular}
\end{center}
\end{table}

\begin{figure}[ht]
\centering
\includegraphics[width=0.45\textwidth]{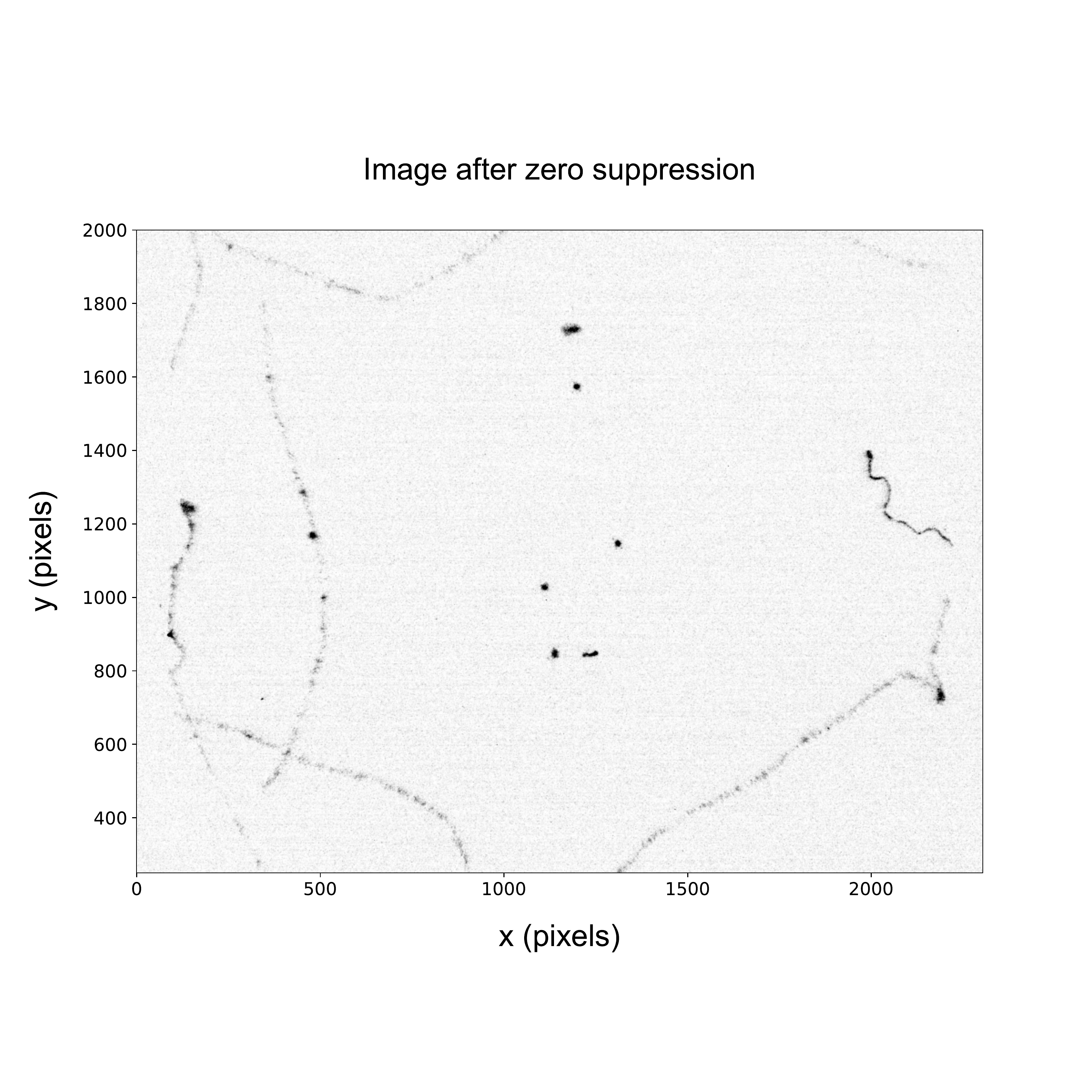}
\includegraphics[width=0.45\textwidth]{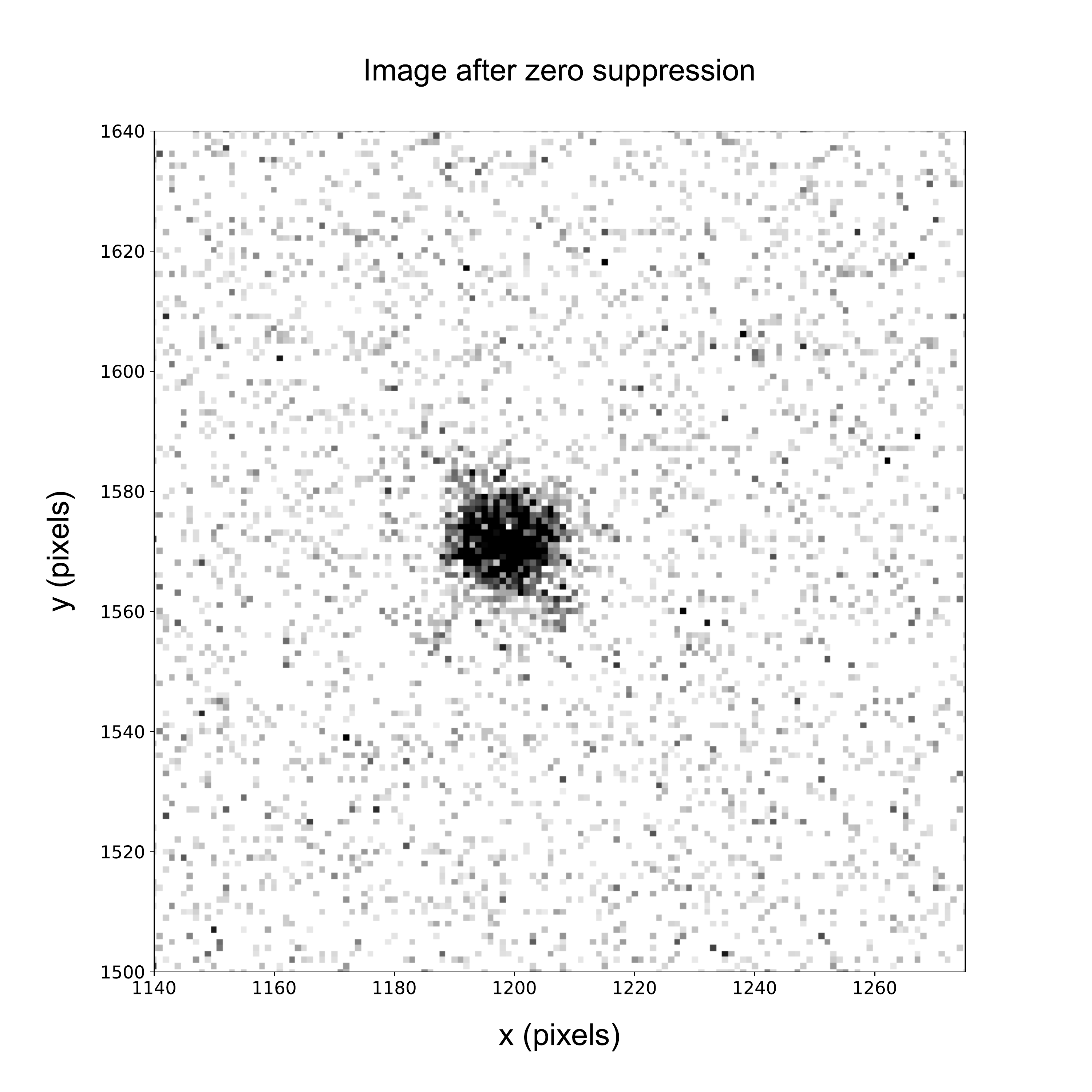}
\caption{Example of an image with natural radioactivity tracks and
  luminous spots indicating the interactions in the gas of \SI{6}{keV}
  X-rays produced by the \fe source. The \fe source is located on the
  top of the sensitive volume and produces spots along the y axis (see Fig.\ref{fig:lime} for the reference frame) of
  the CMOS sensor (top).  A zoom around one of these spots is also
  shown (bottom).
\label{fig:spot}}
\end{figure}

The interaction of the X-ray with the gas molecules produces a
electron recoil with a kinetic energy very similar to the X-ray energy.
According to a SRIM simulation  \cite{Ziegler1985}  in our gas mixture at atmospheric pressure the expected range of the
electron varies from about $\SI{250}{\mu m}$ for a \SI{4}{keV} energy to
about \SI{15}{mm} for a \SI{40}{keV} energy \cite{instruments6010006}.
These electron recoils produce a primary electron-ion pair at the cost
of \SI{42}{eV} \cite{bib:rolandiblum,bib:garfield,bib:garfield1} Along the drift path longitudinal and transversal
diffusion affect the primary ionization electrons distribution.  Once
they reach the GEM surface, these electrons start multiplication
processes yielding an avalanche, producing at the same time also
photons that are visible as tracks in the CMOS sensor image. These
tracks from artificial radioactive sources are shown superimposed to
tracks from natural radioactivity in a typical image (
Fig.~\ref{fig:spot}). The tracks are reconstructed as 2D clusters of
pixels by grouping the pixels with a non-null number of photons above
the pedestal level.

\begin{figure}[ht]
\centering
\includegraphics[width=0.45\textwidth]{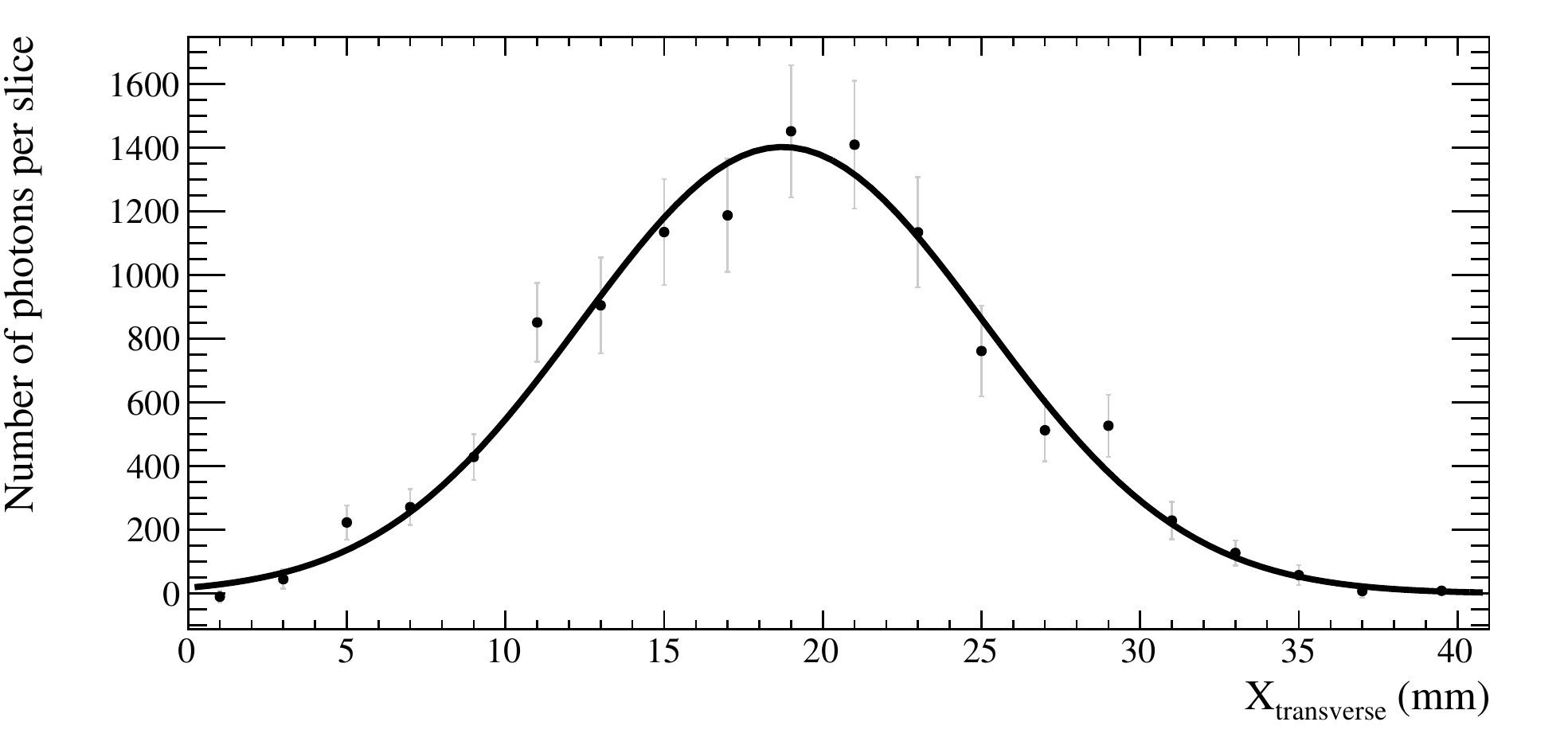}
\includegraphics[width=0.45\textwidth]{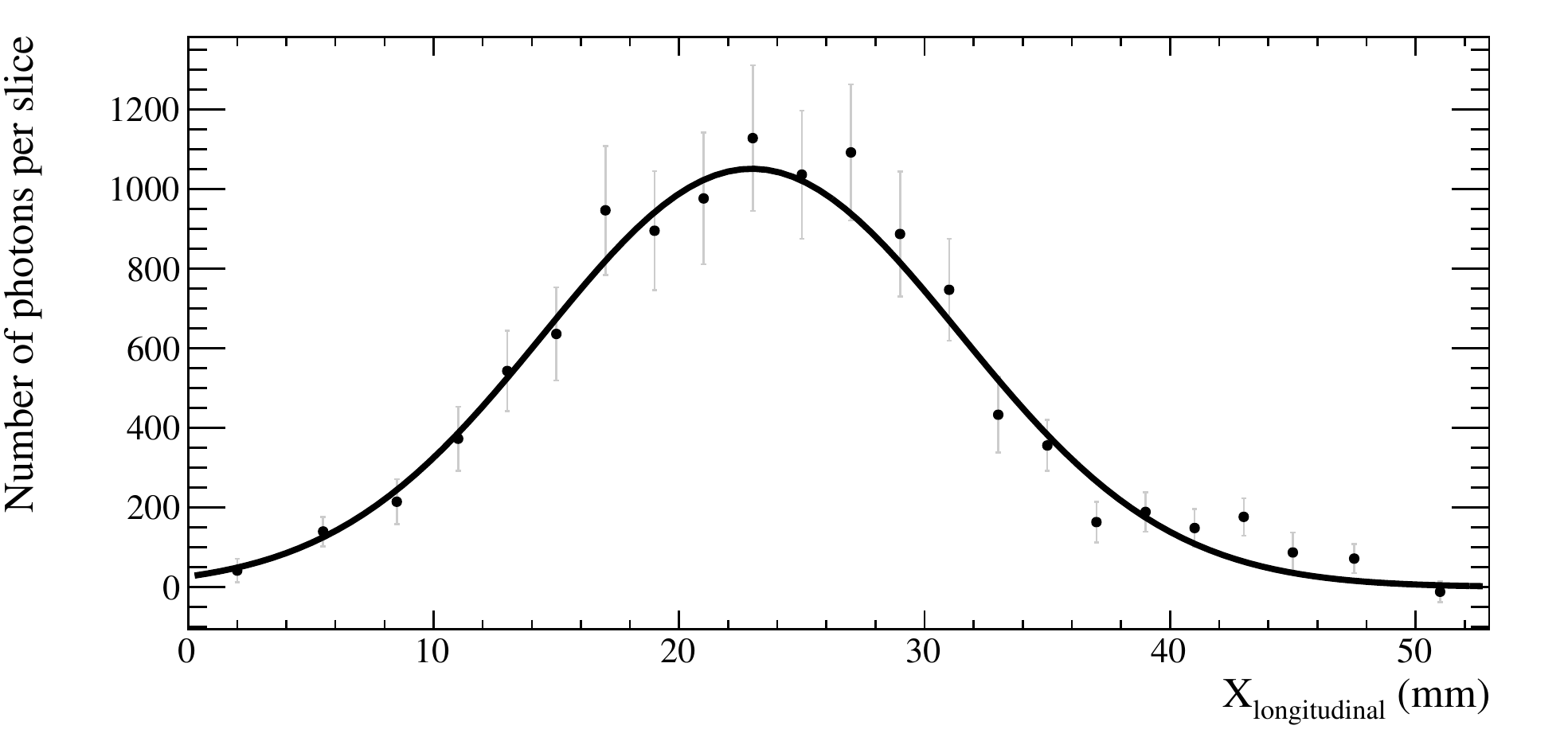}
\caption{Example of transverse (top) and longitudinal (bottom)
  profiles of one luminous spot generated by the interactions in the
  gas of \SI{6}{keV} X-rays produced by the \fe source. From the
  Gaussian fits the values of \St and \Sl are obtained along with the
  amplitudes $A_\mathrm{T}$ and $A_\mathrm{L}$ respectively.
\label{fig:profiles}}
\end{figure}

Once projected to the 2D GEM plane the spherical cloud of the drifting
electrons from the \fe X-ray interaction produces a
$\approx\SI{5}{mm}$ wide light profile along both the orthogonal axes
of the cluster.  The exact span of the profile depends on the running
conditions of the detector and on the $z$ position of the X-ray
interaction. In the following we refer to the longitudinal
(transverse) direction as the orientation of the major (minor) axis of
the cluster, found via a principal component analysis of the 2D
cluster.  The two profiles for a typical cluster are shown in
Fig.~\ref{fig:profiles} with a Gaussian fit superimposed. From these
fits the values of \Sl and \St are obtained along with the amplitudes
$A_\mathrm{L}$ and $A_\mathrm{T}$ respectively In general for
non-spherical cluster due larger energy electron recoil we determine
and utilizes only the \St value.

\section{Reconstruction of electron recoils }
\label{sect:reco}

The energy deposit in the gas through ionization is estimated by
clustering the light recorded in the camera image with a dynamic
algorithm.  The method is developed with the aim to be efficient with
different topologies of deposits of light over the sensors.  It is
able to recognize small spots whose radius is determined by the
diffusion in the gas, or long and straight tracks as the ones induced
by cosmic rays traversing the whole detector, or long and curly tracks
as the ones induced by various types of 
radioactivity. Radioactivity is in fact present in both the
environment surrounding the detector or in the components of the
detector itself.

\subsection{The reconstruction algorithm}
\label{sec:reconstruction}

The reconstruction algorithm consists of four steps: \begin{enumerate*}[label=(\roman*)]
\item a zero suppression to reject the electronics noise of the sensor \label{reco:zs}
\item the correction for vignetting effect described in Sec.~\ref{sec:vignetting} and two steps of iterative clustering \label{reco:vignetting}
\item a super-clustering step to reconstruct long and smooth tracks parameterizing them as
  polynomial trajectories, \label{reco:supercluster} and
\item a small clustering step to find residual short deposits.\label{reco:dbscan}
\end{enumerate*}
The iterative approach is necessary for disentangling possibly
overlapping long tracks recorded in the \SI{50}{ms} time interval of
the exposure of the camera.

As a further noise reduction step, the resolution of the resulting image is
initially reduced by forming \textit{macro-pixels}, by averaging the
counts in $4{\times}4$ pixel matrices, on which a median
filter  is applied, which is effective in
suppressing the electronics noise fluctuations, as it is described in
more details in Ref.~\cite{coronello}. 

\begin{figure}[ht]
\centering
\includegraphics[width=0.45\textwidth]{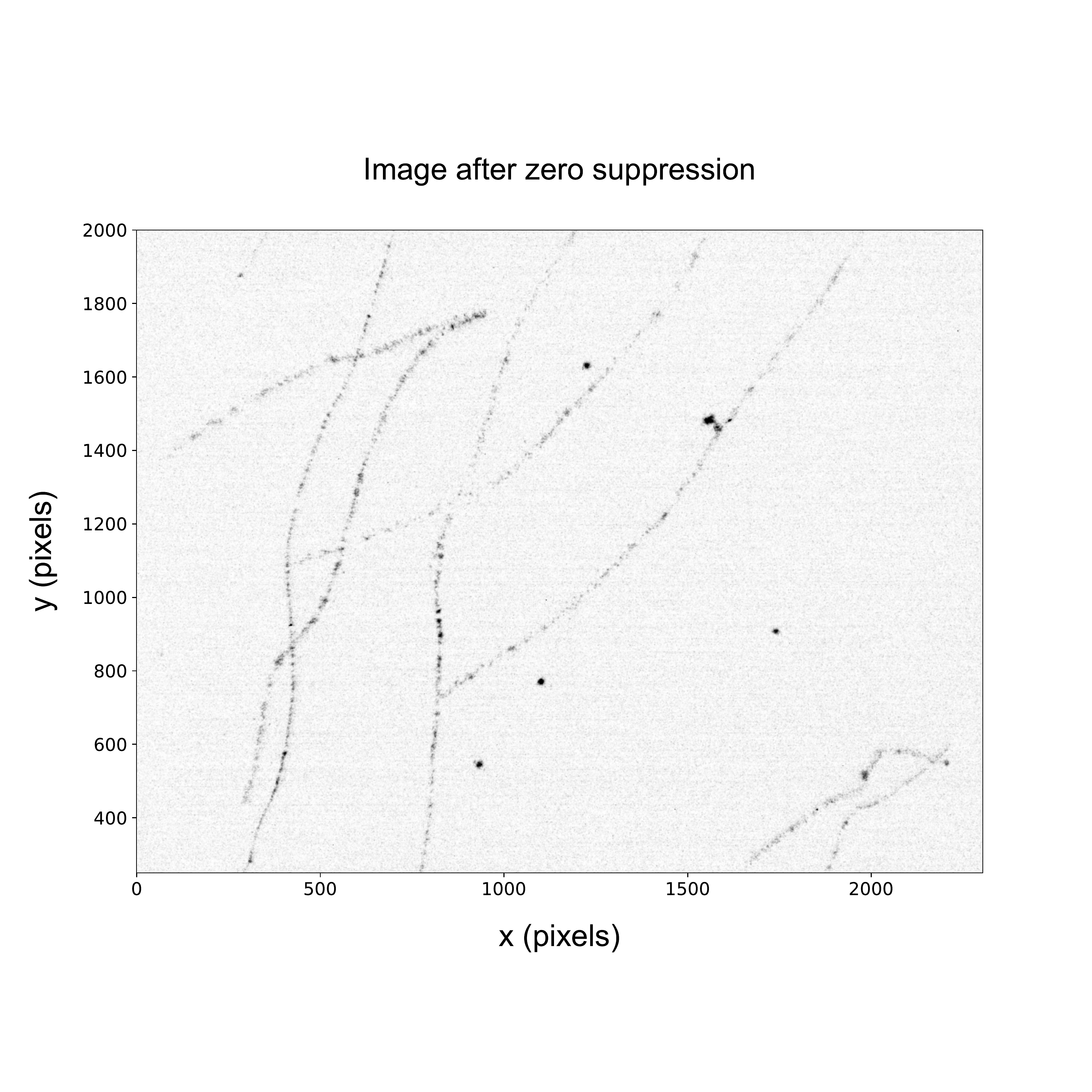}
\includegraphics[width=0.45\textwidth]{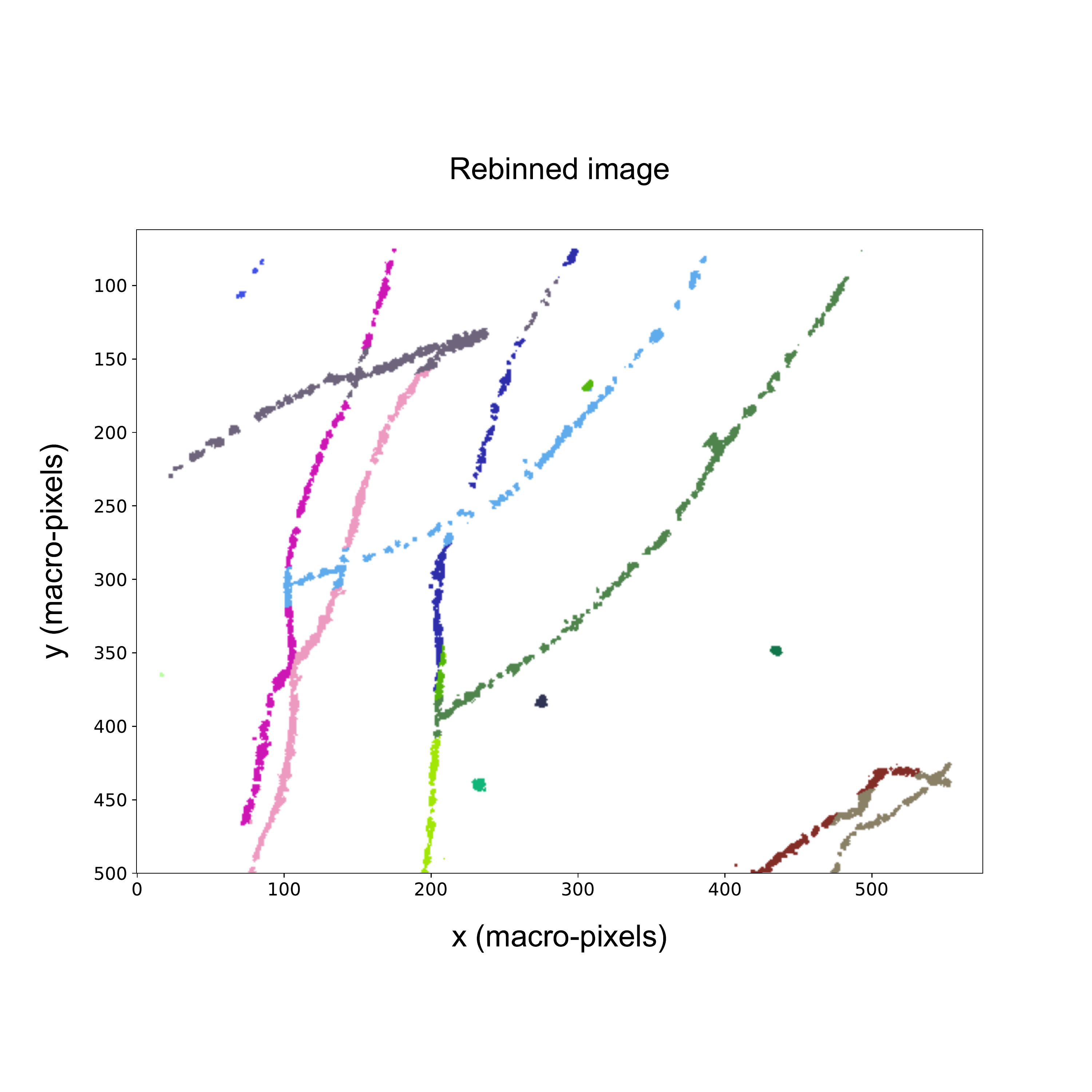}
\caption{Top: image with an exposure of \SI{50}{ms}. Bottom:
  reconstructed clusters after the two step procedure described in the
  text.
\label{fig:clustering}}
\end{figure}

In order to first clean the picture from the long tracks originating
from the ambient radioactivity, the iterative procedure of step
\ref{reco:supercluster} is started, looking for possible candidate
trajectories compatible with polynomial lines of increasing orders,
ranging from 1 (straight line) to 3 as a generalization of the
\textsc{ransac} algorithm~\cite{ransac}.  If a good fit is found, then
the supercluster is formed, and the pixels belonging to all the seed
basic clusters are removed from the image, and the procedure is
repeated with the remaining basic cluster seeds. The step
\ref{reco:supercluster} is necessary to handle the cases of multiple
overlaps of long tracks, as it can be seen in
Fig.~\ref{fig:clustering}. It can be noticed that in the overlap
region the energy is not shared, i.e. it is assigned to one of the
overlapping tracks. In these cases the tracks can be split, but the
pieces are still long enough not to mimick short deposits for low
energy candidates of our interest for DM searches. When no more
superclusters can be found, the superclustering stops, and the
remaining pixels in the image are passed to step \ref{reco:dbscan},
i.e. the search for small clusters.  For this purpose, small-radius
energy deposits are formed with \idbscan, described in details in
Refs.~\cite{iDBSCAN,coronello}.  The effective gathering radius for
pixels around a seed pixel is 5 pixel long, so small clusters are
formed.  Finally, the clusters from any iteration of the above
procedure are merged in a unique collection, which form the track
candidates set of the image.

The track candidates are then characterized through the pattern of the 2D
projection of the original 3D particle trajectory interacting within the
TPC gas mixture. Various cluster shape variables are studied, and are
useful to discriminate among different types of interactions~\cite{coronello}. For 
example a clear distinction can be made between tracks due to muons from  cosmic rays and electron recoils due to X-rays. Moreover, within a given class of interactions,
the cluster shapes are sensitive to the detector response, for example
gas diffusion, electrical field non uniformities, gain non
uniformities of the amplification stages. Thus  they can be exploited to
partially correct these instrumental effects improving the determination of the
original interaction features, like the deposited energy, or
its $z$-position, which cannot be directly inferred by the 2D
information.

\subsection{The \fe source studies}
\label{sec:fesource}

\begin{figure}[ht]
\centering
\includegraphics[width=0.5\textwidth]{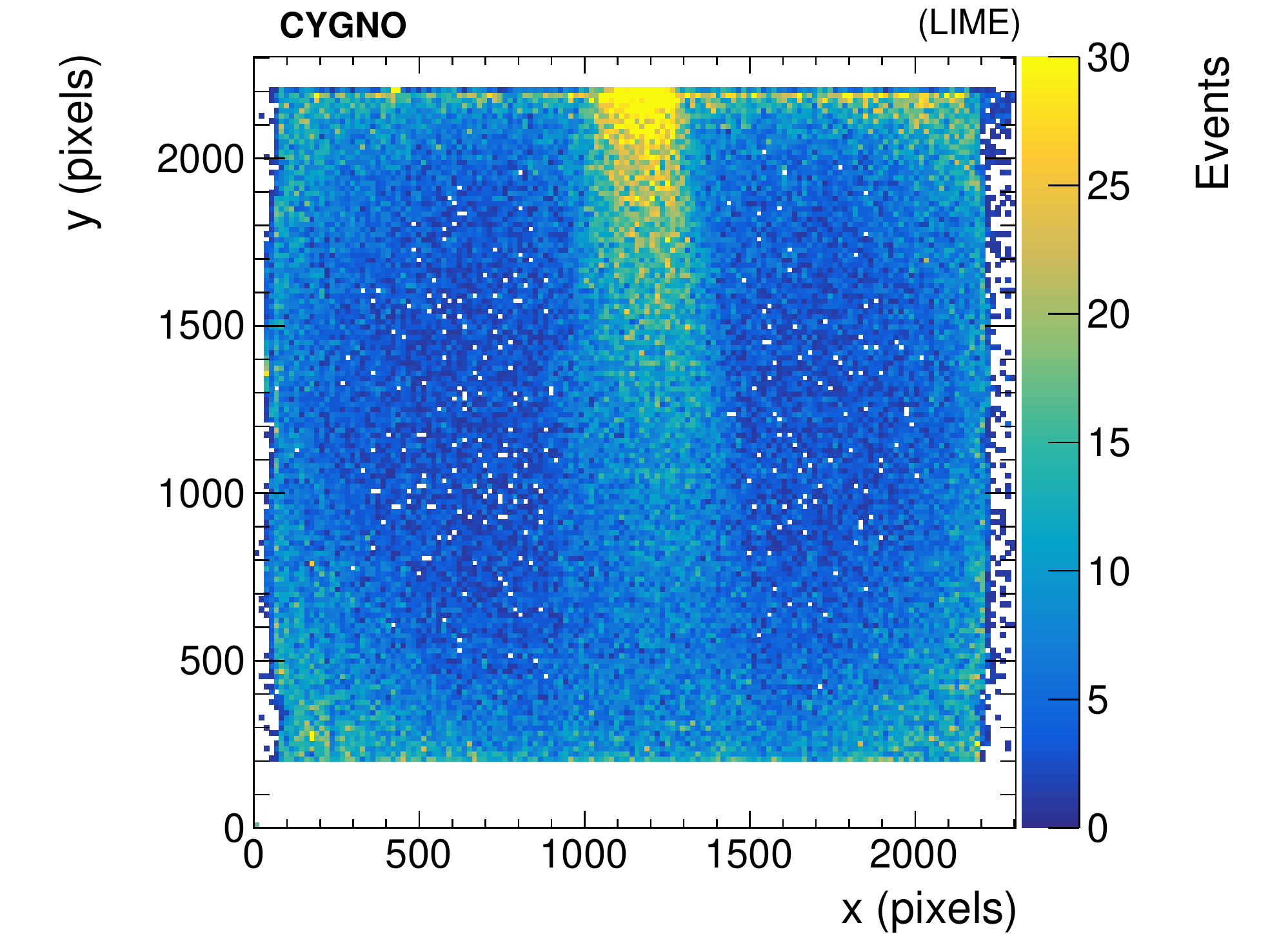}
\caption{Spatial distribution of the reconstructed clusters in data
  collected with \fe source. Only clusters in the central region of
  the GEM plane are selected to remove the noisier regions of the
  sensor. The \fe source is positioned outside the detector at high
  values of $y$.
  \label{fig:shower}}
\end{figure}

The \fe source is able to induce interaction in the gas mixture with
an illumination of the entire vertical span of the detector as shown in
Fig.~\ref{fig:shower}. Due to the collimation of the source, only a
slice in the horizontal direction has a significant occupancy of
\fe-induced clusters.

 Several variables are used for the track characterization: \St, the
track length, the light density $\delta$ (defined as the integral of
the light collected in the cluster, divided by the number of pixels
over the noise threshold), the RMS of the light intensity residuals of the
pixels \SI{}{I_{rms}}, and other variables described in more details
in Ref.~\cite{coronello}.

\begin{figure}[htbp]
\centering
\includegraphics[width=0.45\textwidth]{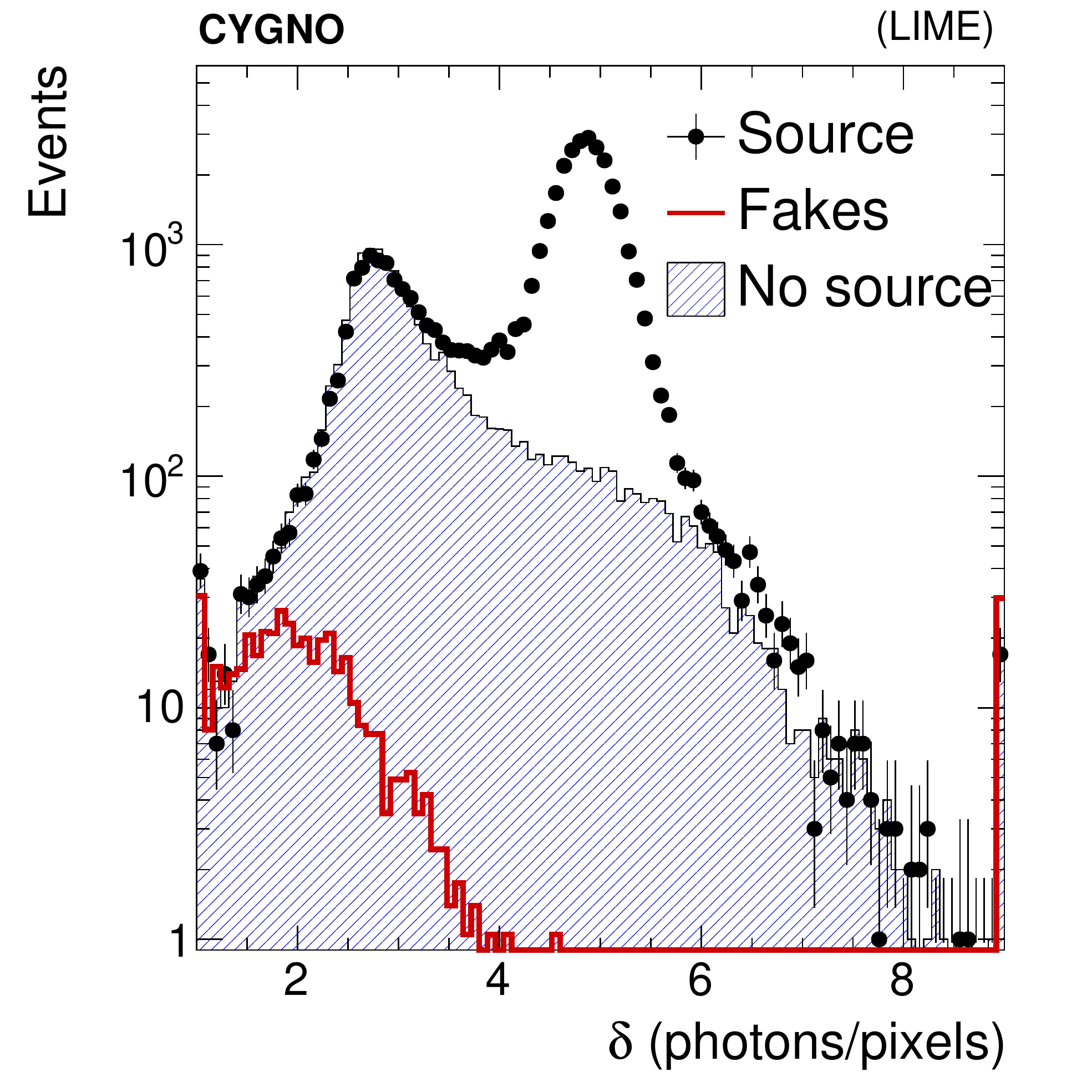}
\includegraphics[width=0.45\textwidth]{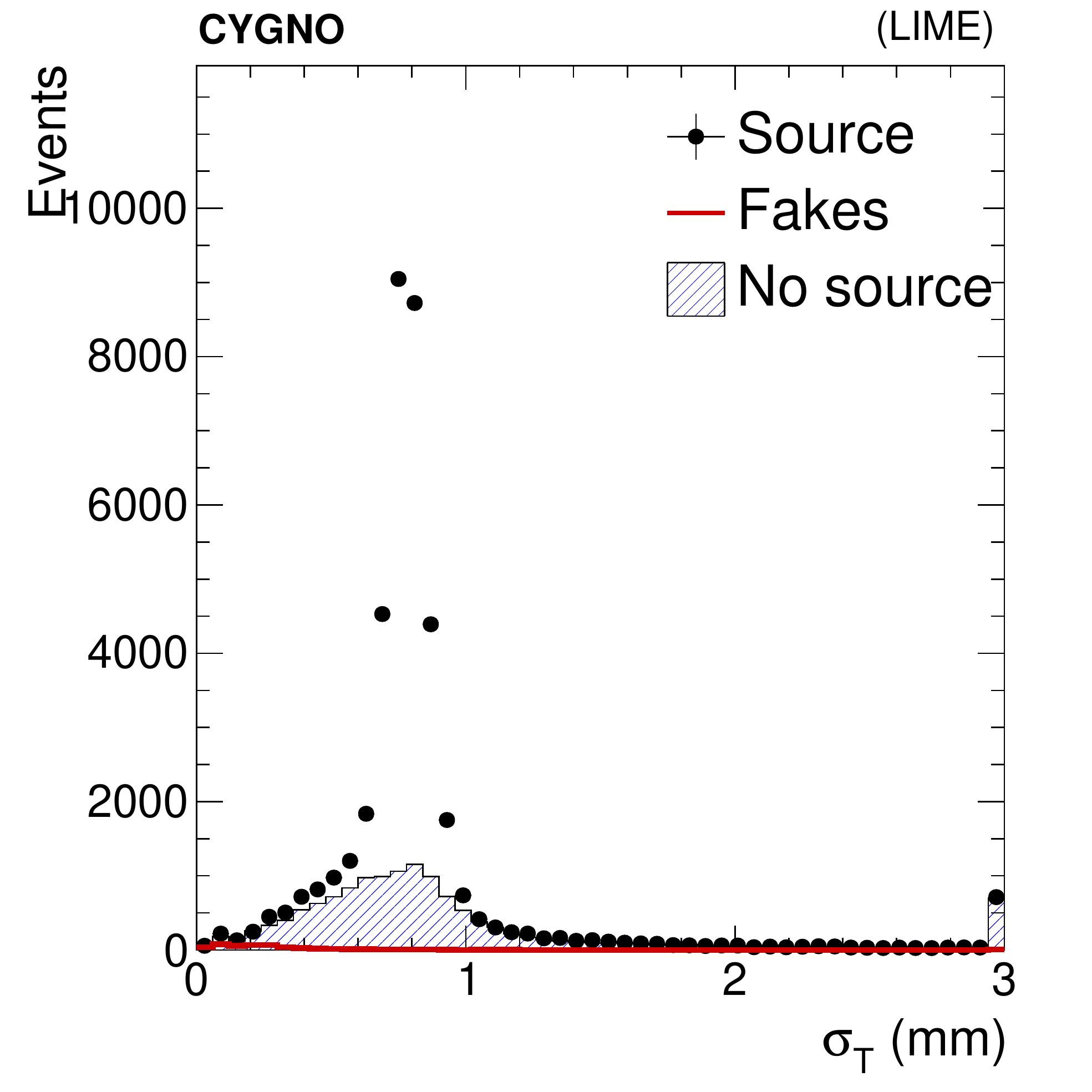}
\caption{Top: light density $\delta$ in the reconstructed clusters,
  as defined in the text. Bottom: transverse dimension of the
  reconstructed cluster \St.  Black points represent data in
  presence of the \fe source, filled histogram represents data without
  the source, while the red hollow histogram represents the
  contribution from mis-reconstructed clusters from electronics
  noise. The latter two are normalized to the live-time of the data
  taking with the \fe source.}
\label{fig:feplots1}
\end{figure}

\begin{figure}[htbp]
\centering
\includegraphics[width=0.45\textwidth]{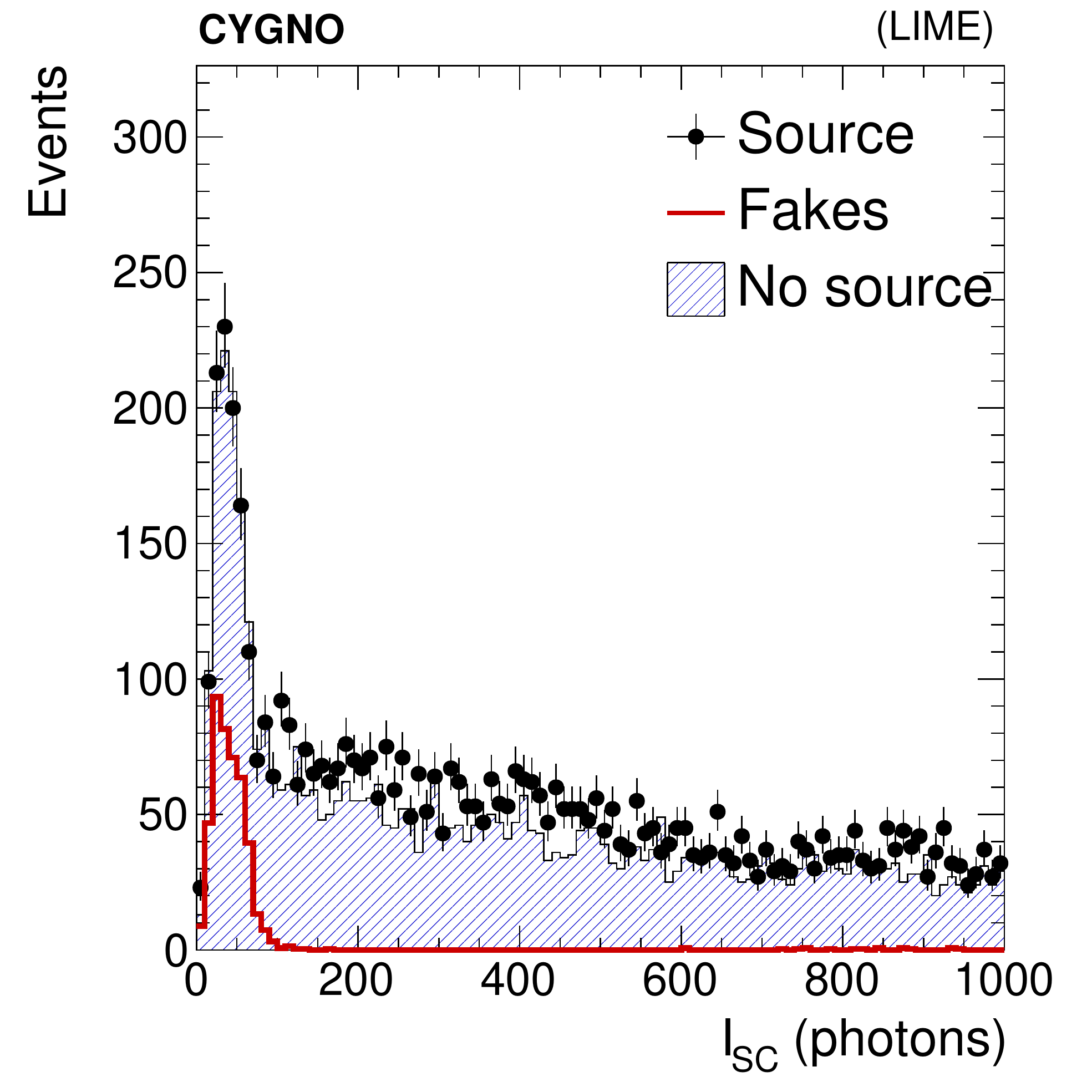}
\includegraphics[width=0.45\textwidth]{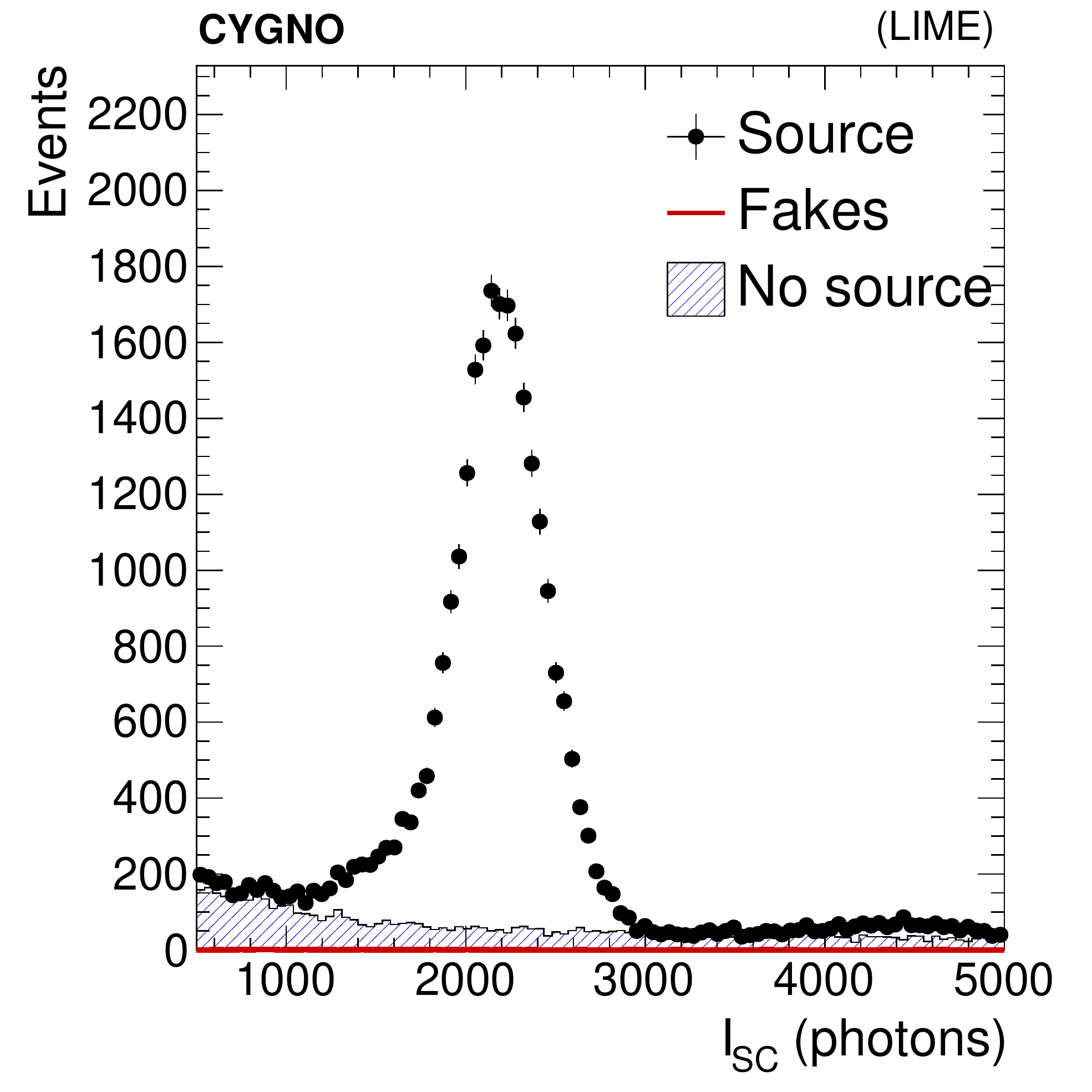}
\caption{Light integral \Isc of the reconstructed clusters, as defined
  in the text. Top (bottom): region below (around) the expected energy
  peak from X-rays interactions from the \fe source.  Black points
  represent data in presence of the \fe source, filled histogram
  represents data without the source, while the red hollow histogram
  represents the contribution from mis-reconstructed clusters from the
  electronics noise. The latter two are normalized to the live-time of
  the data taking with the \fe source.
\label{fig:feplots2}}
\end{figure}

A sample of clusters is obtained applying a very loose selection,
which resembles the one optimized in Ref.~\cite{coronello}.  Examples
of the distributions for $\delta$ and for \St of these clusters are
shown in Fig.~\ref{fig:feplots1}, while the spectrum of \Isc, defined as 
of the sum of the detected light in a cluster, is shown in
Fig.~\ref{fig:feplots2} in a range below and around the expected
deposit from the \fe X-rays.  The distribution of \Isc
also shows a small enhancement at around twice the energy expected by
the \fe X-rays corresponding to the cases when two neighbor deposits
are merged in a single cluster. This can happen because of the
relatively large activity of the employed \fe source. The average size of the spot produced 
by the \fe X-ray interactions is about \SI{20}{mm^2}.


The distributions show the data obtained in data-taking runs both in
presence of the X-ray source and without it, in order to show the
background contribution, after normalizing them at the live-time of
the data taking with the \fe source.  The expected contribution from
{\it fake clusters}, defined as the clusters randomly reconstructed by
neighboring pixels over the zero-suppression threshold, has been also
estimated from the pedestal runs, where no signal contribution of any
type is expected. As can be seen from Fig.~\ref{fig:feplots2} (top), this
contribution becomes negligible for $\Isc \gtrsim \SI{400}{photons}$.

\subsection{Energy  calibration}
\label{sec:calibration}

Despite the correction of the optical effects of the camera applied
before the clustering, the light yield associated to a cluster \Isc
still depends on the position of the initial ionization site where the
interaction within the active volume happened. Therefore the light
yield \Isc must be converted in an energy \Erec by a calibration
factor and then corrected to infer the original energy deposit E.

\begin{figure}[ht]
\centering
\includegraphics[width=0.5\textwidth]{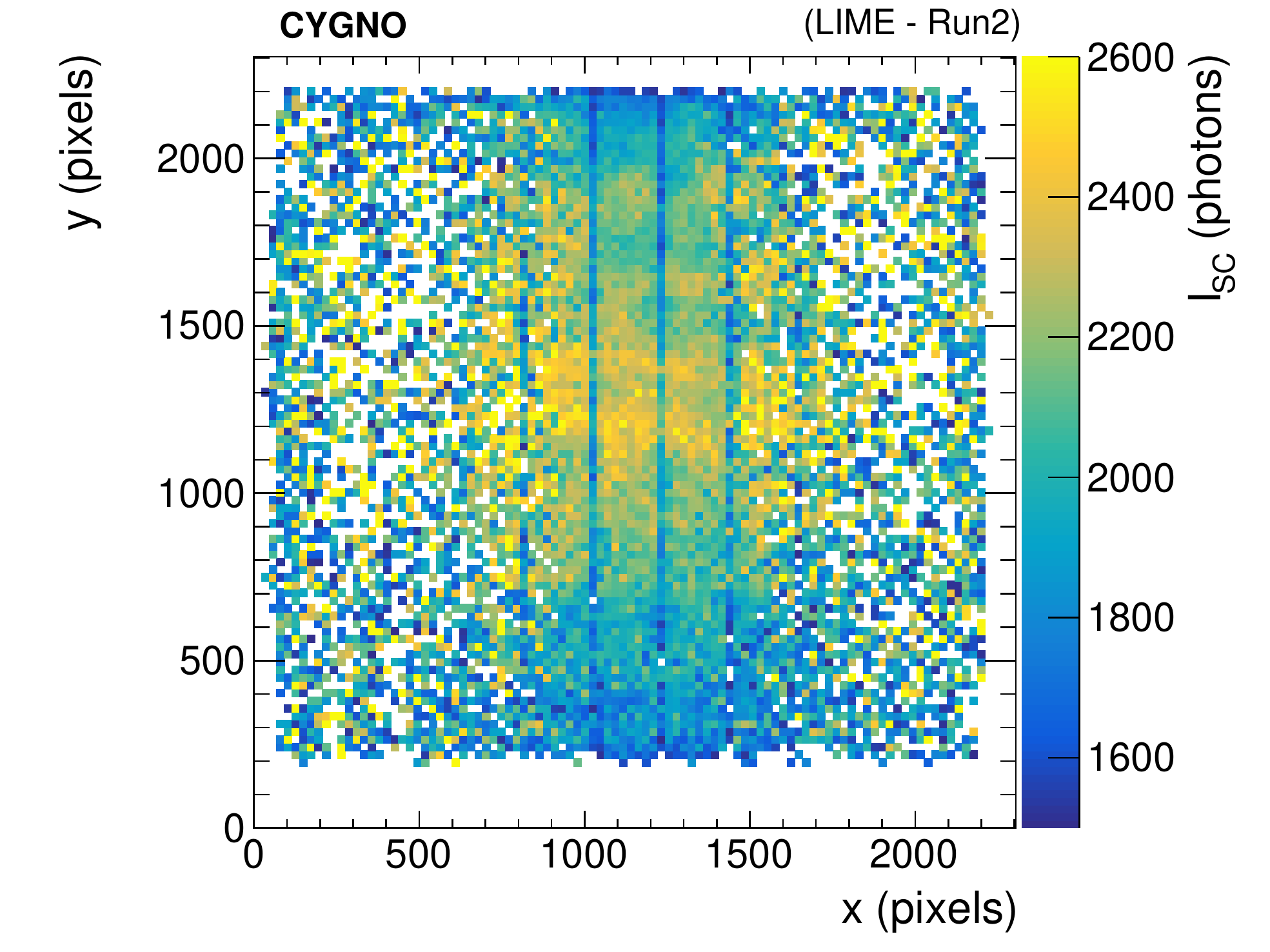}
\caption{Average light yield, \Isc, for the clusters as a function of
  the $x$--$y$ position in the 2D projection, for data collected with
  the \fe source positioned at a $z=\SI{26}{cm}$. \label{fig:ivsxy}}
\end{figure}

The \Erec dependence on the $x$--$y$ position of the initial
interaction can be affected by possible imperfect correction of the
vignetting effect, non uniformities of the drift field and of the
amplification fields, especially near the periphery of the GEM planes, as shown
in Fig.~\ref{fig:ivsxy}.

\begin{figure}[ht]
\centering
\includegraphics[width=0.50\textwidth]{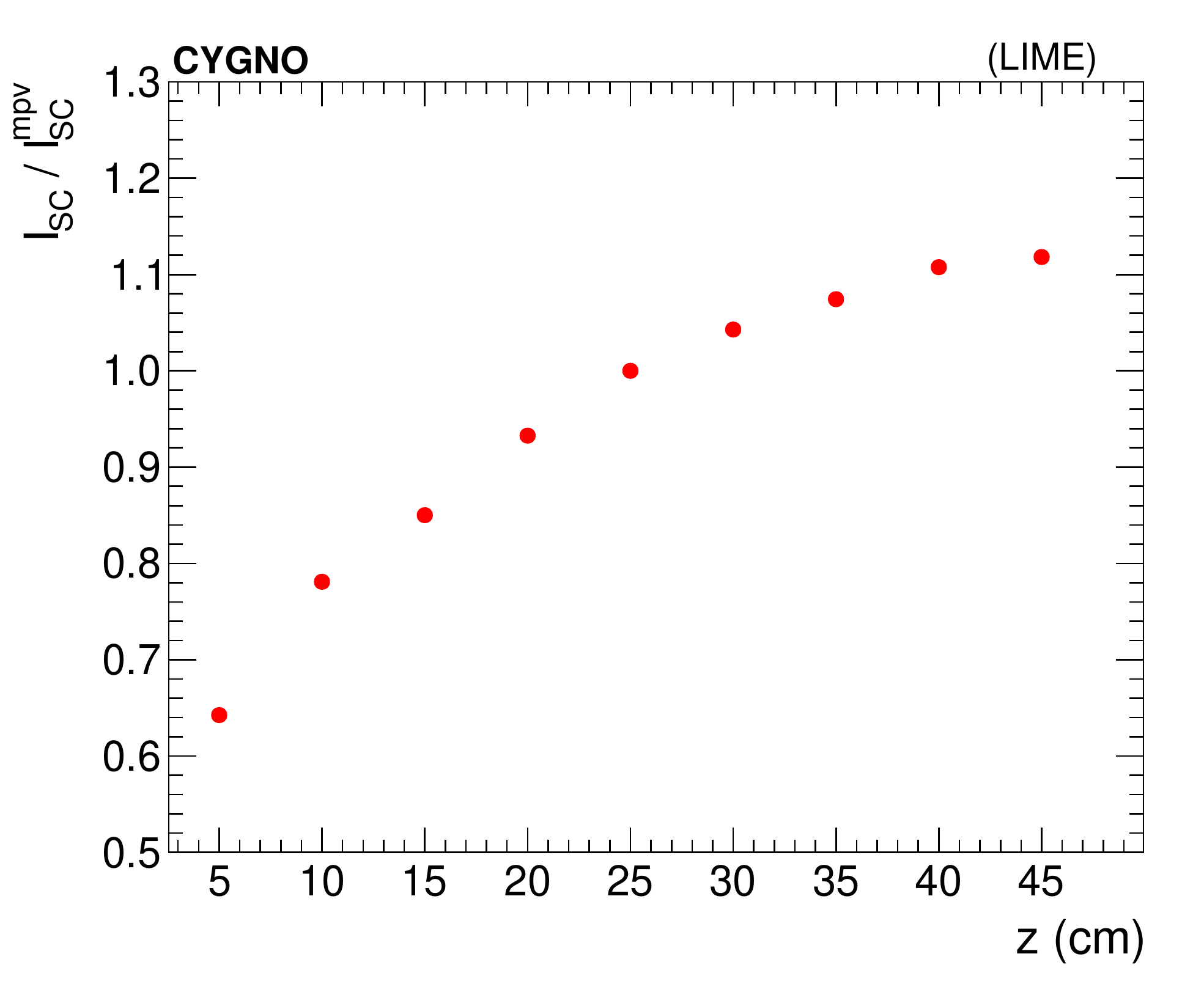}
\caption{Average light yield, \Isc, normalized to its most
  probable value, $\SI{}{I_{SC}^{mpv}}$, for clusters reconstructed in
  presence of the \fe source as a function of the $z$ distance with
  respect the GEM planes. \label{fig:ivsz}}
\end{figure}

Moreover, inefficiency in the transport of the primary ionization
electrons due to attachment during their drift in the gas would result
in a monotonic decrease of \Isc as a function of $z$ of the initial
interaction. However, as shown in Fig.~\ref{fig:ivsz}, a continuous increase
of \Isc with the $z$ of the initial interaction is observed.

This effect can be interpreted in the following way. During the
amplification process, the channels across the GEM foils are filled
with ions and electrons produced in the avalanches, but thanks to
their small size they can rapidly drain.
In recent years, however, several studies \cite{bib:highrate} have
shown that for high-gain (10$^6$--10$^7$) operations, the amount of
charge produced by a single avalanche is already sufficient to change
locally the electric field. In general this has the effect to reduce
the effective gain of the GEM, causing a \textit{saturation effect}.
This also makes the response of the GEM system dependent on the amount
of charge entering the channels and - in the case of many primary
electrons from the gas ionization - on the size of the surface over
which these electrons are distributed.  In LIME, the
diffusion of the primary ionization electrons over the 50 cm drift
path can almost quadruple the size of the surface involved in the
multiplication, thus reducing the charge density and therefore
reducing the effect of a gain decrease.

We think this to be the cause of the observed behavior of the spots
originated by the \fe X-rays over the whole drift region: the light
yield \Isc for spots originated by interactions farther from the GEM
is larger than for spots closer to the GEM. Thus, the overall trend of
\Isc as a function of the $z$ position of the ionisation site
therefore presents an initial growth followed by an almost plateau
region, as shown in Fig.~\ref{fig:ivsz}.

These effects partially impact the observed cluster shapes. However,
they can be used as a handle, together with the $x$--$y$ measured
position in the 2D plane, to infer E. Since multiple effects impact
different variables in a correlated way, corrections for the non
perfect response to the true energy deposits have been optimized using
a multivariate regression technique, also denoted as multivariate analysis (MVA), 
based on a Boosted Decision Tree
(BDT) implementation, following a strategy used in Ref.~\cite{cmsecal}.


The training has then been performed on data recorded with the various
X-rays deposits described in Table~\ref{tab:multitarget} and
Table~\ref{tab:multitarget_custom}. The target variable of the
regression is the mean value of the ratio $\Isc/\SI{}{I_{SC}^{mpv}}$,
where the most probable value \SI{}{I_{SC}^{mpv}} is the most probable
value of the \Isc distribution for each radioactive source.  The
performance of the regression using the median of the distribution
instead of the mean have been checked and found giving a negligible
difference.

The clusters were selected by requiring their \St to be consistent
with the effect of the diffusion in the gas and their length not
larger than what is expected for an X-ray of energy E. In addition it
is required that \Isc falls within $5\,\sigma_G$ from the expected E
for a given source, where $\sigma_G$ is the measured standard
deviation of the peak in the \Isc distribution (estimated through a
Gaussian fit).  The background contamination of the training samples
after selection, estimated by applying the selection on the data
without any source, is within 1--5\% of the total number of selected
clusters.

The input variables to the regression algorithm are the $x$ and $y$
coordinates of the supercluster, and a set of cluster shape variables,
among which the most relevant are the ratio
$\frac{\sigma_\mathrm{T}}{A_\mathrm{T}}$, \SI{}{I_{rms}} and $\delta$.  Variables
that are proportional to \Isc are explicitly removed, in order to
derive a correction which is as independent as possible on the true
energy E. In order to be sensitive to the variation of the inputs
variables as a function of $z$, and possibly correct for the
saturation effect, data with the \fe source have been collected with
the source positioned at different values of $z$ uniformly
distributed, with a step of \SI{5}{cm} from the GEM to the cathode.
The data collected with the other sources of
Tables~\ref{tab:multitarget} and~\ref{tab:multitarget_custom} instead
were only taken at $z=\SI{26}{cm}$.

A sanity check on the output of the regression algorithm is performed
on the data without any source, where the energy spectrum of the
reconstructed clusters extends over the full set of $K_{\alpha}$ and
$K_{\beta}$ lines used for the training.  No bias or spurious bumps
induced by the training using only few discrete energy points is
observed. 

The $K_\alpha$ line expected for the \fe X-rays, when the source is
positioned at $z=\SI{26}{cm}$, is used to derive the absolute energy
calibration conversion, which equals is approximately
$\kappa~=~\SI{0.38}{photons/eV}$. The absolute reconstructed raw
energy is thus defined as $\Erec~=~\Isc\,/\,\kappa$. The absolute
energy, after the multivariate regression correction described above,
is denoted as E in the following.

\begin{figure}[ht]
\centering
\includegraphics[width=0.45\textwidth,height=0.45\textwidth]{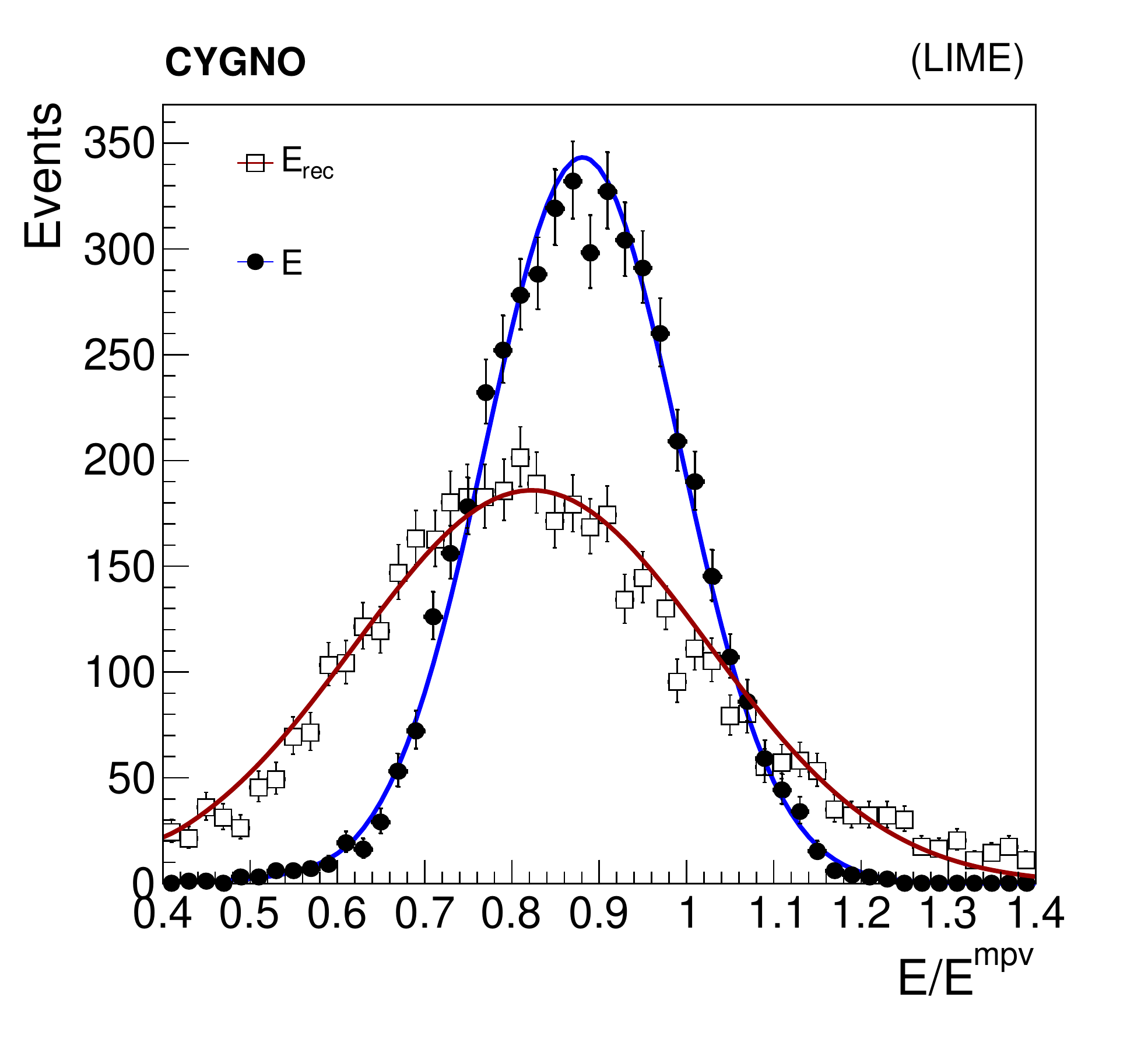}
\includegraphics[width=0.45\textwidth,height=0.45\textwidth]{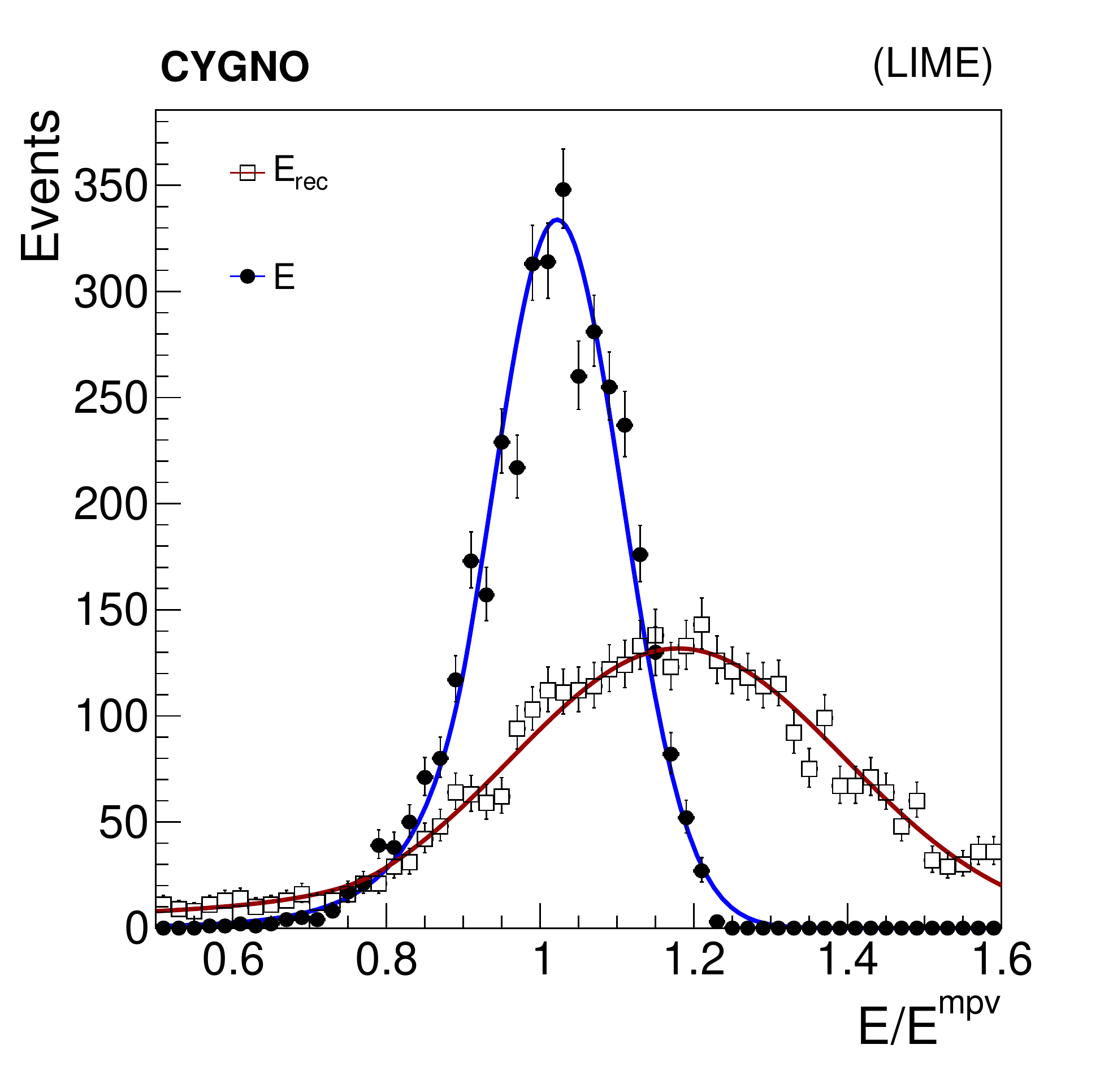}
\caption{Comparison between \Erec (open squares) and E (filled
  circles) normalized by the most probable value of the corresponding
  distribution for $z=\SI{26}{cm}$, on data collected with \fe source
  at a distance of $z=\SI{11}{cm}$ (top) or $z=\SI{41}{cm}$ (bottom)
  from the GEM planes. A fit with a Crystal Ball function, as
  discribed in the text, is superimposed to each
  distribution.\label{fig:regrcomp}}
\end{figure}

The comparison of the distributions for the raw supercluster energy,
\Erec, and E, using data collected in presence of the \fe radioactive
source is shown in Fig.~\ref{fig:regrcomp} for two extreme distances
from the GEM planes, $z=\SI{11}{cm}$ and $z=\SI{41}{cm}$.  The
improvement in the energy resolution is substantial. The distribution
after the correction shows a small tail below the most probable value
of the distribution, indicating a residual non-perfect containment of
the cluster, that systematically underestimates the energy and should
be corrected by improving the cluster reconstruction.

The efficacy of the MVA regression in correcting for the saturation
effect and other response non uniformities is estimated with the data
sample collected with \fe source.  The $\Erec/\SI{}{E_{rec}^{mpv}}$
and E/\SI{}{E^{mpv}} distributions are fit with a Crystal Ball
function~\cite{CrystalBall}, which describes their tails:
$f(E;m_G,\sigma_G,\alpha,n)$, where the parameters $m_G$ and
$\sigma_G$ describe the mean and standard deviation of the Gaussian
core, respectively, while the parameters $\alpha$ and $n$ describe the
tail.

\begin{figure}[ht]
\centering
\includegraphics[width=0.45\textwidth]{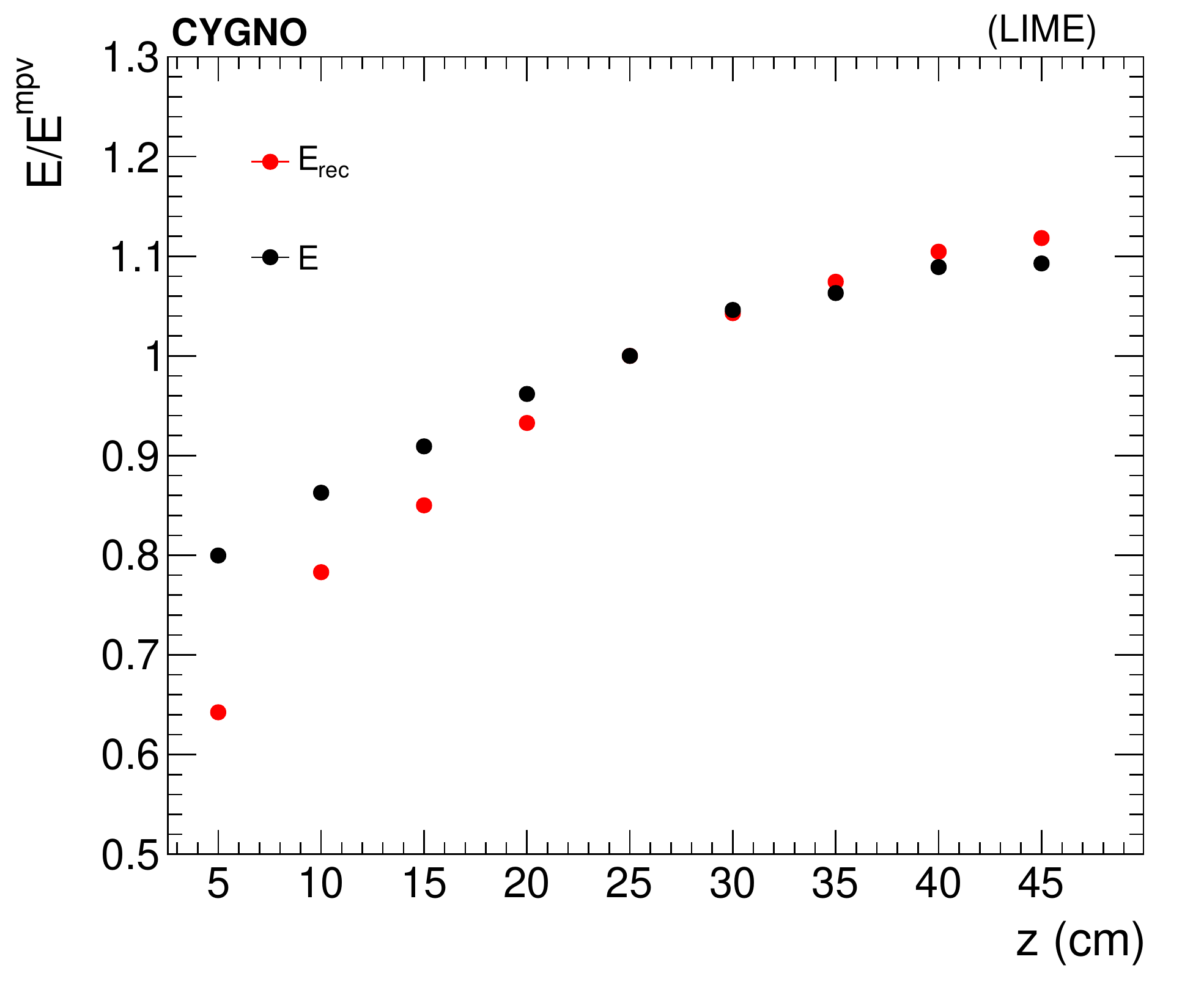}
\includegraphics[width=0.45\textwidth]{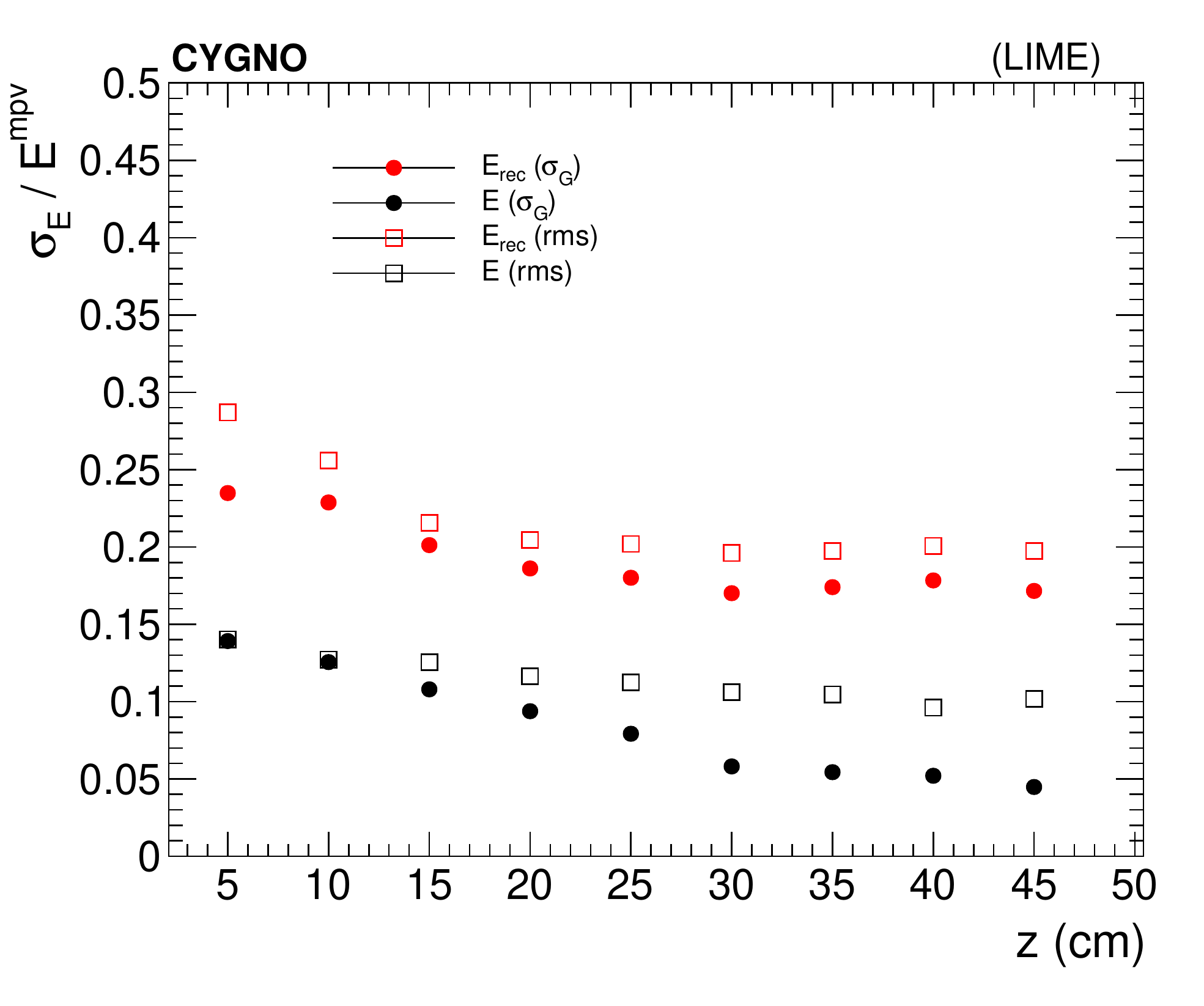}
\caption{Top: average energy response to X-rays from \fe source,
  normalized to the most probable value of the distribution of the
  sample with $z=\SI{26}{cm}$, estimated from the raw supercluster
  energy \Erec (red points) and including the correction with the MVA
  regression, E (black points), as a function of the $z$ distance from
  the GEM planes.  Bottom: energy resolution in the same data,
  estimated either as the RMS of the full distribution (open squares)
  or from the fitted $\sigma_G$ of the Crystal Ball function described
  in the text (filled circles), as a function of the $z$ distance from
  the GEM planes.\label{fig:regression_vsz}}
\end{figure}

The average response is estimated with the fitted value of $m_G$. Its
value, as a function of the $z$ position, is shown in
Fig.~\ref{fig:regression_vsz} (top).  The effect of the saturation is
only partially corrected through this procedure: the consequence of
the gain loss is reduced by about 15\% in correspondence of the
smallest distance tested, $z=\SI{5}{cm}$.  Yet, this small improvement
indicates that it is possible to roughly infer the $z$ position
through a similar regression technique, where the target variable is
$z$, instead of E. This procedure will be discussed in
Sec.~\ref{sec:zestimate}. The same procedure, applied on data samples with 
variable energy and variable $z$ position, would allow to build the model of the 
correction with larger sensitivity to $z$, thus resulting in an improved correction 
of the saturation effect.

On the other hand, it is evident that the MVA regression improves the
energy resolution for any $z$, by correcting effects distinct from the
saturation.  The standard deviation of the Gaussian core of the
distribution is estimated by $\sigma_G$, representing the resolution
of the best clusters.  Clusters belonging to the tails of the
distribution, for which the corrections are suboptimal, slightly worsen
the average resolution. Its effective value for the whole sample is
then estimated with the standard deviation of the full distribution. The values of both
estimators are shown in Fig.~\ref{fig:regression_vsz} as a function of
the $z$ position of the \fe source: for the clusters less affected by
the saturation ($z\gtrsim\SI{15}{cm}$) the RMS value improves from
$\approx 20\%$ to $\approx 12\%$. The best clusters, whose resolution
is estimated with $\sigma_G$, have a resolution smaller than $10\%$
for $z\gtrsim\SI{25}{cm}$, when the saturation effect is small.

\subsection{Study of the response linearity }
\label{sec:linearity}
\begin{figure*}[htbp]
\centering
\includegraphics[width=0.75\textwidth]{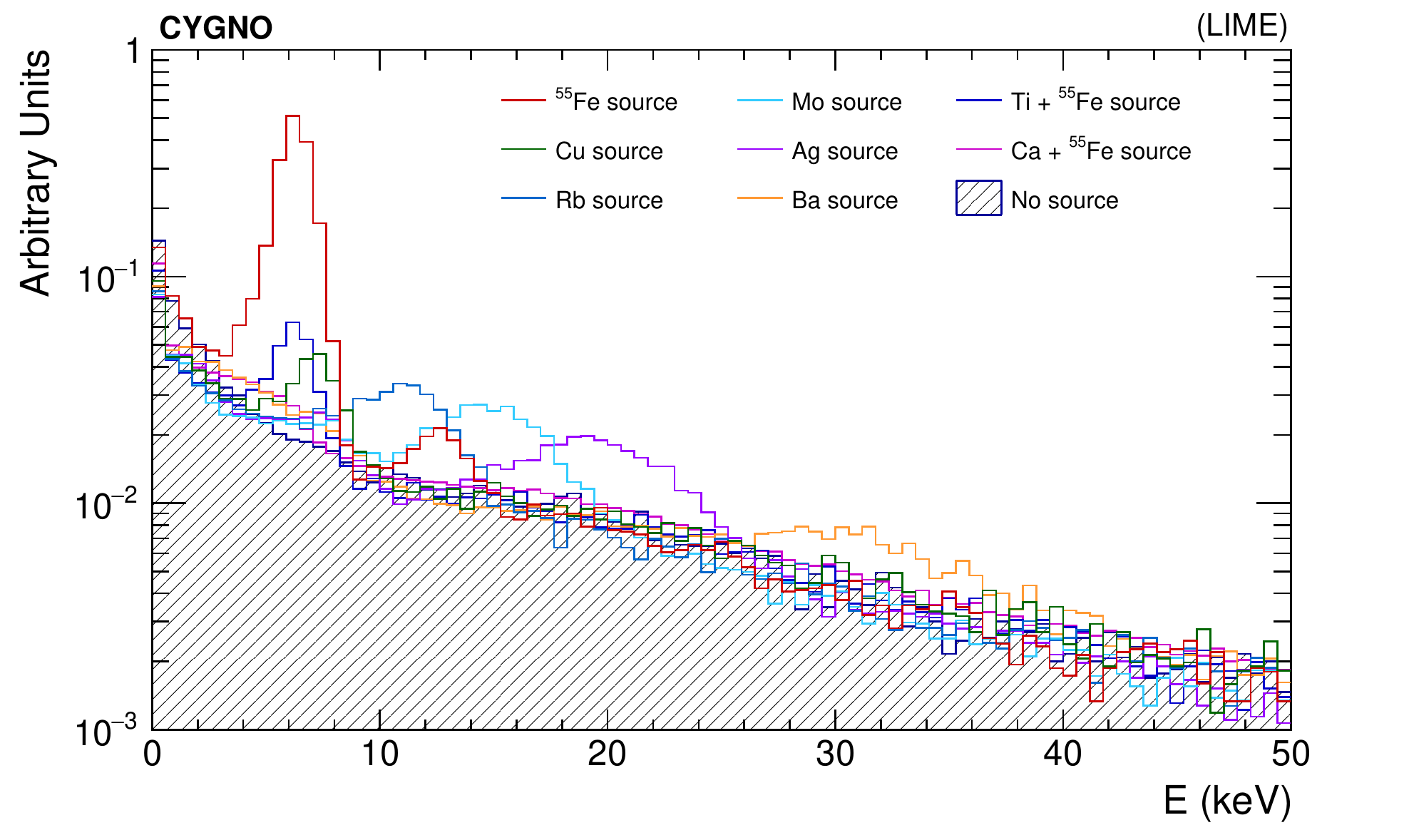}
\caption{Spectra of the calibrated energy E for data collected in
  presence of the radioactive sources, placed at $z=\SI{25}{cm}$, listed in
  Table~\ref{tab:multitarget}, compared to the spectrum of clusters
  reconstructed in a data sample without any source. The distributions
  are normalized to the same live-time. \label{fig:spectra}}
\end{figure*}

The energy response of the detector as a function of the impinging
X-ray energy is studied by selecting clusters reconstructed in
presence of the different radioactive sources enumerated in
Table~\ref{tab:multitarget}, in addition to the large data sample
recorded with the \fe source positioned at the same distance from the
GEM plane. The data used were recorded placing the radioactive source at $z=\SI{25}{cm}$. 
The average energy response of the latter is used to derive
the absolute energy scale calibration constant. The distributions of
the cluster energy E, for the data collected with any of the
radioactive sources used, are shown in Fig.~\ref{fig:spectra}. The
samples are selected with a common loose preselection, and the spectra,
normalized to the live-time, are compared to the one measured in data
acquired without any source. This proves that the shape of the
background is common to all the data samples, thus will be estimated
from this control sample in what follows.

For each data sample a loose cluster selection, slightly optimized for
each source with respect to the loose common preselection, is applied
to increase the signal over background ratio.  As it is shown in
Fig.~\ref{fig:feplots2}, the energy spectrum of the underlying
background from natural radioactivity deposits is in general a
smoothly falling distribution, while the response to fixed-energy
X-rays is a peak whose position represents the mean response to that
deposit, while the standard deviation is fully dominated by the
experimental energy resolution. Deviations from a simple Gaussian
distribution are expected especially as an exponential tail below the
peak, due to non perfect containment of the energy in the
reconstructed clusters.

The average energy response is estimated through a fit of the energy
distribution, calibrated using the one to the \fe source, using two
components: one accounting for the non-peaking background from natural
radioactivity, and one for true X-rays deposits.  The background shape
is modeled through a sum of Bernstein polynomials~\cite{bernstein} of
order $n$, with $n=[1 \dots 5]$: the value of $n$ and its coefficients
are found fitting the energy distribution of clusters selected on data
without the \fe source.  The value of $n$ is chosen as the one giving
the minimum reduced $\chi^2$ in such a fit. The signal shape is fitted
using the sum of two Cruijff functions, each of one is a centered
Gaussian with different left-right standard deviations and exponential
tails~\cite{cruijff}. The two functions represent the contribution of
the $K_{\alpha}$ and $K_{\beta}$ lines listed in
Table~\ref{tab:multitarget}: the energy difference between the two
(denoted main line and $2^{nd}$ line in the figures) is fixed to the
expected value, thus in each fit only one scale parameter is fully
floating.  The remaining shape parameters of the Cruijff functions are
constrained to be the same for the two contributions, since they
represent the experimental resolution which is expected to be the same
for two similar energy values. While the energy difference between the
main and subleading line are well known, the relative fraction of the
two contributions $f_2$ also depends on the absorption rate of low
energy X-rays by the detector walls, so it is left floating in the
fit, with the constraint $f_2<0.3$.  In particular the \fe source was
separately charatecterized with with a Silicon Drift Detectors with
about \SI{100}{eV} resolution on the energy and the fraction of
$K_{\beta}$ transitions was found to be 18\%.  In the case of the
\ce{Rb} target, the range of energy of the reconstructed clusters
covers the region of possible X-rays induced by the $^{241}$Am primary
source impinging the copper rings constituting the field cage of the
detector. Thus a line corresponding to \ce{Cu} characteristic energy
is added: its peak position is constrained to the main \ce{Rb}
$K_{\alpha}$ line fixing the energy difference $\Delta
E=K_{\alpha}^{\ce{Rb}}-K_{\alpha}^{\ce{Cu}}$ to the expected
value. Since only a small contribution is expected from \ce{Cu} with
respect the main \ce{Rb} one, no $K_\beta^{\ce{Rb}}$ is added.  The
normalization of the \ce{Cu} component is left completely floating.

The response to X-rays with lower energies than the \SI{6}{keV}
emitted by \fe have been tested with the \ce{Ti} and \ce{Ca} targets
listed in Table~\ref{tab:multitarget_custom}.  As discussed earlier,
in this setup an unknown fraction of the original \SI{6}{keV} X-rays
also pass through the target, so the fit to the energy spectrum is
performed addding to the total likelihood also the two-components PDF
expected from \fe contribution.  While the shape for the four expected
energy lines is constrained to be the same, except the mean values and
the resolution parameters, the relative normalization is kept
floating. The shape of the natural radioactivity background is fixed
to the one fitted on the data collected without source.

\begin{figure}[htbp]
\centering
\includegraphics[width=0.32\textwidth]{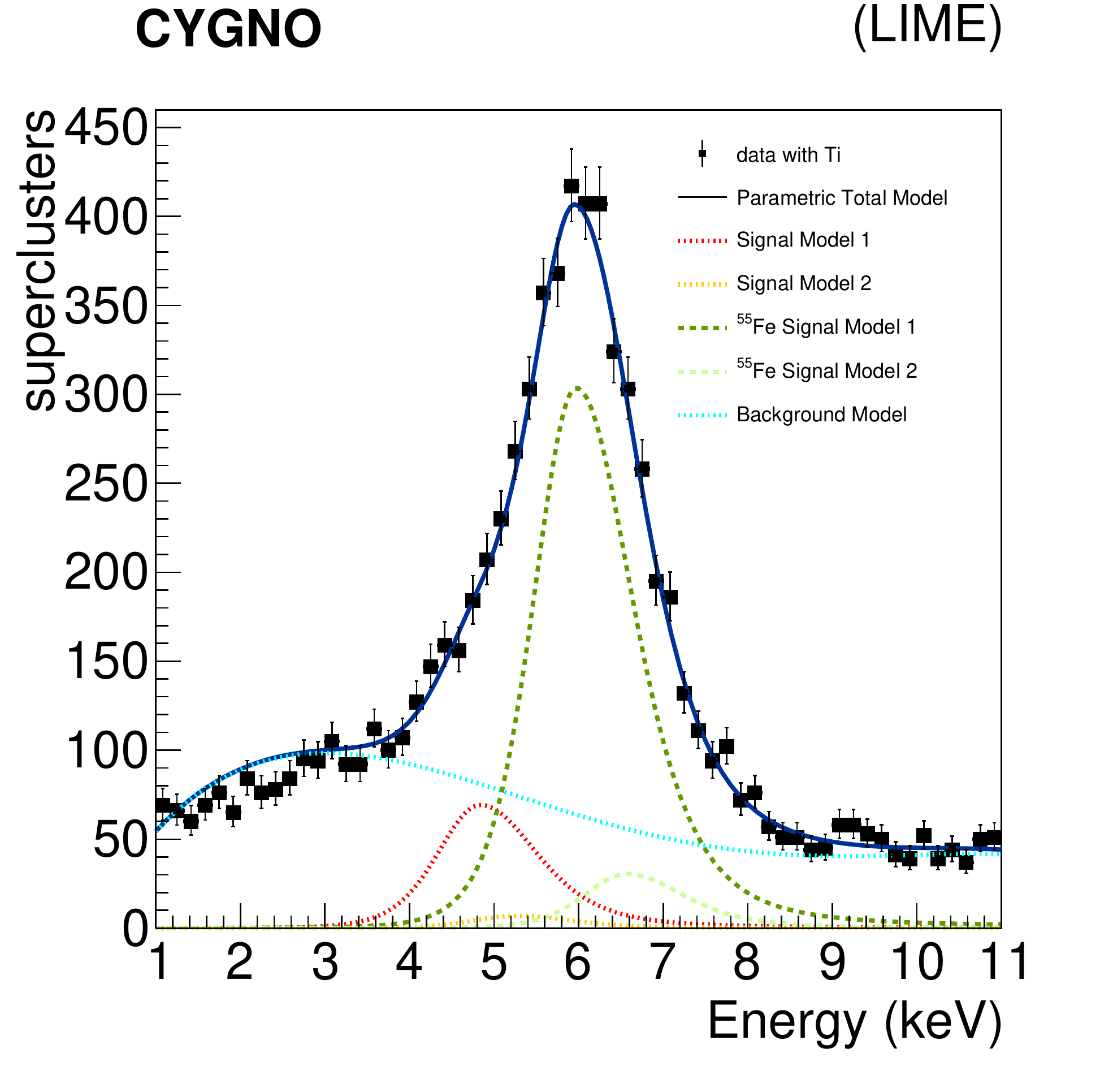}
\includegraphics[width=0.32\textwidth]{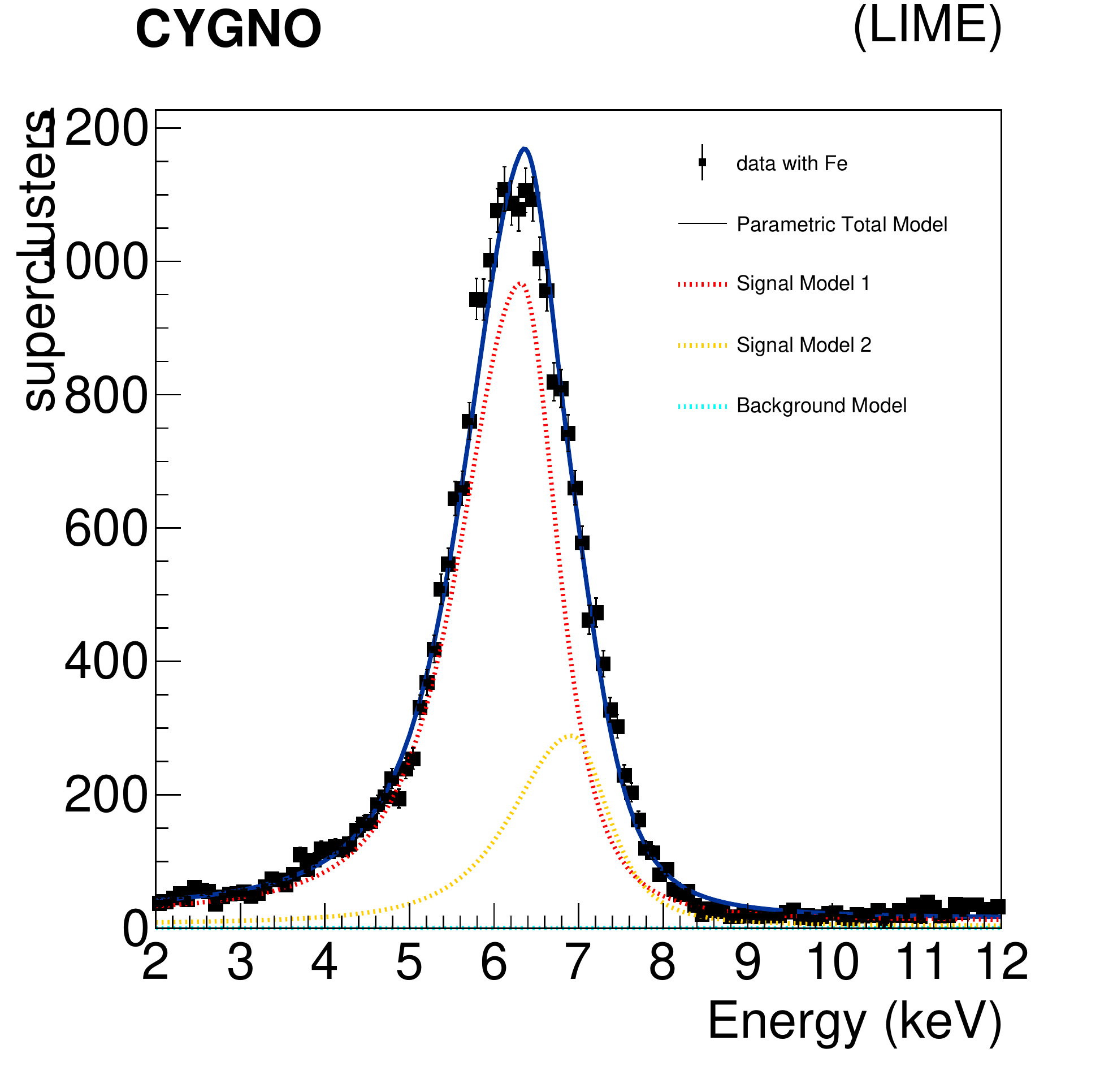}
\caption{Energy spectra of reconstructed clusters in presence of
  different X-ray sources.  Top: \ce{Ti} source. Bottom: \fe source
  (used also to estimate the absolute energy scale calibration
  throughout the paper). Blue dotted line represents the background
  shape, modelled on data without any source; red dotted line
  represent the $K_\alpha$ line signal model; red dotted line
  represents the $K_\beta$ line signal model. The blue continuous line
  represents the total fit function.
\label{fig:multitarget1}}
\end{figure}

\begin{figure}[htbp]
\centering
\includegraphics[width=0.32\textwidth]{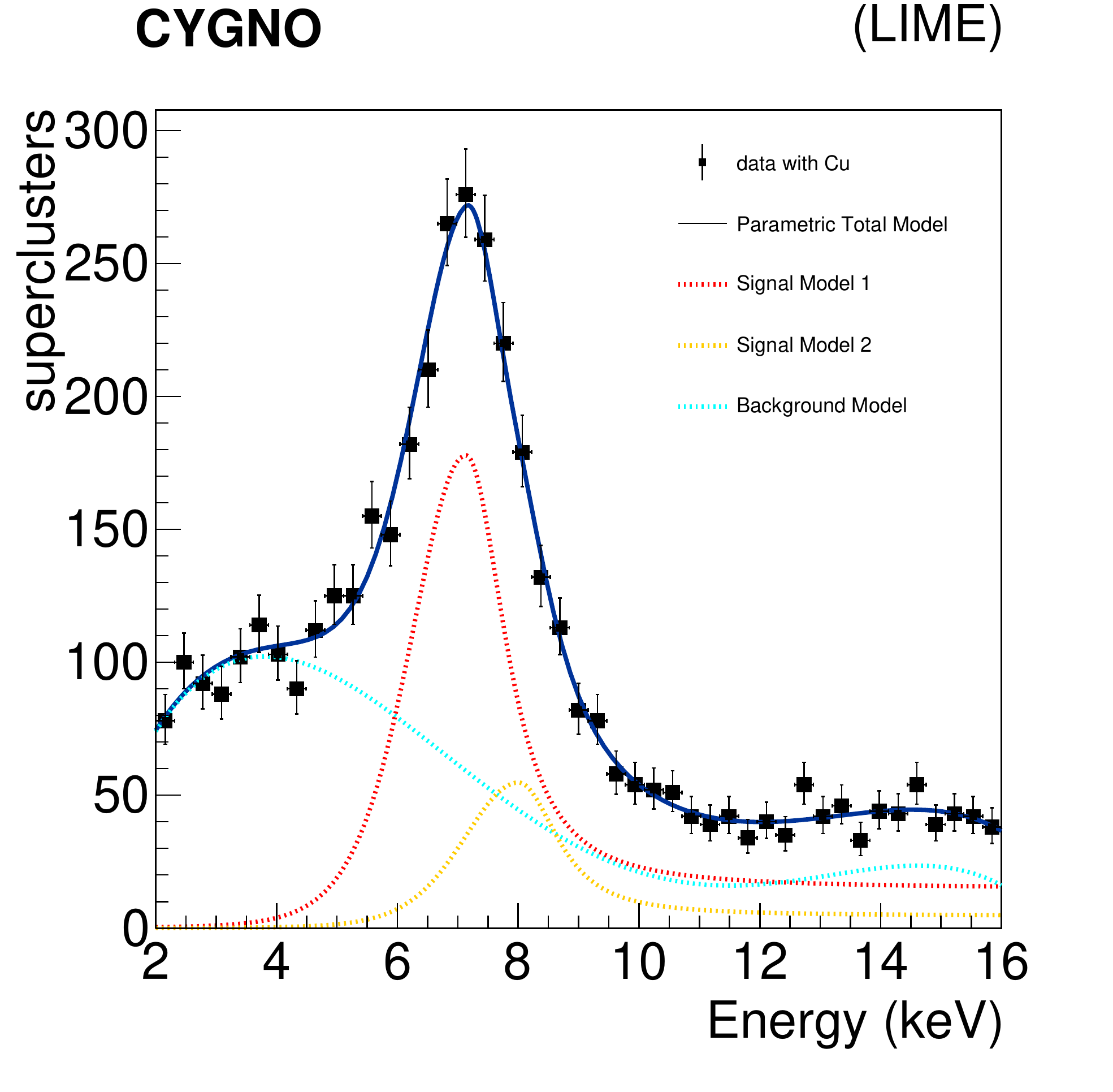}
\includegraphics[width=0.32\textwidth]{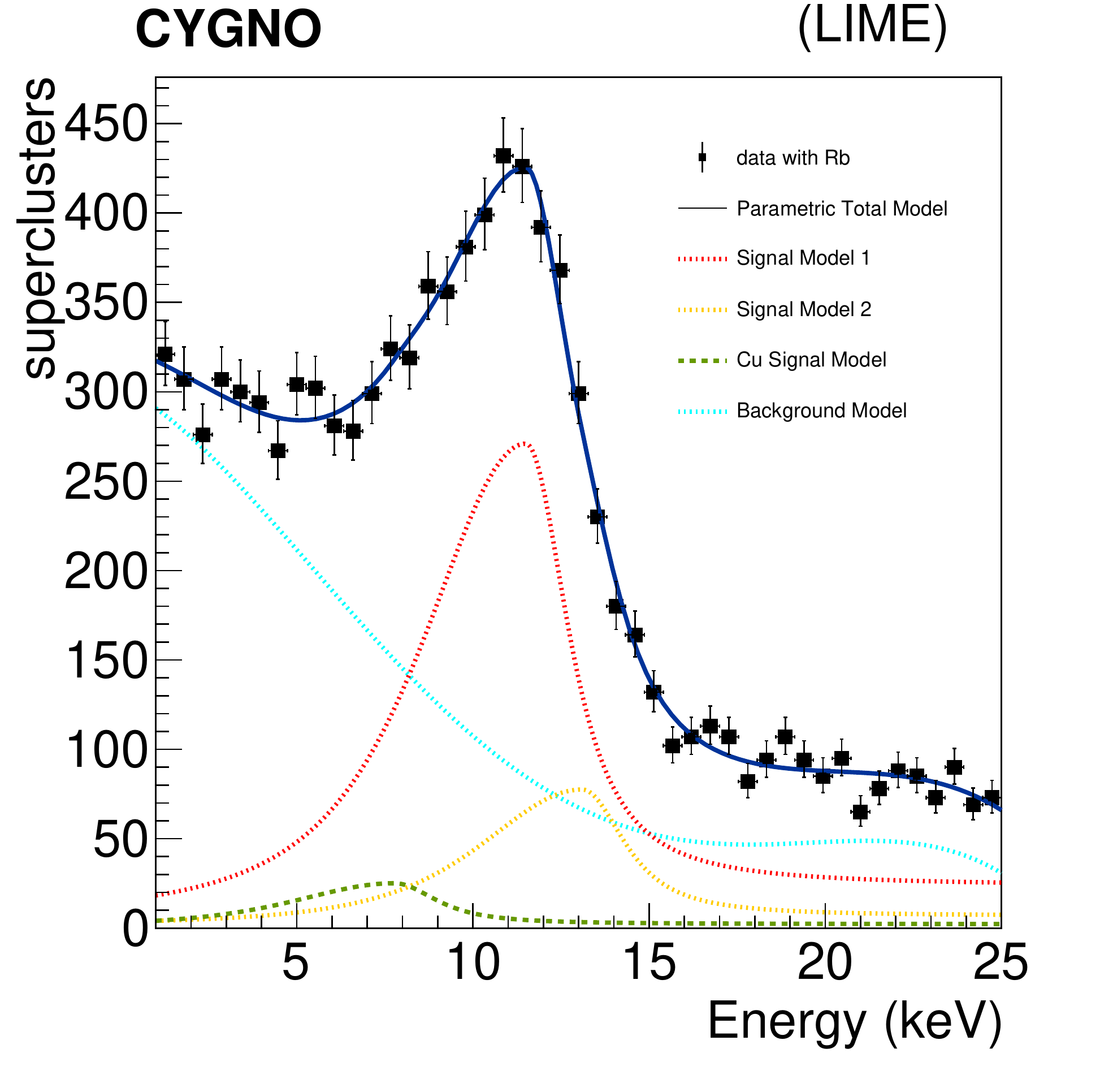}
\caption{Energy spectra of reconstructed clusters in presence of
  different X-ray sources.  Top: \ce{Cu} source. Bottom: \ce{Rb} source.
  Ag source. Bottom: Ba source. Blue dotted line represents the
  background shape, modelled on data without any source; red dotted
  line represent the $K_\alpha$ line signal model; red dotted line
  represents the $K_\beta$ line signal model. The blue continuous line
  represents the total fit function.  As explained in the text, for the
  \ce{Rb} target, a component from the expected contribution of
  \ce{Cu} induced X-rays is added, represented by the green dashed
  line.
\label{fig:multitarget2}}
\end{figure}

Example of the fits to the energy spectra in the data with
different X-ray sources are shown in Fig.~\ref{fig:multitarget1} and
Fig.~\ref{fig:multitarget2}.

\begin{figure}[htbp]
\centering
\includegraphics[width=0.5\textwidth]{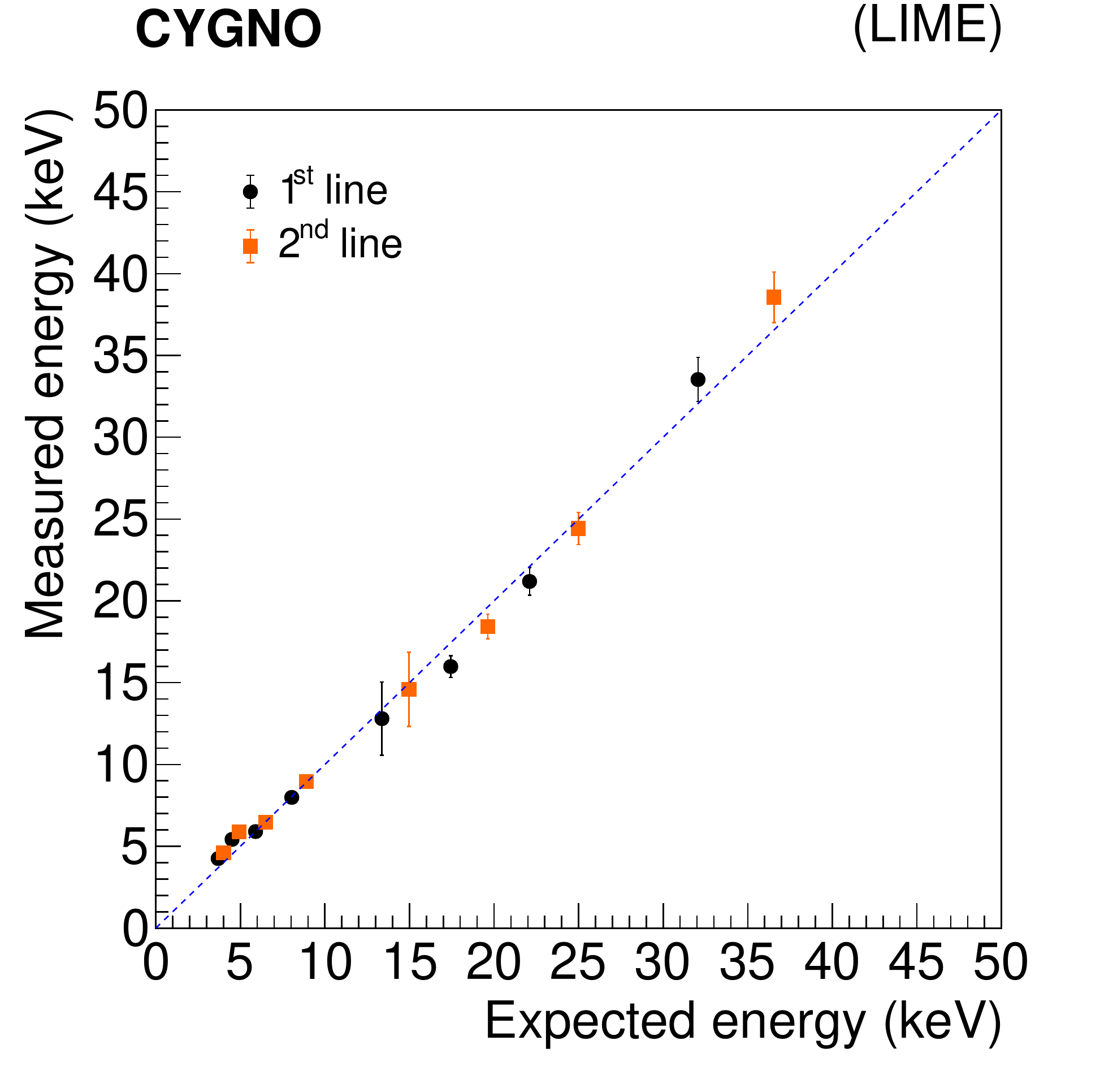}
\caption{Estimated average energy response versus the expected one
  from the $K_\alpha$ (black dots) or $K_\beta$ lines
  contributions. The uncertainties on each point represent the
  statistical contribution and the systematic uncertainty arising from
  the knowledge of the $z$ position. The dotted line represents the a
  perfect linear response of the detector.
\label{fig:linearity}}
\end{figure}

The estimated energy response from these fits, compared to the
expected X-ray energy for each source is shown in
Fig.~\ref{fig:linearity}. In the graph the contributions from both
$K_\alpha$ and $K_\beta$ lines are shown, because both components are
used in the minimization for the energy scale in each fit. The two
values are correlated by construction of the fit model.  A systematic
uncertainty to the fitted value is considered, originating from the
knowledge of the $z$ position of the source. Because of the effect
described in Sec.~\ref{sec:fesource}, a change in this coordinate
results in a change of the light yield: with the source positioned at
$z\approx\SI{21}{cm}$, data with \fe source (shown in
Fig.~\ref{fig:regression_vsz}) allow to estimate a variation $\Delta
E/\Delta z\approx 2\%/\SI{1}{cm}$. An uncertainty $\Delta
z=\SI{1}{cm}$ is assumed for the position of the X-ray source, and the
resulting energy uncertainty is added in quadrature to the statistical
one from the fit.

\section{Evaluation of the $z$ coordinate of the ionization point}
\label{sec:zestimate}

The ability to reconstruct the three-dimensional position in space of events within the detector allows, as has been shown in \cite{Daw:2013waa},  the rejection of those events  too close to the edges of the sensitive volume and therefore probably due to radioactivity in the detector materials (GEM, cathode, field cage).
As shown in other work, the optical readout allows submillimeter accuracy in reconstructing the position of the spots $x$--$y$ plane \cite{NIM:Marafinietal,bib:jinst_orange1}.
The   $z$ coordinate can be evaluated by exploiting the effects of electron diffusion in the gas during the drift path.
The diffusion changes the distribution in space of the electrons in the cluster produced by the ionization and therefore it modifies the shape of the light spot produced by the GEM and collected by the CMOS sensor.
Based on this, a  simple method was developed for ultra-relativistic particle  tracks \cite{bib:relativistic}, relying on $\sigma_T$ (see for example ~Fig.~\ref{fig:profiles}). 

We evaluated the $z$-reconstruction performance  by studying the behavior of several shape variables of the spots produced by the \fe source, and therefore at a fixed energy, as a function of the $z$ coordinate of the source (z$_{\fe}$ in the following).

\begin{figure}[htbp]
\centering
\includegraphics[width=0.45\textwidth]{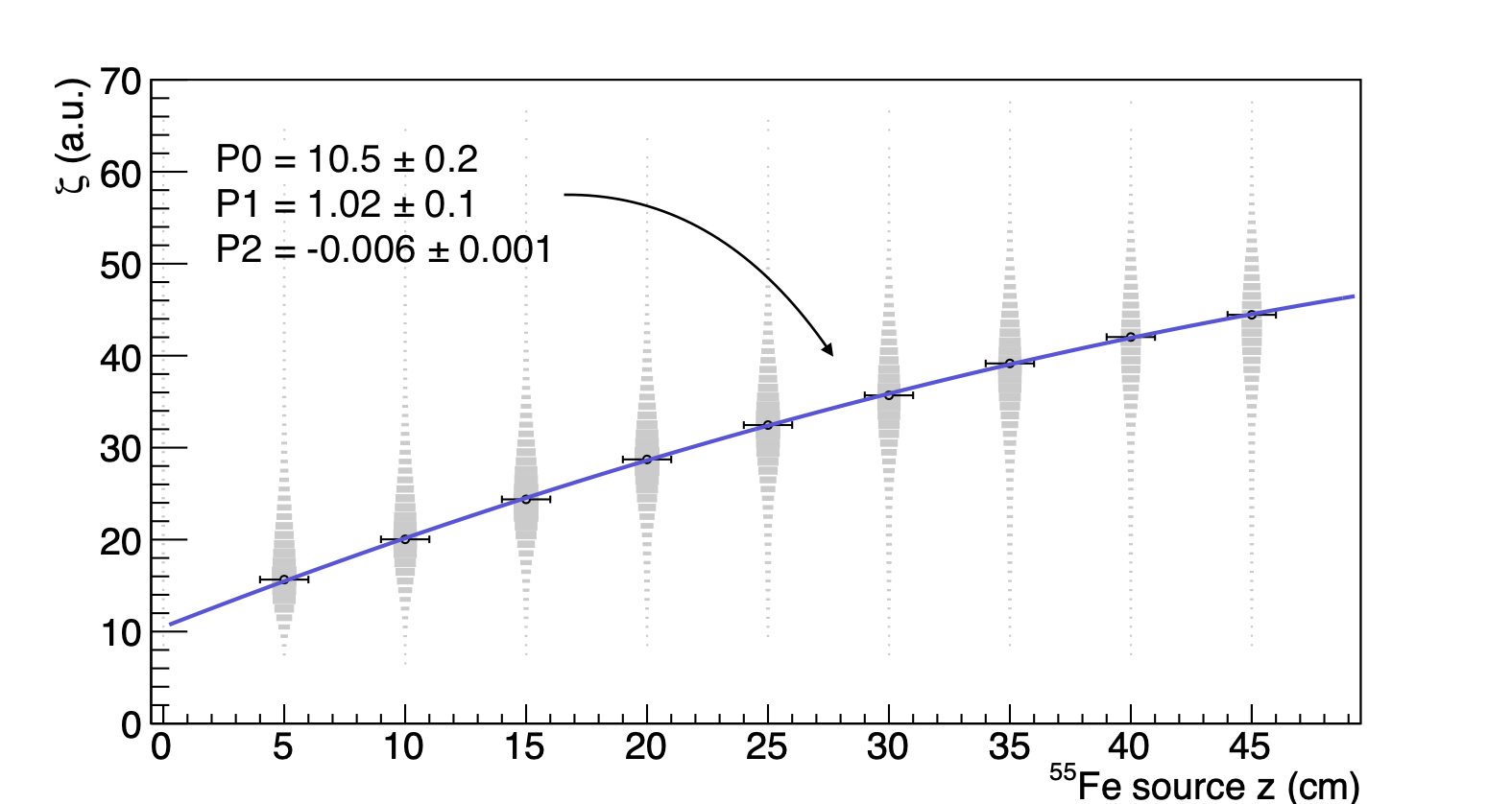}
\includegraphics[width=0.45\textwidth]{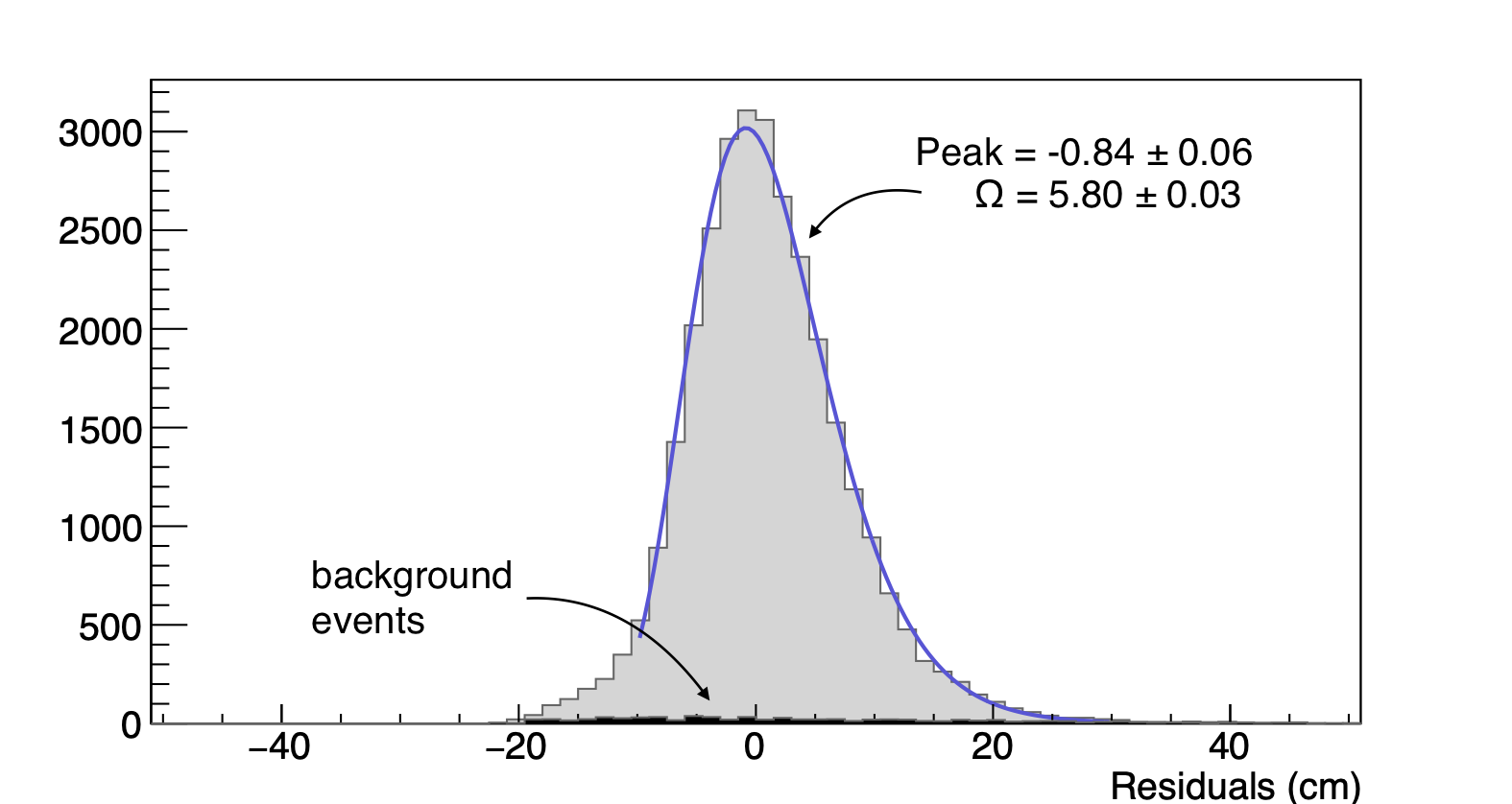}
\caption{Top: distribution of the values of $\zeta$ (see text) in the runs with the \fe source at different z$_{\fe}$. Bottom: distribution of the $z$-residuals  at  z$_{\fe}$ =  20 cm with a superimposed fit to the Novosibirsk function.} 
\label{fig:absz}
\end{figure}

The variable that showed  a better performance is $\zeta$  defined as the product of the gaussian sigma fitted to transverse profile of the spots (see Fig. \ref{fig:profiles}) $\sigma_T$ and the standard deviation of the counts per pixel inside the spots \SI{}{I_{rms}}.
Figure~\ref{fig:absz} shows on the left the distribution of $\zeta$ of all reconstructed spots as a function of nine values of z$_{\fe}$ (in the range from \SI{5}{cm} to \SI{45}{cm}). For each value of z$_{\fe}$ the mean value of the distribution of $\zeta$ is superimposed together with a quadratic fit to the trend of these averages as a function of z$_{\fe}$.

As can be seen, although there are large tails in all cases, the main part the spots provide values of $\zeta$ increasing as $z$ increases.

\begin{figure}[htbp]
\centering
\includegraphics[width=0.5\textwidth]{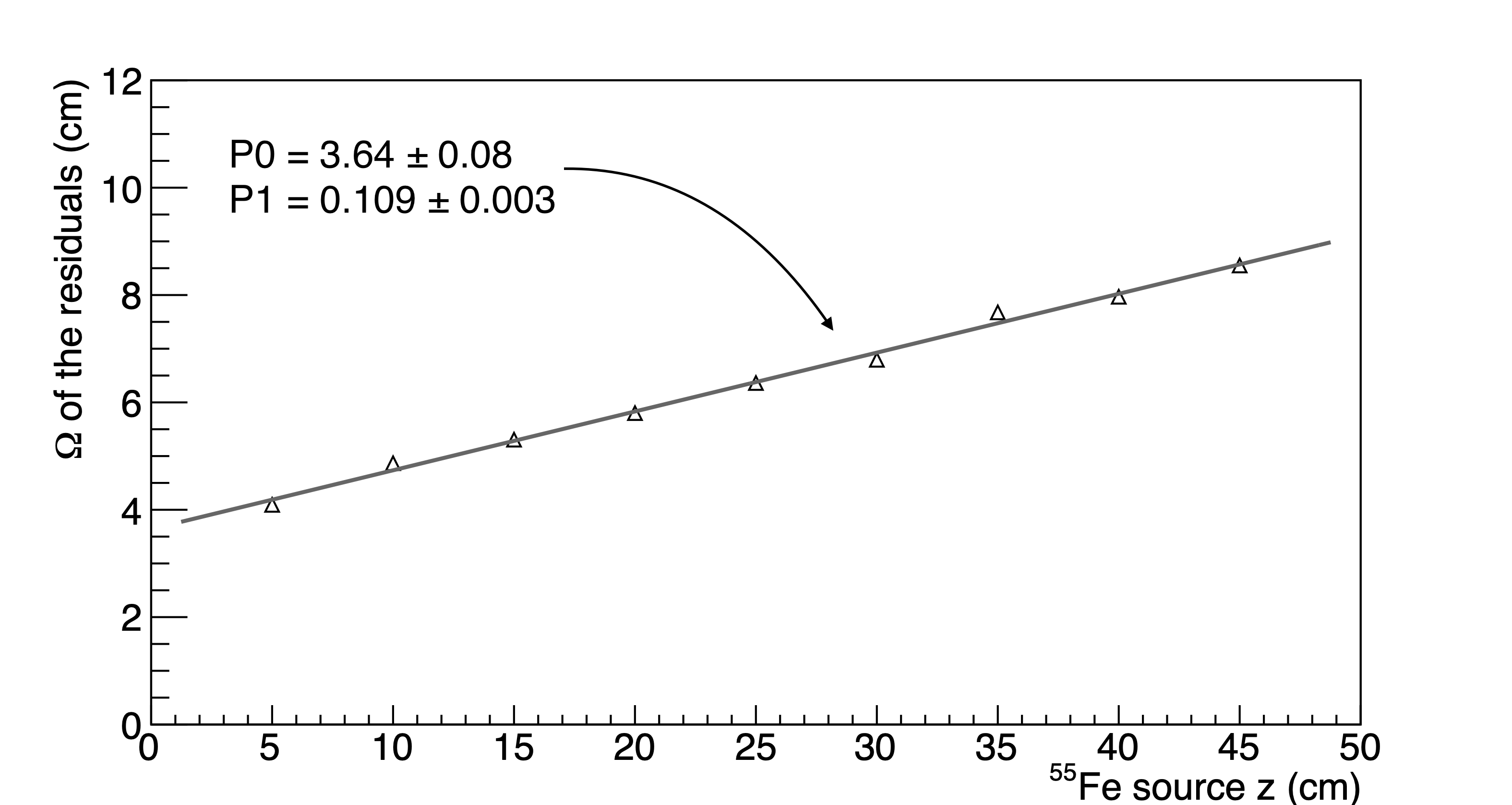}
\caption{Behaviour of the values of the $\Omega$ evaluated from the Novosibirsk function on the residuals distributions as a function of z$_{\fe}$ with a superimmposed linear fit.} 
\label{fig:res}
\end{figure}

Shown on the bottom side of the figure there is the distribution of the $z$ residuals of the clusters reconstructed from the measured $\zeta$ for a z$_{\fe}$ value of 20 cm. The distribution  of the residual was fit with a Novosibirsk function \cite{bib:novo} and from this fit, the value of the parameter $\Omega$  \footnote{$\Omega$ is defined as FWHM/2.36 }  was extracted.
The $\Omega$ values obtained for the nine datasets at the various positions are plotted as a function of the  nine z$_{\fe}$ in Fig \ref{fig:res}. 

As can be seen, although the absolute uncertainty worsens slightly as the distance of the spots from the GEM increases, this method showed to be able to provide an estimate of $z$ of \fe photons interactions, with an uncertainty of less than \SI{10}{cm} even for events occurring near the cathode.

\section{Study of the absorption length of \fe X-rays}

From the above studies the overall LIME performance is  found to be excellent to detect low energy electron recoils. 
We then analyzed the  \fe  data   to measure the average absorption length $\lambda$ of the  \fe  X-rays.
As we have seen, the source mainly emits photons of two different energies (5.9 keV and 6.5 keV).
For these two energy values the  absorption lengths $\lambda$  in a 60/40 He/CF$_4$ mixture  at atmospheric pressure were estimated (from \cite{pressure1,pressure2}  to be \SI{19.5}{cm} and \SI{25.6}{cm}, respectively. 
A variation of the order of 10\%  of CF$_4$ fraction reflects in a variation of the $\lambda$ value of about 2.0 cm. 
In particular, an higher  amount of CF$_4$  results in a lower  $\lambda$ value.

A Monte Carlo (MC) technique  was then used  to evaluate the spatial distribution of the interaction points of a mixture of photons of the two energies (in the  proportions reported in Sec.\ref{sec:linearity}).
Being the $z$ coordinate uncertainty relatively large,  we used only the $x$ and $y$ coordinates to infer $\lambda$. With this MC we then evaluated the effect of  the missing  $z$ coordinate information on the measurement of $\lambda$.   In this MC we  took into account the angular aperture of the X-rays  exiting the collimator,  estimated to be 20$^{\circ}$. For each  simulated interaction point,  the distance $d$ from the source (located above the LIME vessel) was then calculated. From the exponential fit of the $d$ distribution,  we obtained a simulated expected value of the effective absorption length $ \lambda_{eff}$ = \SI{20.4}{cm}.

 In data we then studied the reconstructed  $d$ values  in runs taken with the \fe source at the nine different distances from the GEM. Some variation of the reconstructed value of  $\lambda$ as a function of the range of $y$ studied was found, with large uncertainties in the regions far from the GEM centre where optical distortions are  more important. For this reason, our study was carried out  eliminating the bands of the top and bottom \SI{6}{cm} in $y$.

The background distribution in the region of interest was obtained from runs taken without the source. The distribution of $d$ values in this case was found to be substantially flat. The  distribution in \fe events  was then fitted to an exponential function summed to a constant term fixed to account for the background events.

\begin{figure}[htbp]
\centering
\includegraphics[width=0.5\textwidth]{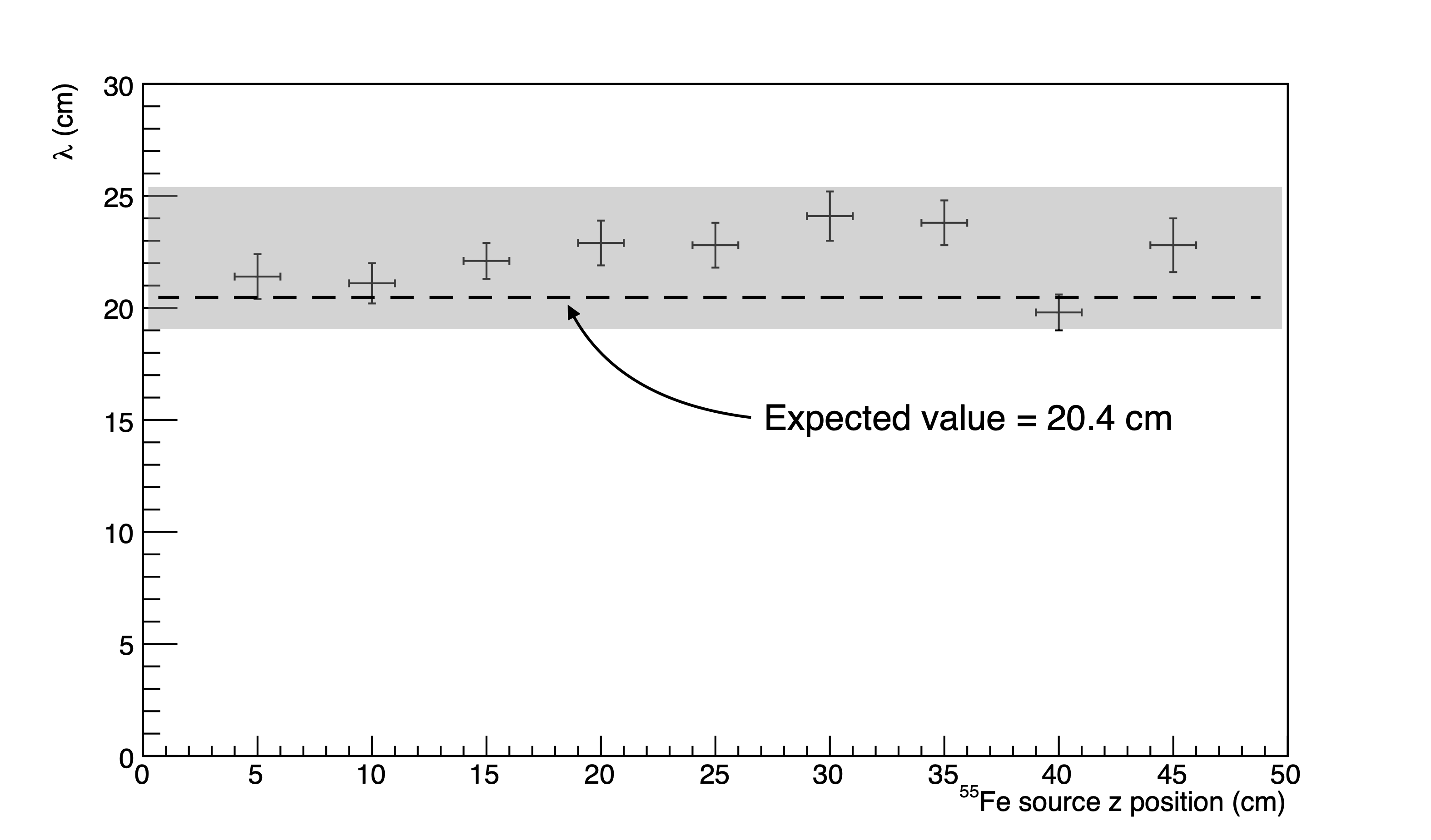}
\caption{$\lambda$ values resulted from exponential fits to $d$ distribution in  data taken with the \fe source at different $z_{\fe}$ positions.} 
\label{fig:lambda_vs_z}
\end{figure}

 To study possible systematic effects introduced by the charge transport along the drift field, the reconstructed $\lambda$ was first evaluated at different \fe source positions along the $z$-axis and  shown  in Fig.~\ref{fig:lambda_vs_z}. 
 
 Variations  of the order of 3.0~cm around the mean value, which is estimated to be \SI{22.4}{cm}, are visible, however no clear systematic trend is present. 

\begin{figure}[htbp]
\centering
\includegraphics[width=0.5\textwidth]{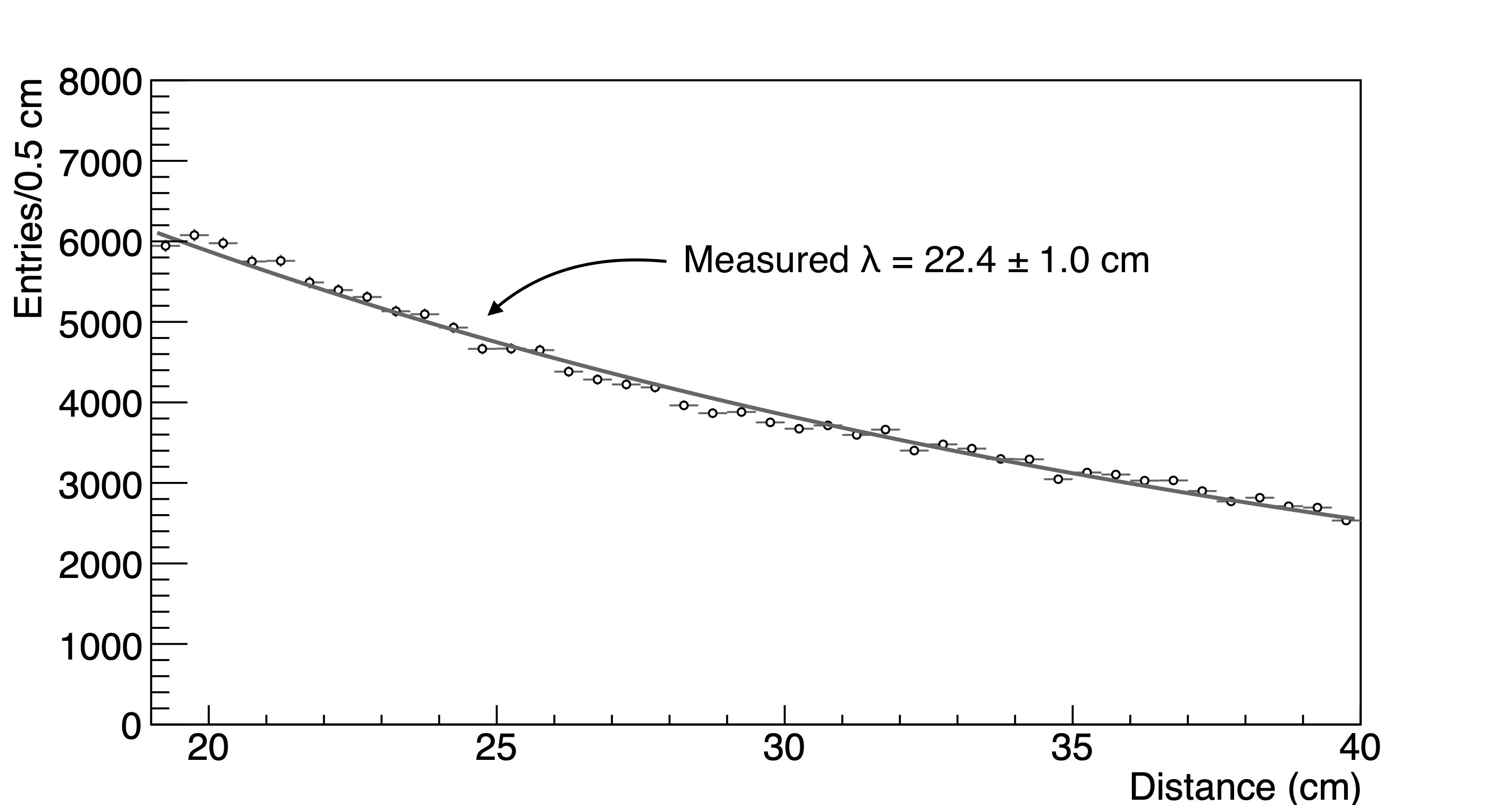}
\caption{Distribution of $d$  with superimposed exponential fit for all the data at all distance of \fe source from the GEM plane.} 
\label{fig:lambda_tot}
\end{figure}

Figure \ref{fig:lambda_tot} shows the distribution of the values of $d$ evaluated at all the $z_{\fe}$     with a superimposed fit.

This analysis  provides a value reasonably in agreement with the expected one, given the statistical fluctuations and possible systematic errors not accounted here.

A more relevant result lies in the fact that in this measurement no
systematic effects due to the position of the spots were revealed,
either in the $x$--$y$ plane  of the
image or versus $z_{\fe}$.  This allows us to conclude that the charge
transport and detection efficiency within the sensitive volume of the
detector shows good uniformity.

\section{Long term stability of detector operation}
\label{sect:stability}

A DM search is usually requiring long runs of data-taking of months or even years. This imposes the capability to monitor the stability of the performance of the detector over time. 
We then evaluate  the stability of  the LIME prototype by  maintaining  the detector running for two weeks at LNF.  Without any direct  human intervention, runs of pedestal events and \fe  source runs were automatically collected. In two occasions, data were not properly saved because of an issue with the internal network of the laboratory.

\begin{figure}[htbp]
\centering
\includegraphics[width=0.50\textwidth]{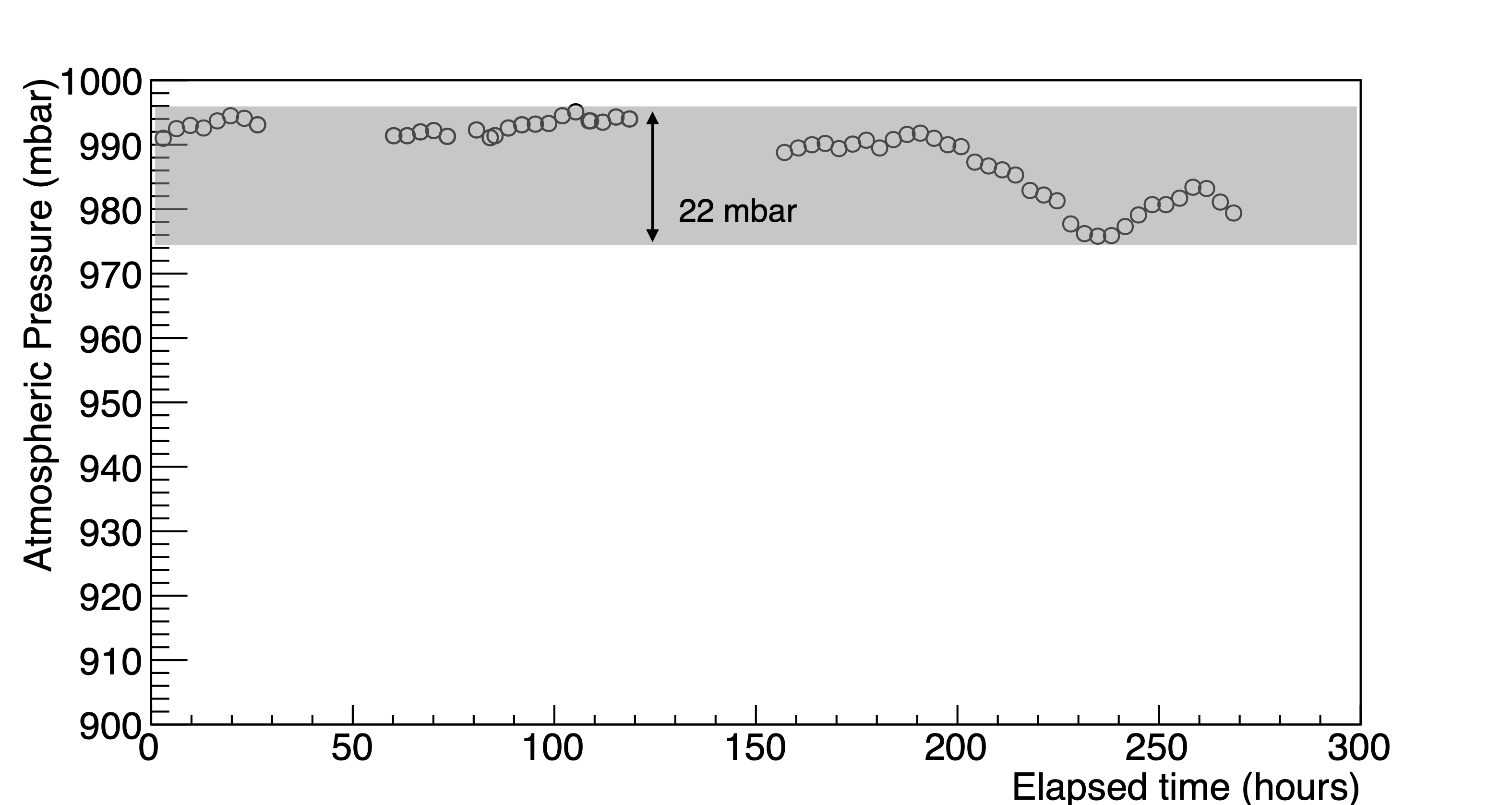}
\caption{Atmospheric pressure recorded during the runs acquired for the test on the LIME's response stability.} 
\label{fig:pt}
\end{figure}

The laboratory is equipped with a heating system to keep the temperature under control. Therefore in this period the room temperature was found to be quite stable with an average value of 298.7 $\pm$ 0.3~K.
In the same period the atmospheric pressure showed visible variations with an important oscillation of about \SI{15}{mbar} in the latest period of the test as it is shown on the bottom in Fig.~\ref{fig:pt}.

\begin{figure}[htbp]
\centering
\includegraphics[width=0.45\textwidth]{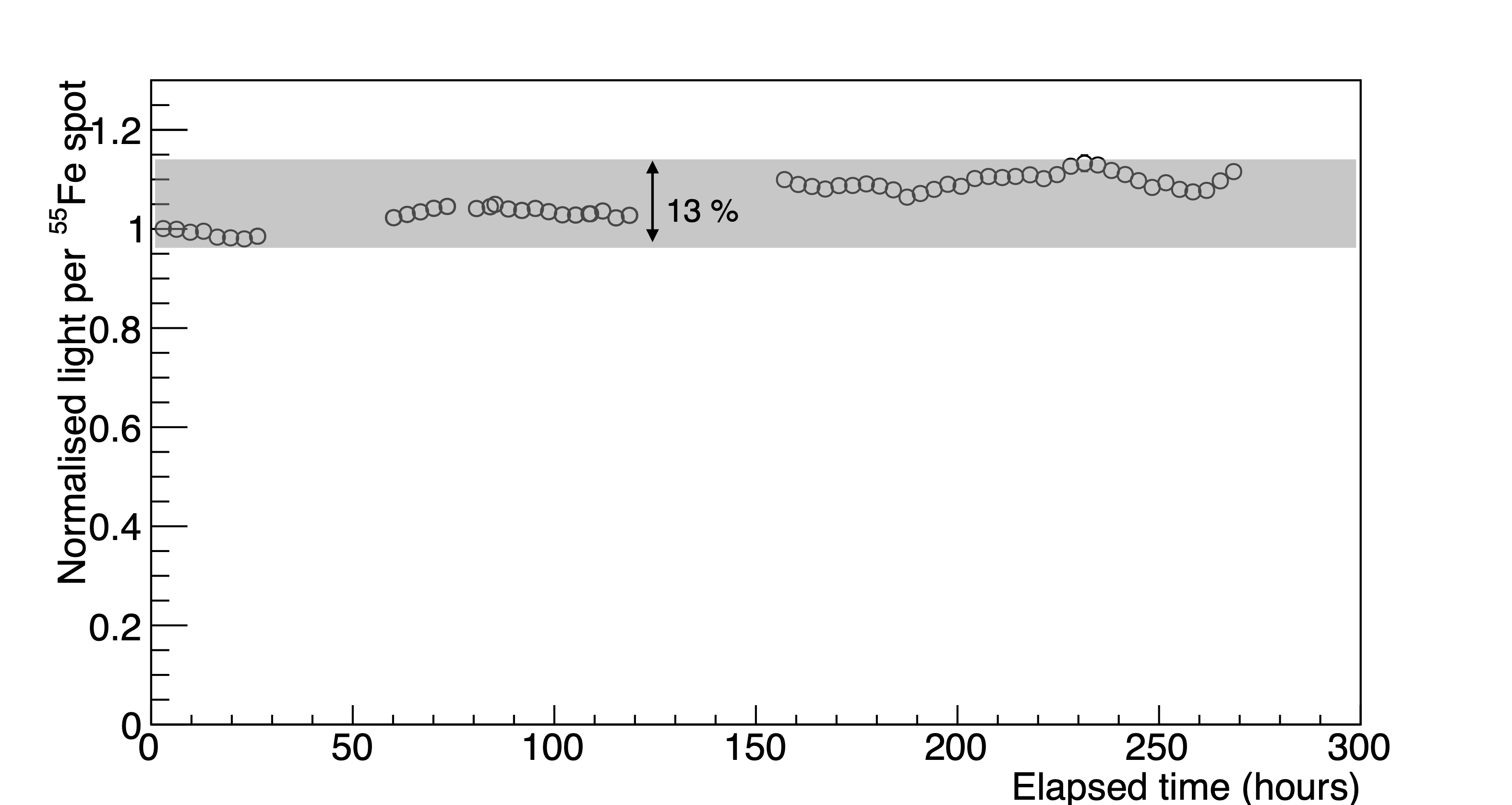}
\includegraphics[width=0.45\textwidth]{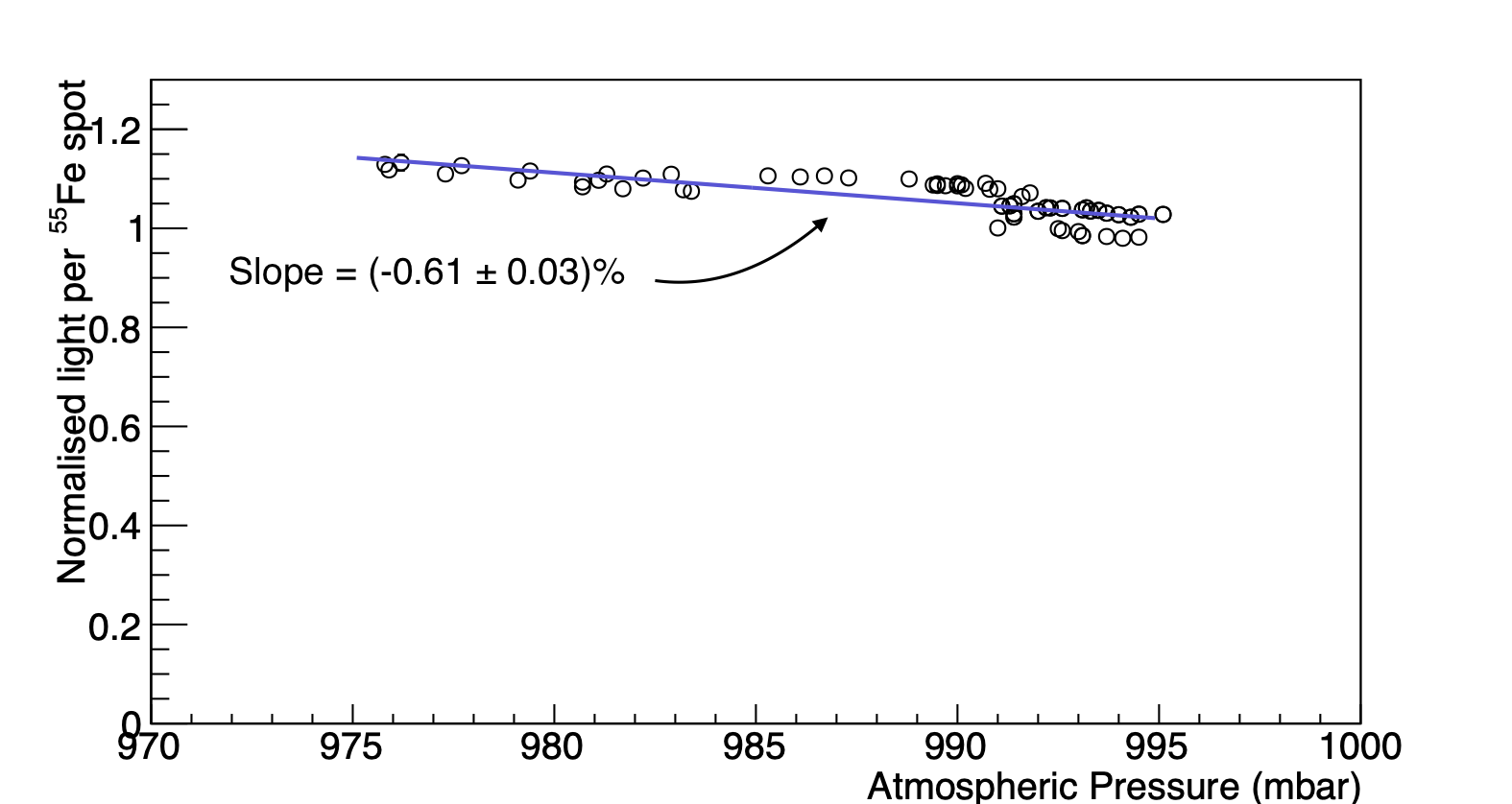}
\caption{Behavior of the number of photons as a function of elapsed time normalised to the initial value (top) and as a function of the atmospheric pressure (bottom) with a superimposed linear fit.} 
\label{fig:respo}
\end{figure}

The average number of photons in the spots of \fe X-ray  interactions was evaluated and its behavior (normalised to the initial value) is shown on the top in  Fig. \ref{fig:respo}.

 The detector light yield shows an almost constant increase during the whole data-taking period. This behavior can be directly correlated with the variation of the gas pressure as shown on the bottom of Fig. \ref{fig:respo}.

From the result of the superimposed linear fit, we evaluated a light yield decrease of about 0.6\% per millibar due to the expected decrease of the gas gain with the increasing of the gas density \cite{bib:blum}.

\section{Background evaluation at LNF}
\label{sect:spectrum}

The data taken  with the LIME prototype at LNF in absence  of any artificial source were analyzed. A number of  interactions of particles in the active volume were detected. The origin of these particle can be ascribed to various sources, primarily the decays of radioactive elements present in the materials of the detector itself and of  the surrounding environment and cosmic rays. Those interactions are to be considered as a background in searches for ultra-rare events as the interaction of a DM particle in the detector. A first assessment of this background is therefore necessary to understand how to improve in future  the radiopurity of the detector itself. Shielding  against cosmic rays can be achieved by operating the detector in an underground location (as INFN LNGS) while the effect of the radioactivity of the surrounding environment can be largely mitigated by using high radiopurity passive materials (as water or copper) around the active volume of the detector.

The analysis of the images reveals the presence of several
interactions that the reconstruction algorithm is able to identify
with a very good efficiency. Due to the fact that LIME was not built
with radiopure materials and given the overground location of the
data-taking, crowded images are usually acquired and
analyzed. Sometimes, because of the piling-up of two or more tracks in
the image, the reconstruction can lead to an inaccurate estimate of
the number of tracks. Because the iterative procedure of the step
\ref{reco:supercluster} of the reconstruction, described in
Sec.~\ref{sec:reconstruction}, when a long cluster is reconstructed
all the pixels belonging to it are removed. This implies that   in the next iteration
the pixels in the overlap region with another track are no more available and the other 
overlapping track is typically split in two pieces. This results in a number of
reconstructed long clusters systematically higher than the true one.
 
\begin{figure}[ht!]
\centering
\includegraphics[width=0.32\textwidth]{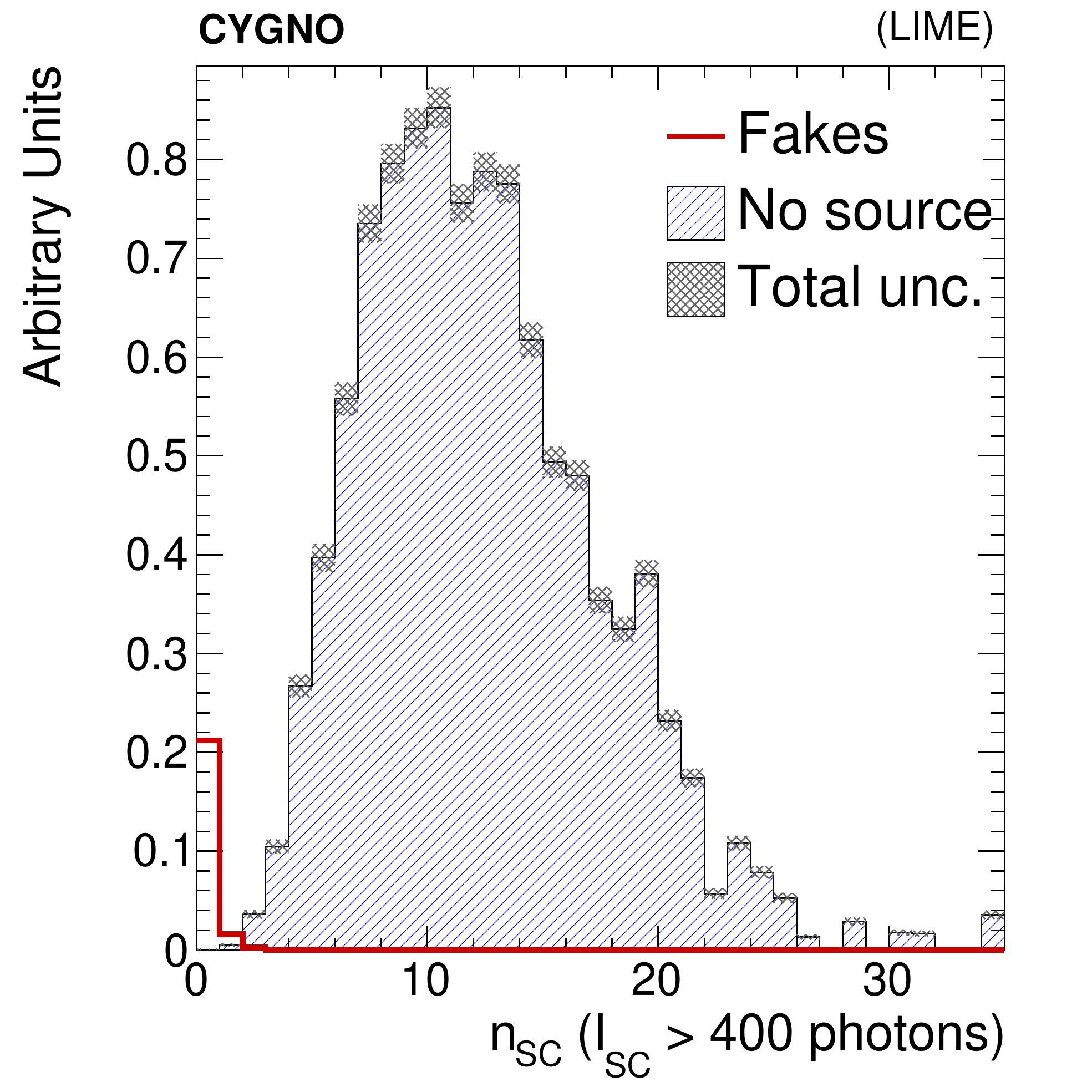}
\includegraphics[width=0.32\textwidth]{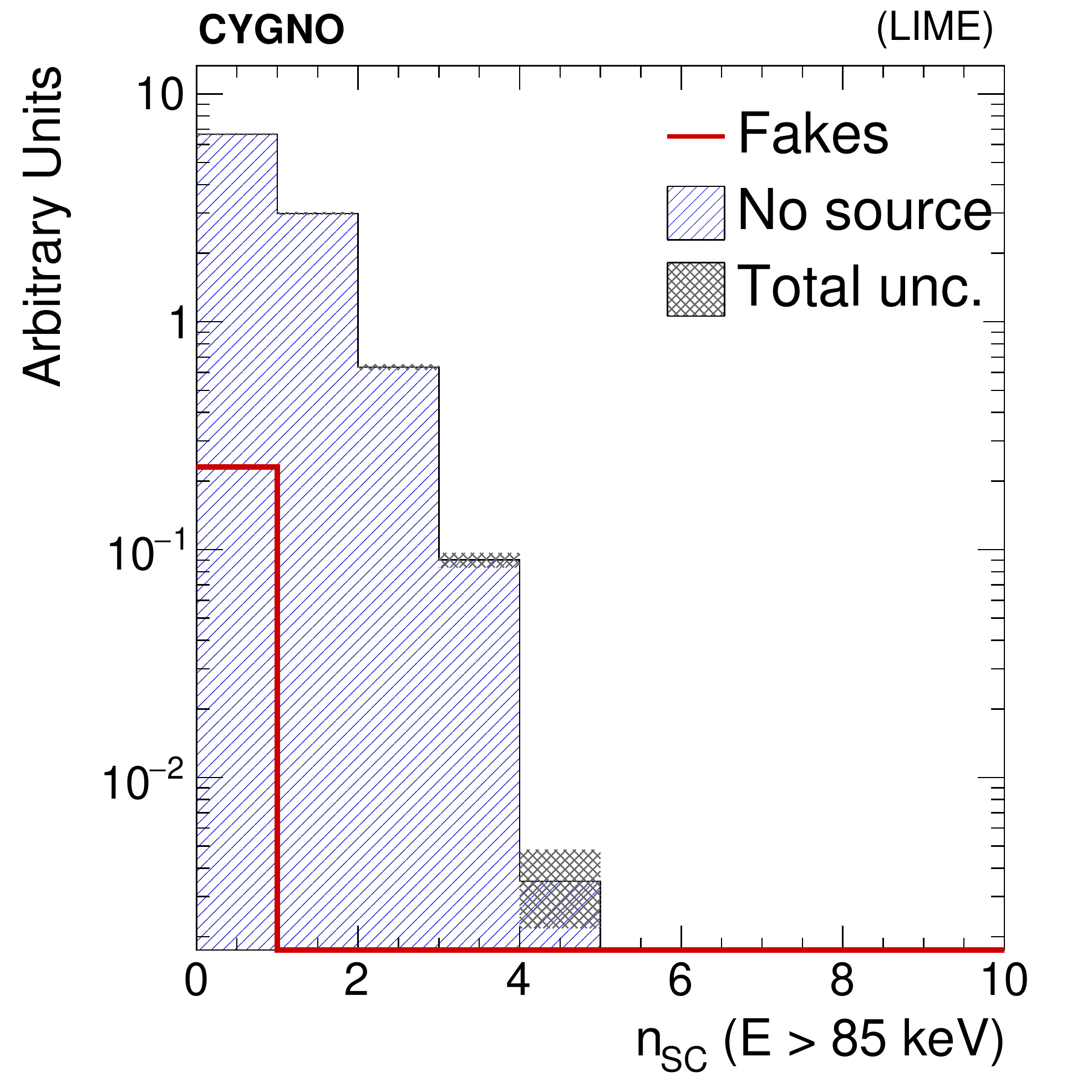}
\includegraphics[width=0.32\textwidth]{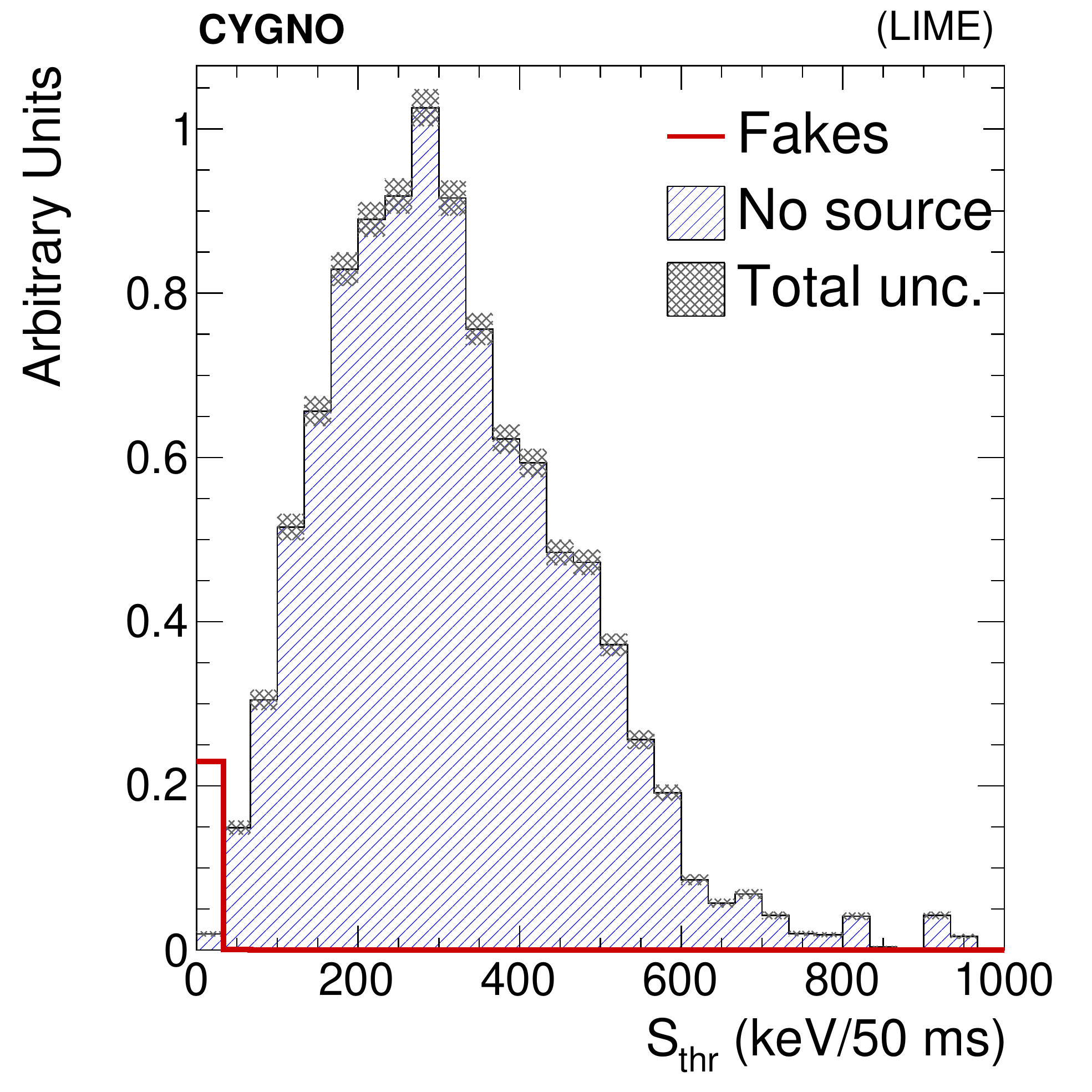}
\caption{Top: number of clusters reconstructed
  in each image with a minimal threshold on the
  light yield to remove fake clusters, $\Isc>\SI{400}{photons}$
  (corresponding to an energy $E\gtrsim\SI{300}{eV}$).  Middle: number of clusters with energy
  $E>\SI{85}{keV}$ reconstructed in each image. Bottom: distribution
  of $S_\mathrm{thr}$, sum of the energy for all the reconstructed
  clusters in one image with energy $E\gtrsim\SI{300}{eV}$. The filled
  histogram represents data without the source, while the red hollow
  histograms represents the estimated contribution from fake
  clusters. All the images have been acquired with an exposure of
  \SI{50}{ms}.\label{fig:cmos_energy}}
\end{figure}

In Fig.~\ref{fig:cmos_energy} (top) the distribution of the number of
reconstructed super-cluster per image  in a sample of $\approx 2000$ images is shown. Each
image corresponds to a live-time (i.e. the total exposure time of the
camera) of \SI{50}{ms} and   these images were acquired in a period of
about 10 minutes.  The  requirement $\Isc>\SI{400}{photons}$ is applied
on the minimal energy of the cluster, in order to remove the
contribution of the fake clusters, as shown in Fig.~\ref{fig:feplots2}
(top), which corresponds to a threshold of
$E\gtrsim\SI{300}{eV}$. This corresponds to an average rate of detected
interaction of $r\approx\SI{250}{Hz}$.  Figure~\ref{fig:cmos_energy}
(bottom) shows the distribution of the energy sum for all the clusters
satisfying the above minimum energy threshold in one image, defined as
$S_\mathrm{thr}$. The average $S_\mathrm{thr}$   per unit time is 
$ \approx \SI{6.3}{MeV/s}$.

During the data taking a 3x3 inches NaI crystal scintillator detector
(Ortec 905-4) was used to measure the environmental radioactivity in
the LNF location of LIME. The lowest threshold to operate this NaI
detector was \SI{85}{keV}. A rate of \SI{350}{Hz} of energy deposits
was measured.  By scaling this NaI rate to the mass of the LIME active
volume a rate of \SI{11}{Hz} is predicted. This can be compared with
the average rate of $\approx \SI{20}{Hz}$ measured by counting the
number of reconstructed cluster with $E > \SI{85}{keV}$ in LIME whose
distribution is shown in Fig.~\ref{fig:cmos_energy} (middle). For this
comparison we selected only the clusters in a central region of the
active volume where the signal to noise ratio is larger. This
corresponds to a geometrical acceptance of about 50\%. This
demonstrated that at the LNF location only part of the contribution to
background is due to the external radioactivity.

The overground location of the LIME prototype implies that a
significant flux of cosmic rays traverses the active volume, releasing
energy with their typical energy pattern of straight lines.  This
allows to define a cosmic rays sample with excellent purity by applying
a simple selection on basic cluster shapes. The track   length can be estimated  as the major axis of the cluster  and compared with  the length of a curved path interpolating the cluster shape. By requiring the ratio of the these two variable to be  larger than 0.9, straight tracks are selected against curly tracks due to natural radioactivity. Further requirements are  the track length being larger than
\SI{10}{cm} and the ratio between the \St and the length lower than 0.1  in order to avoid tracks with small branches due to mis-reconstructed overlapping clusters. 
 The ratio between the energy $E$ associated to the cosmic ray cluster and its length can be described in terms of  the specific ionization of a minimum ionizing particle. 
 Using the standard cosmic ray flux at sea level of~$\approx  \SI{70}{Hz \, m^{-2} \, sr^{-1}}$~\cite{Workman:2022ynf} we predict a maximum rate of interaction in the active volume of  $\approx $ \SI{24}{Hz} to be compared with a measured rate of   $\approx $ \SI{15}{Hz}.

\begin{figure}[ht!]
\centering
\includegraphics[width=0.40\textwidth]{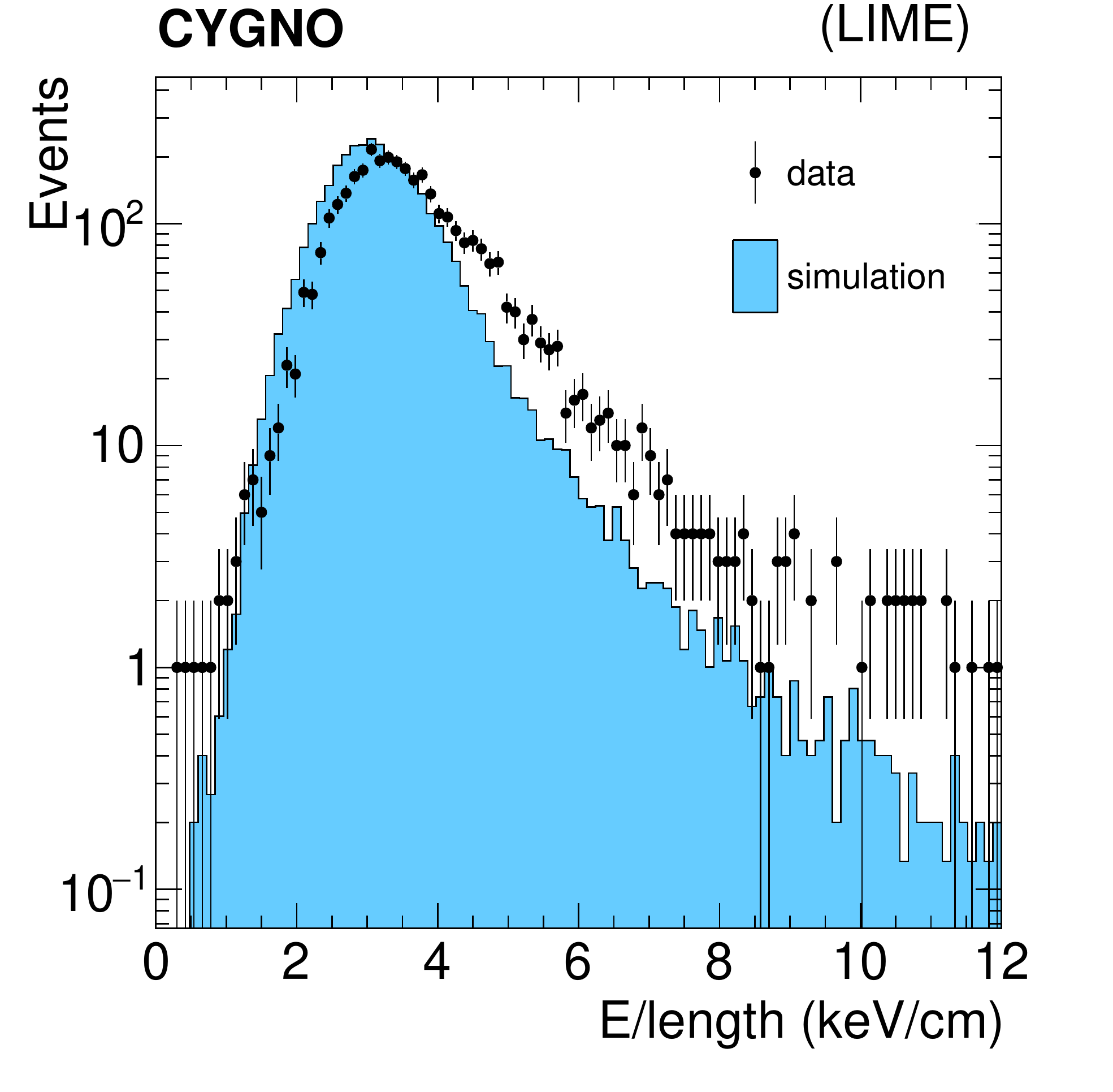}
\caption{Distributions of energy divide by the total length, of
  clusters identified as cosmic rays. Black points represent data, filled histogram represents a   Monte
  Carlo  simulated sample. \label{fig:cosmdedx}}
\end{figure}

 The track length of  the cosmic ray clusters reconstructed by the camera images  is in fact the $x$--$y$ projection  of the actual trajectory length in 3D of the cosmic ray particles. 
 Therefore a MC simulation of the interaction of cosmic rays with momenta in the range $[1-100]\,{\rm GeV}$  in the LIME active volume taking into account their  angular distribution has been carried out. A comparison of the specific ionization evaluated on  the data and MC for the cosmic rays is reported in   Fig.~\ref{fig:cosmdedx} showing a good agreement.

\section{Conclusion and perspective }
\label{sect:conclusion}

The search for  DM particles requires a vast experimental program with different strategies being put forward. A sensitivity to DM masses below 10 GeV  might be  useful to test alternative model to WIMPs. Experimental tools to infer  the DM direction would represent a powerful ingredient to reject background events in  the context of future  DM searches. The \cygno project aim at  demonstrating that a gaseous TPC with GEM amplification and optical readout,  operating  at atmospheric pressure with a He/CF$_4$ mixture might represent a viable candidate for a future generation of DM direct searches with directional sensitivity.

 In this paper we have fully described the calibration and reconstruction techniques developed for a 50 liters prototype  - named LIME - with a mass of 87 g in its active volume that represents 1/18    of a 1 m$^3$ detector. LIME was operated in an overground location at INFN LNF  with no shielding against environmental radioactivity. 

 With LIME we   studied the interaction of X-ray in the energy range from few keV to tens of keV with artificial radioactive source.  The use of a scientific CMOS camera with single photon sensitivity allowed to identify spots of light originated by the electron recoil energy deposit in the active gas volume.
 A very good linearity over two decades of energy was demonstrated with a $\approx 10$\% energy resolution thanks a regression algorithm exploiting at best all the topological information of the energy deposits.  A   position reconstruction was possible  in the plane transverse to the ionization electron drift  thanks to  the high granularity of the CMOS readout and  with an algorithm based on the  ionization electrons diffusion  to measure the longitudinal coordinate. 
 
 Moreover the absorption length of \fe X-ray was measured and found compatible with the expectation demonstrating a good control of the uniformity and efficiency of the detector. Also during a more than a week long data-taking a remarkable stability of the detector was achieved. 
 
  Cosmic rays were also easily identified and their specific ionization results very compatible with the usual prediction in gas. 
  
 An analysis of the events detected in absence of any artificial source showed that the detected  photon interaction rate (about 20 Hz) can be partly understood in terms of the ambient radioactivity. However given the long integration time (50 ms)  of the sCMOS camera the pile-up of interaction in the active volume can lead to an overestimate of the number of interaction.  This implies the necessity to operate LIME in a shielded environment as INFN LNGS with a tenfold reduction of the external background.  This will reduce to a negligible levle  the pile-up in images and will allow an assessment of the level of radiopurity of the materials used for  LIME. This measurements will be the basis for the design of   a future   \cygno DM detector. 

 In future a direct evaluation of the capability of LIME to identify nuclear recoils induced by neutron will be performed with dedicated calibration data-taking.
 Given the performance of LIME in reconstructing in details the topology of the energy deposit  a  very good nuclear recoil identification down to few keV is foreseen  \cite{coronello}. This will represent the fundamental element of a  competitive DM detector.


%
%

\begin{acknowledgements}
This project has received fundings under the European Union’s Horizon 2020 research and innovation programme from the European Research Council (ERC) grant agreement No 818744 and is supported by the Italian Ministry of Education, University and Research through the project PRIN: Progetti di Ricerca di Rilevante Interesse Nazionale “Zero Radioactivity in Future experiment” (Prot. 2017T54J9J).
We want to thank General Services and Mechanical Workshops of Laboratori Nazionali di Frascati (LNF) and Laboratori Nazionali del Gran Sasso (LNGS) for their precious work and L. Leonzi (LNGS) for technical support.
\end{acknowledgements}

\bibliographystyle{spphys}       
\bibliography{mybiblio}   

%
%

\end{document}